\documentclass[twoside,12pt]{article}
\usepackage{amssymb,amsmath}
\usepackage[dvipdfmx]{graphicx}
\usepackage{color}

\def\q{\vec q}

\newcommand{\be}{\begin{equation}}
\newcommand{\ee}{\end{equation}}
\newcommand{\bea}{\begin{eqnarray}}
\newcommand{\eea}{\end{eqnarray}}

\newcommand{\psiL}{\psi_{\text{L}}}
\newcommand{\psiR}{\psi_{\text{R}}}
\newcommand{\bpsiL}{\bar{\psi}_{\text{L}}}
\newcommand{\bpsiR}{\bar{\psi}_{\text{R}}}
\newcommand{\alphas}{\alpha_{\text{s}}}
\newcommand{\UA}{\mathrm{U(1)_A}}
\newcommand{\LQCD}{\Lambda_{\text{QCD}}}
\newcommand{\MeV}{\;\text{MeV}}
\newcommand{\GeV}{\;\text{GeV}}
\newcommand{\fm}{\;\text{fm}}

\newcommand{\calA}{\mathcal{A}}

\newcommand{\calL}{\mathcal{L}}

\newcommand{\calN}{\mathcal{N}}
\newcommand{\calO}{\mathcal{O}}

\newcommand{\muB}{\mu_{\text{B}}}
\newcommand{\muq}{\mu_{\text{q}}}

\newcommand{\MN}{M_{\text{N}}}
\newcommand{\Mq}{M_{\text{q}}}
\newcommand{\mmu}{m_{\text{u}}}
\newcommand{\mmd}{m_{\text{d}}}
\newcommand{\mms}{m_{\text{s}}}
\newcommand{\mq}{m_{\text{q}}}
\newcommand{\mn}{m_{\text{N}}}
\newcommand{\sigmaN}{\sigma_{\pi\text{N}}}

\newcommand{\Nc}{N_{\text{c}}}
\newcommand{\Nf}{N_{\text{f}}}
\newcommand{\Tc}{T_{\text{c}}}
\newcommand{\Td}{T_{\text{d}}}
\newcommand{\Th}{T_{\text{H}}}

\newcommand{\nF}{n_{\text{F}}}
\newcommand{\kF}{k_{\text{F}}}

\newcommand{\tr}{\text{tr}}
\newcommand{\Tr}{\text{Tr}}

\newcommand{\bx}{\boldsymbol{x}}

\newcommand{\boldm}{\boldsymbol{m}}
\newcommand{\bp}{\boldsymbol{p}}
\newcommand{\bq}{\boldsymbol{q}}
\newcommand{\br}{\boldsymbol{r}}
\newcommand{\bnabla}{\boldsymbol{\nabla}}
\newcommand{\bsigma}{\boldsymbol{\sigma}}
\newcommand{\bpi}{\boldsymbol{\pi}}
\newcommand{\btau}{\boldsymbol{\tau}}

\newcommand{\ukk}{u_{\text{KK}}}
\newcommand{\rhosym}{\rho_{\text{sym}}}
\newcommand{\rhobr}{\rho_{\text{broken}}}
\newcommand{\rhoq}{\rho_{\text{q}}}
\newcommand{\feyn}[1]{
  \setbox0=\hbox{\ensuremath{#1}}
  \hbox to\wd0{\hbox to0pt{\hbox to\wd0{\hss/\hss}\hss}\box0}}

\begin{document}
\title{\vspace{1cm} The phase diagram of nuclear and quark matter\\
 at high baryon density}
\author{Kenji Fukushima,$^1$ and Chihiro Sasaki$^2$\\
\\
$^1$Department of Physics, Keio University,
Kanagawa 223-8522, Japan\\
$^2$Frankfurt Institute for Advanced Science, J.W.\ Goethe University,\\
D60498 Frankfurt, Germany}
\maketitle

\begin{abstract}
 We review theoretical approaches to explore the phase diagram of
 nuclear and quark matter at high baryon density.  We first look over
 the basic properties of quantum chromodynamics (QCD) and address how
 to describe various states of QCD matter.  In our discussions on
 nuclear matter we cover the relativistic mean-field model, the chiral
 perturbation theory, and the approximation based on the large-$\Nc$
 limit where $\Nc$ is the number of colors.  We then explain the
 liquid-gas phase transition and the inhomogeneous meson condensation
 in nuclear matter with emphasis put on the relevance to quark matter.
 We commence the next part focused on quark matter with the bootstrap
 model and the Hagedorn temperature.  Then we turn to properties
 associated with chiral symmetry and exposit theoretical descriptions
 of the chiral phase transition.  There emerge some quark-matter
 counterparts of phenomena seen in nuclear matter such as the
 liquid-gas phase transition and the inhomogeneous structure of the
 chiral condensate.  The third regime that is being recognized
 recently is what is called quarkyonic matter, which has both aspects
 of nuclear and quark matter.  We closely elucidate the basic idea of
 quarkyonic matter in the large-$\Nc$ limit and its physics
 implications.  Finally, we discuss some experimental indications for
 the QCD phase diagram and close the review with outlooks.
\end{abstract}
\tableofcontents

\section{Introduction}
\label{sec:intro}

The existence of our world as it stands today relies on peculiar
properties of nuclei and ultimately the dynamics of quarks and gluons
in quantum chromodynamics (QCD) at the microscopic level.  Research on
nuclear and quark matter at high baryon (quark) density is expected to
anchor our empirical understanding of the origin of matter in the
Universe to a more fundamental language.  Without strong medium
effects quarks and gluons are confined inside of protons and neutrons
(or hadrons in general) and this property is generally called ``color
confinement,'' which is a consequence of non-perturbative and
non-linear dynamics of QCD.\ \ Another important feature of QCD is the
generation of dynamical mass due to a condensate of quarks and
anti-quarks, i.e.\ the chiral condensate.  Theoretical understanding
of color confinement and mass generation is one of the unanswered
challenges in modern physics.

In extreme environments such as the high temperature $T$, the high
baryon density $\rho$, strong external fields, etc, the color
confinement and/or the dynamical mass may be lost and a new state of
matter out of quarks and gluons, namely, the quark-gluon plasma (QGP)
could be formed.  The relativistic heavy-ion collision experiments
have aimed to create QGP in the laboratory and it is almost doubtless
that Relativistic Heavy-Ion Collider (RHIC) at Brookhaven National
Laboratory discovered QGP at high enough $T$ and Large Hadron Collider
(LHC) at CERN confirmed it at higher energy.  With the axis of the
baryon chemical potential $\muB$ in addition to $T$, one can draw the
QCD phase diagram in the $\muB$-$T$ plane.  The beam-energy-scan of
the heavy-ion collision is expected to explore the QCD phase diagram
experimentally.  The Facility for Antiproton and Ion Research (FAIR)
at GSI and the Nuclotron-based Ion Collider Facility (NICA) at JINR
are under construction to investigate the baryon-rich state of QCD
matter as well as RHIC at lower collision energies.  The theoretical
understanding of the phase diagram, on the other hand, appears
stalled.  The purpose of this review is not to cover as many topics as
possible on the whole phase diagram but to assemble theoretical
discussions in the high density region accessible by the experiment.

The study of QGP intrinsic properties is an interesting subject on its
own, and furthermore, we could have inferred a deeper insight to
confinement and mass generation mechanisms by approaching them not
only from the vacuum but from the QGP side.  From the theoretical
point of view, it is quite non-trivial how to address ``confinement''
in the language of the quantum field theory, while it is possible to
formulate ``quark confinement'' unambiguously in finite-$T$ QCD with
quarks made infinitely heavy (quench limit).  This implies that the
``order parameter'' of quark confinement cannot have a strict meaning
in the presence of light quarks and that of ``gluon confinement''
cannot be given in a simple way.  In a gluonic medium gluons are
screened by themselves which makes the meaning of confinement blurred.
Mathematically speaking, quark confinement is well-defined only for a
static color-charge in the fundamental representation, but gluons are
associated with a color-charge in the adjoint representation.
Indeed, it has been a long standing problem how to find a rigorous
characterization of confinement and deconfinement for arbitrary
systems with dynamical quarks and gluons.  The strict order parameter
for color confinement is, if any, still unknown, and it could be even
conceivable to interpret the lack of the order parameter as indicating
that confinement and deconfinement are connected.  As we will see
later, such a connection through smooth crossover between confinement
and deconfinement may make it possible to approximate the QCD
thermodynamics near crossover in terms of the hadronic degrees of
freedom alone or of the quasi-particles of quarks and gluons

As a matter of fact, it is still a challenging problem to extract
analytical information directly from theory in order to investigate
QCD matter even at asymptotically high temperature and/or baryon
density.  The running coupling constant, $\alphas=g^2/(4\pi)$, of the
strong interaction becomes smaller at high energies owing to the
asymptotic freedom, and this seems to suggest that confinement might
be lost at high enough $T$ or $\muB$.  Then quark matter could have
been realized in high-density environments such as inner cores of the
neutron star~\cite{Itoh:1970uw,Collins:1974ky}.  For the purpose to
investigate QCD matter at high $T$ and $\muB$ perturbative methods
have been developed (see
Refs.~\cite{kapusta2006finite,bellac2000thermal} for textbooks).  It
has been understood by now, however, that the perturbative expansion
of QCD thermodynamics breaks down badly at the ultra-soft magnetic
scale $\sim\calO(g^2 T)$.  One can immediately pin down the source of
this incapability.  Only the Matsubara zero-mode is dominant in the
high-$T$ limit and the dimensional reduction
occurs~\cite{Appelquist:1981vg}, so that hot QCD is translated into
magnetostatic QCD (MQCD), i.e.\ three-dimensional QCD with the
magnetic coupling constant, $g_{\rm M}^2=g^2 T$.  Because such
three-dimensional QCD is a confining theory, non-perturbative
information should be required for the QCD thermodynamics even in the
high-$T$ limit.  Linde's problem of infrared
catastrophe~\cite{Linde:1978px,Linde:1980ts} is the most typical
manifestation of the breakdown of the perturbation theory, and the
magnetic screening mass, $m_{\rm M}\sim g^2 T$, should be generated
non-perturbatively from MQCD (see Ref.~\cite{Zwanziger:2006sc} for a
concrete evaluation of the magnetic mass in a confining model).  The
introduction of $\muB$ would not make this situation better.  The
magnetic mass is as vanishing perturbatively, which is seen concisely
in the form of the hard-dense-loop (HDL) effective
action~\cite{Manuel:1995td}.

We note that the perturbation theory can be successful in a
color-superconducting (CSC) phase~\cite{Barrois:1977xd,Bailin:1983bm}
in which the unscreened transverse gluon enhances the gap energy
significantly~\cite{Son:1998uk} and also the gap energy lead to
interesting confining properties~\cite{Rischke:2000cn}.  Although CSC
has been an important element of the QCD phase diagram since two
seminal papers appeared~\cite{Alford:1997zt,Rapp:1997zu}, we will
restrict the density region considered in this review before CSC is
turned on.  This is partially because it is unlikely to detect CSC in
the heavy-ion collision experiment and partially because the phase
structure involving CSC (including inhomogeneous states) is sensitive
to model uncertainties at intermediate density and it is difficult to
identify the robust part from such model dependent results.
Interested readers can consult Ref.~\cite{Alford:2007xm} for general
features of CSC, Refs.~\cite{Rischke:2003mt,Fukushima:2010bq} for CSC
in the context of the phase diagram, and Ref.~\cite{Anglani:2013gfu}
for inhomogeneous CSC states.

Although QCD has been a well-established theory, there is no reliable
way to obtain any information on high-density QCD directly.  The most
powerful non-perturbative method, i.e.\ numerical Monte-Carlo
simulation on the lattice, is of no practical use for the system at
low-$T$ and high-$\muB$ because of the notorious sign problem.  The
Dirac determinant takes a complex value at $\muB\neq0$ and neither its
real nor imaginary part is positive semi-definite.  The Monte-Carlo
simulation based on the importance sampling therefore breaks down (see
Refs.~\cite{Muroya:2003qs,Fukushima:2010bq} for various ideas to evade
the sign problem and obstacles).  There are many QCD-like models
designed to mimic some part of QCD dynamics but we should keep in mind
that each model has its validity limit.  Nevertheless, it should be
feasible to reach a consistent picture by collecting various pieces of
knowledge from different approaches.  This is actually, if not the
best, the only possible strategy to construct the most presumable
scenario for the QCD phase structure, unless the sign problem will be
resolved.  In this sense, we believe, it should be of paramount
importance to gather known results widely from nuclear matter, quark
matter, and also a new paradigm in between, namely, quarkyonic
matter~\cite{McLerran:2007qj}.  Because the notion of quarkyonic
matter is relatively new as compared to traditional nuclear and quark
matter, we shall pay attention to its definition and formulate it in a
form of the ``McLerran-Pisarski conjecture'' aiming to clarify its confusing and
sometimes misunderstood interpretation.

To make this review article as self-contained as possible, we begin
with basics of QCD in Sec.~\ref{sec:QCD} and explain its global
symmetries.  Some of QCD symmetries are spontaneously broken or
restored depending on external parameters such as $T$ and $\muB$,
which determines the location of the phase boundary on the $\muB$-$T$
plane.  In the theoretical research on the QCD phase diagram,
\textit{center symmetry} in the gauge sector and
\textit{chiral symmetry} in the flavor sector play the most important
role.  In particular, if the phase transition is of second order, the
classification according to the universality is at excellent work to
make model-independent statements for the physical properties in the
vicinity of the critical point.  The universality argument predicts
that, when the temperature is raised, the deconfinement phase
transition is expected to be of first order if the quark mass is
infinitely heavy, while the chiral phase transition should be of
second order (first order) with two (three, respectively) massless
flavors.  In reality the quark masses are non-zero and neither center
symmetry nor chiral symmetry is exact.  We overview some highlights of
the recent lattice-QCD results to find that QCD undergoes crossover
for deconfinement and chiral restoration nearly simultaneously.

We then proceed to the discussions on nuclear matter in
Sec.~\ref{sec:nuclear}.  For the instructive purpose we revisit the
calculations using a relativistic mean-field model, where most of the
technical procedures are analogous to the mean-field treatment of
quark matter.  In fact, contemporary challenges in the quark matter
research, i.e.\ the possibilities of the QCD critical point and the
inhomogeneous condensates, were studied long ago in the context of
nuclear matter, and it should be useful to flash back those
theoretical considerations.  We note, at the same time, that nuclear
matter is a more complicated environment than quark matter;  for
example, it is not obvious at all how chiral symmetry should be
assigned in terms of hadronic degrees of freedom.  In other words, it
is an unanswered question what kind of mesons and baryons would become
lighter if chiral symmetry is (partially) restored.  We elaborate
theoretical background and some possible scenarios.  In the final part
of Sec.~\ref{sec:nuclear} we introduce an approximation based on
infinitely large number of colors ($\Nc\to\infty$), which is a crucial
ingredient for the correct understanding of quarkyonic matter.  In
this context we shall follow holographic nuclear physics according to
Ref.~\cite{Bergman:2007wp} and explain the most promising holographic
QCD model called the Sakai-Sugimoto model~\cite{Sakai:2004cn}.  In
this review we will take a close look at the derivation of the phase
diagram from this model since the Sakai-Sugimoto model is a genuine
QCD dual having the correct physical degrees of freedom and the chiral
symmetry.  Besides, this holographic QCD model is so unique that it
can describe baryonic and quark matter within the single framework.
Any other (and more conventional) chiral models are not quite
successful in incorporating baryonic and quark matter in a unified
way.  Section~\ref{sec:SSM} may look a bit technical with many
equations since we have tried not to skip details too much for
convenience for readers who would intend to go beyond a sketchy
knowledge.

In Sec.~\ref{sec:quark} we consider the phase diagram beyond the
territory of nuclear physics.  The very first prototype of the QCD
phase diagram was drawn by Cabibbo and Parisi in 1975 based on the
physical interpretation of the Hagedorn limiting
temperature~\cite{Cabibbo:1975ig}.  Since the exponentially growing
spectrum of hadrons plays an important role in the interpretation of
experimental data and the nature of deconfinement crossover, we take a
quick look at the statistical bootstrap model and the derivation of
the Hagedorn spectrum.  Then, we discuss the fate of chiral symmetry
adopting well-used chiral models --- the Nambu--Jona-Lasinio (NJL)
model and the quark-meson (QM) model.  These chiral models are
designed to give rise to the spontaneous breaking of chiral symmetry
and can properly reproduce critical phenomena if the chiral phase
transition is of second order.  Furthermore, the effect of the
deconfinement crossover can be partially taken into account in the
extended version of these chiral models.  We next elucidate the
underlying physics of the liquid-gas transition and the critical point
in a parallel way to the previous section on nuclear matter.  Also, we
develop a qualitative argument on the mechanism to induce
inhomogeneous condensates such as the chiral spiral configurations.

The last part of Sec.~\ref{sec:quarkyonic} is devoted to the
clarification of the characterization and the properties of quarkyonic
matter~\cite{McLerran:2007qj}.  To understand quarkyonic matter
correctly, we need to know how nuclear matter should behave in the
large-$\Nc$ limit.  Quarkyonic matter is, in a sense, not a new state
of matter but just a possible view of large-$\Nc$ nuclear matter.
There are several theoretical proposals for the identification of
quarkyonic matter, but what really features quarkyonic matter is the
strength of inter-baryon interactions.  It is not quite
straightforward to give a clear definition of quarkyonic matter in
real QCD with $\Nc=3$.  The intuitive interpretation would lead us to
a picture of quarkyonic matter as (\textit{baryonic}) matter whose
pressure is sustained mainly by \textit{quarks}.  This explains how a
nomenclature, quarkyonic (= quark + baryonic), makes sense.

In Sec.~\ref{sec:experiment} we make a quick overview of what has been
confirmed in the relativistic heavy-ion collision experiment.  The
so-called beam-energy scan over the QCD phase diagram is still ongoing
and there are already many suggestive results available.  We will not
(and cannot) cover all of interesting data.  In addition to the
heavy-ion collision data, neutron star physics has brought useful
information on the equation of state (EoS) of dense
matter~\cite{Demorest:2010bx};  the discovery of the heaviest neutron
star with almost two solar-mass enabled us to dig out possible EoS's.
It is, however, still an open question whether quark matter may exist
in the neutron star.  The repulsive vector interaction could make the
EoS hard enough to be consistent with such a heavy neutron star.
Interestingly this issue of the vector interaction is closely related
to the question of the QCD critical point.

We close this review with outlooks in Section~\ref{sec:outlook}.  In
theory, as long as we cannot be equipped with any versatile tool, we
must continue making a patchwork of various approaches that complement
each other.  Recent theoretical works indeed suggest that the
baryon-rich state of matter may have rich contents than believed, but
there are always counter-arguments that would favor less structures.
We illustrate two possible scenarios with and without the first-order
phase boundary and emphasize that the QCD phase diagram can be still
non-trivial enough even without the first-order phase transition and
the QCD critical point.

\section{Symmetries of the Strong Interaction}
\label{sec:QCD}

The aim of this review is to discuss the phase diagram of nuclear and
quark matter with solid boundaries associated with first-order,
second-order phase transitions, or rather vague borders of smooth
crossover.  In most cases of the phase transition the manifestation of
the global symmetries changes in accord with the state of matter.  In
this section we summarize the global symmetries of the strong
interaction based on the fundamental theory.

QCD is a non-Abelian gauge theory with $\Nc$ colors and $\Nf$ flavors.
It is known up to now that $\Nc=3$ and $\Nf=6$ in nature.  Since only
\textit{up}, \textit{down}, and \textit{strange} quarks are relevant
to the thermodynamic properties at the QCD energy scale,
$\LQCD\sim 200\MeV$, we will limit ourselves to the case with two
light and one heavy flavors, which is commonly denoted as
``(2+1) flavors'' in the convention.  Furthermore, the system with
(2+1) flavors is sometimes approximated by only 2 light flavors, which
may look like a crude approximation, but can be absolutely legitimate
to investigate the (pseudo) critical phenomena governed by the softest
modes only.  It is also a useful limit to take $\Nf=0$ or to make all
quarks infinitely heavy (though the $\Nf\to0$ limit should be
carefully taken~\cite{Stephanov:1996ki}).  Such a limit is often
called the quench approximation, and utilized in the lattice-QCD
simulation.  In the quench limit the system of the pure Yang-Mills
theory has no excitation of dynamical quarks but only gluon loops are
allowed.

The Lagrangian density of QCD consists of the pure Yang-Mills part,
the quark part, and the CP-violating part, respectively, i.e.
\begin{equation}
 \calL = -\frac{1}{4}F_{\mu\nu}^a F^{\mu\nu a} +\bar{\psi}_{\alpha}^i
  (i\gamma^\mu\partial_\mu \delta_{ij} + g\gamma^\mu A_\mu^a T^a_{ij}
  -m_\alpha\delta_{ij})\psi_{\alpha}^j + \theta \frac{g^2}{32\pi^2}
  \widetilde{F}_{\mu\nu}^a F^{\mu\nu a} \;.
\label{eq:QCD_Lag}
\end{equation}
Here, $a$ refers to the adjoint color index from $1$ to $\Nc^2-1$ and
$i,j$ refer to the fundamental color indices from $1$ to $\Nc$
(namely, in the $\Nc=3$ case, $1=$\textit{red}, $2=$\textit{green},
$3=$\textit{blue}, particularly).  Also, $\alpha$ is the flavor index
($1=$\textit{up}, $2=$\textit{down}, $3=$\textit{strange}, etc).
The field strength tensor is
$F_{\mu\nu a}=\partial_\mu A_{\nu a}-\partial_\nu A_{\mu a}
+gf^{abc}A_\mu^b A_\nu^c$ and its dual is defined as
$\widetilde{F}^{\mu\nu a}=\frac{1}{2}
\varepsilon^{\mu\nu\rho\sigma}F_{\rho\sigma}^a$ with the convention
$\varepsilon^{0123}=1$.  This Lagrangian density involves 5 parameters
for $\Nc=3$ and $\Nf=3$.  The latest estimate for the current quark
masses is;
\begin{equation}
 \mmu = 2.3^{+0.7}_{-0.5}\MeV\;,\quad
 \mmd = 4.8^{+0.7}_{-0.3}\MeV\;,\quad
 \mms = 95\pm 5\MeV\;,
\end{equation}
in a mass-independent subtraction scheme at a scale $\sim
2\GeV$~\cite{Beringer:1900zz}.  The strong coupling constant,
$\alphas$, is one of the physical constants and it runs with the
energy scale.  The world average value at present
is~\cite{Beringer:1900zz},
\begin{equation}
 \alphas(m_Z) = 0.1184(7)\;.
\end{equation}
The CP-violating part in Eq.~\eqref{eq:QCD_Lag} originates from the
structure of the $\theta$-vacuum and there is no reason why $\theta$
should be vanishingly small.  The experiment for the neutron electric
dipole moment yields an upper limit,
$|\theta|<10^{-10}$~\cite{Baker:2006ts}, and $\theta$ is so far
consistent with zero.  It is still a big mystery in theoretical
physics whether $\theta\simeq0$ is just an accidental fact or a
consequence from some unknown dynamics such as axions.  Because
$\theta$ is such small, the CP-violating term does not have a
phenomenological impact in normal circumstances.  In far
non-equilibrium situations, however, $\theta$ may take a non-zero
value locally (corresponding to pseudo-scalar condensation) and then
the strong CP-violation could be possibly detectable in the
relativistic heavy-ion
collision~\cite{Kharzeev:2007jp,Fukushima:2008xe}.  Furthermore, even
though $\theta$ itself is vanishing, the energy curvature with respect
to $\theta$ (i.e.\ the topological susceptibility) is non-zero and
closely related to the $m_{\eta'}$
mass~\cite{Witten:1979vv,Veneziano:1979ec}.  Thus, the in-medium
$m_{\eta'}$ mass could provide us with information on the topological
structure of the QCD vacuum.

\subsection{Gauge symmetry}

The QCD Lagrangian density is invariant under the gauge transformation
by construction.  The gauge transformation changes the quark and the
gluon fields, respectively, as
\begin{equation}
 \psi \to V \psi \;,\qquad
 A_\mu \to V\Bigl(A_\mu - \frac{1}{ig}\partial_\mu\Bigr)V^\dagger \;,
\label{eq:gauge_transform}
\end{equation}
where $V\in \mathrm{SU}(\Nc)$ (i.e.\
$V\cdot V^\dagger=V^\dagger\cdot V=1$ and $\det V=1$).  The local gauge
symmetry is never broken spontaneously owing to Elitzur's
theorem~\cite{Elitzur:1975im}, while the global one can be which is
realized in superconductivity leading to the Meissner mass for the
gauge bosons as a result of the
Englert-Brout-Higgs-Guralnik-Hagen-Kibble
mechanism~\cite{Englert:1964et,Higgs:1964ia,Guralnik:1964eu}.  The
Meissner effect on gluons plays an important role in color
superconductivity, but it is not the main subject of this article.

\subsubsection{Center symmetry}

A particularly important part of the gauge symmetry at finite $T$ is
center symmetry that characterizes deconfinement of quark degrees of
freedom in a gluonic medium.  In the pure Yang-Mills theory without
quark field $\psi$, the genuine gauge symmetry is not
$\mathrm{SU}(\Nc)$ but $\mathrm{SU}(\Nc)/\mathrm{Z}_{\Nc}$.  One can
understand this from Eq.~\eqref{eq:gauge_transform} by choosing
$V=z_k=e^{2\pi i k/\Nc}\times{\boldsymbol{1}_{\Nc\times\Nc}}$ where
$e^{2\pi i k/\Nc}~(k=0,\dots,\Nc-1)$ is an element of
$\mathrm{Z}_{\Nc}$ (i.e.\ $(e^{2\pi i k/\Nc})^{\Nc}=1$).  This $z_k$
is certainly an element of $\mathrm{SU}(\Nc)$ and commutes with all
elements of $\mathrm{SU}(\Nc)$.  In other words, $z_k$ belongs to the
center subgroup of $\mathrm{SU}(\Nc)$ that is nothing but
$\mathrm{Z}_{\Nc}$.

Let us focus on the pure Yang-Mills part first and put aside the quark
field $\psi$ for the moment.  At finite $T$ the imaginary-time
direction is compact with a period $\beta=1/T$.  To maintain the
periodicity of the gauge field, $V$ should be a pseudo-periodic
function with a twist at most by $z_k$, that is;
$V(\bx,\tau+\beta)=z_k V(\bx,\tau)$.  For a concrete example, one can
choose $V$ as
\begin{equation}
 V(\tau) = \text{diag}(e^{2\pi i k\tau/\Nc\beta},
  e^{2\pi ik\tau/\Nc\beta},\dots,e^{-2\pi i(\Nc-1)k\tau/\Nc\beta}) \;,
\end{equation}
where the first $(\Nc-1)$ elements are identical and the last one is
chosen to satisfy $\det V(\tau)=1$ for any $\tau$.  This $V(\tau)$
obviously belongs to $\mathrm{SU}(\Nc)$ and satisfies
$V(\tau+\beta)=z_k V(\tau)$.  Likewise other forms with
$e^{-2\pi i(\Nc-1)k\tau/\Nc\beta}$ permutated with different diagonal
components can also be the center transformation.

The point is that the QCD Lagrangian density is invariant under
general gauge transformation~\eqref{eq:gauge_transform}, but the
boundary condition of the manifold on which the theory is defined may
not.  (One may prefer to adhere the general gauge
transformation~\eqref{eq:gauge_transform}, and it is indeed possible
to change the prescription of the imaginary-time formalism; see
Ref.~\cite{James:1990it} for this example.)  Center symmetry at finite
$T$ is thus the gauge subsymmetry with an aperiodicity by $z_k$.

Center symmetry controls the behavior of the temporal Wilson line or
the Polyakov loop.  We will use the following notation throughout this
article;
\begin{equation}
 L(\bx) \equiv \mathcal{P} e^{ig\int_0^\beta d\tau\, A_4(\bx,\tau)} \;,\qquad
 \ell(\bx) \equiv \frac{1}{\Nc}\tr L(\bx) \;,\qquad
 \Phi \equiv \langle \ell \rangle \;,
\label{eq:Polyakov}
\end{equation}
where $L(\bx)$ is called the Polyakov loop, $\ell(\bx)$ is the traced
Polyakov loop, and $\Phi$ represents the Polyakov loop expectation
value.

For the physical interpretation, $\Phi$ is usually considered
to be related to a single-quark free energy $f_{\text{q}}(\bx)$ as
$\Phi\propto \exp[-\beta f_{\text{q}}(\bx)]$, but there are
theoretical subtleties on this (see
Refs.~\cite{Fukushima:2011jc,Smilga:1996cm} for details).  The
Polyakov loop correlation function is related to the heavy quark
potential $f_{\bar{q}q}(r)$ (in the singlet channel) in the following
way;
\begin{equation}
 \langle \ell^\dagger(\bx_1) \ell(\bx_2) \rangle = C
  \exp[- \beta f_{\bar{q}q}(|\bx_1-\bx_2|)] \;.
\end{equation}
The inter-quark potential is an important measure to characterize
whether the system is in the confined phase or in the deconfined
phase.  The quark confined phase should have a linearly rising
potential, $f_{\bar{q}q}(r) = \sigma r$ with a finite string tension
$\sigma$, so that the Polyakov loop correlation function decays
exponentially at large separation.  In the deconfined phase, on the
other hand, the inter-quark potential is thermally screened and
$f_{\bar{q}q}(r\to\infty) \to \text{(const.)}$ leading to a
non-vanishing correlation function of the Polyakov loop.  In summary,
in the pure Yang-Mills theory, when we take the limit of
$|\bx_1-\bx_2|\to\infty$, the behavior of the correlation function
should be
\begin{equation}
 \begin{split}
 \text{(Confined Phase)}~~~ &\qquad
  \langle \ell^\dagger(\infty) \ell(0) \rangle = 0 \quad\rightarrow\quad
  \Phi = 0 \qquad \text{(Symmetric Phase)}\;,\\
 \text{(Deconfined Phase)} & \qquad
  \langle \ell^\dagger(\infty) \ell(0) \rangle \neq 0 \quad\rightarrow\quad
  \Phi \neq 0 \quad ~~~ \text{(Broken Phase)}\;,
 \end{split}
\label{eq:phases}
\end{equation}
where we inferred $\Phi$ postulating the clustering decomposition,
$\langle\ell^\dagger(\infty)\ell(0)\rangle\to|\langle\ell\rangle|^2$.

It is easy to make sure that $\Phi$ is an order parameter for the
spontaneous breaking of center symmetry, as labeled in
Eq.~\eqref{eq:phases}.  Under the general gauge transformation $V$,
the Polyakov loop changes as $L(\bx)\to V(\beta)L(\bx)V^\dagger(0)$.
If the physical state bears symmetry under the gauge transformation
that satisfies $V(\beta)=z_k V(0)$, the Polyakov loop expectation
value transforms as
\begin{equation}
 \Phi \;\to\; e^{2\pi ik/\Nc} \Phi \;,
\end{equation}
that means that $\Phi=0$ is concluded.  In other words, the
center-symmetric state corresponds to the confined phase, whilst the
deconfined phase is accompanied by the spontaneous breaking of center
symmetry.

\begin{figure}
 \begin{center}
 \includegraphics[width=0.5\textwidth]{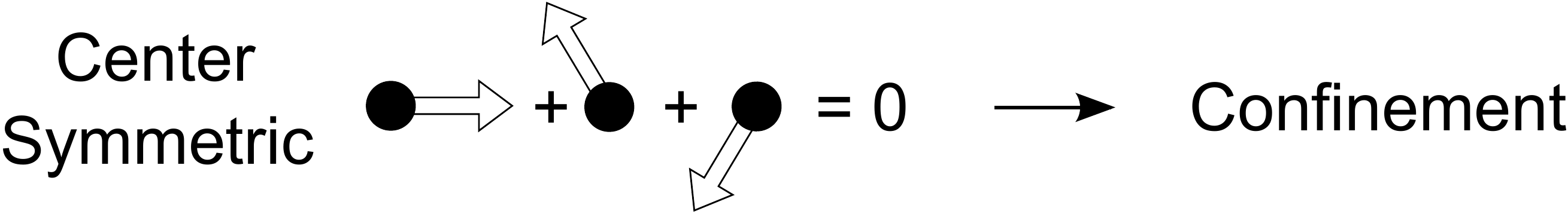}
 \caption{Schematic picture of quark confinement in a gluonic medium.
 The color orientation is randomly distributed and the average is
 vanishing which is interpreted as the prohibition of single-quark
 excitation.}
 \label{fig:polyakov}
 \end{center}
\end{figure}

In the definition according to the finite-temperature field theory,
the Polyakov loop might look a mystical quantity.  Intuitively, the
Polyakov loop represents the screening factor of the fundamental color
charge in a gluonic medium.  Contrary to the na\"{i}ve picture, the
confined phase is a highly disturbed state as exemplified in the
strong coupling expansion on the lattice~\cite{Wilson:1974sk}, and
there, the color does not have any preferred direction.  Then all
different $\mathrm{Z}_{\Nc}$ sectors equally appear and the thermal
weight for quark excitation in a gluonic medium picks up all
$\mathrm{Z}_{\Nc}$ phase factors, that leads to
$\sum_k e^{2\pi i k/\Nc}=0$.  The single-quark excitation is thus
screened in the confined phase as a result of symmetric average in
color space.  In the $\Nc=3$ case, for instance, this confining
situation is illustrated in Fig.~\ref{fig:polyakov}.

In the above, we have considered only the clean environment without
dynamical quarks, but the situation becomes drastically different
once light quarks are included.  It is evident from
Eq.~\eqref{eq:gauge_transform} that center symmetry is lost because
the boundary condition of quark field $\psi$ directly reflects the
aperiodicity by $z_k$.  Hence, the Polyakov loop $\Phi$ is only an
approximate order parameter for deconfinement and it always takes a
non-zero value regardless of the state of matter.  Nevertheless,
$\Phi$ effectively works well to indicate crossover behavior as a
function of $T$ as we will see in Sec.~\ref{sec:highlights}.

We should emphasize that it is a highly non-trivial statement that
$\Phi$ or any other operator cannot be an order parameter for
deconfinement.  Despite tremendous amount of theoretical efforts,
there has been no way found to construct an exact order parameter for
deconfinement in the presence of dynamical quarks, and all proposed
candidates turned out to be unsuccessful (see
Refs.~\cite{Detar:1982wp,Meyer:1983hm} for several examples).  One
logical interpretation for the lack of the order parameter seems to be
the absence of strict distinction between confinement and
deconfinement.  In other words, for any $T\neq0$ in principle, the
probability to find colored excitations in a thermal bath may not be
strictly zero (though exponentially small like
$\sim e^{-\beta f_{\text{q}}}$).  This may sound like a radical idea,
but otherwise, one should question the theoretical framework of
finite-$T$ QCD itself.

\subsubsection{Deconfinement phase transition and critical phenomena}
\label{sec:deconf}

The phase transition is described by the dynamics in terms of the
order parameter and soft modes in general.  Concerning the
deconfinement phase transition in the absence of dynamical quarks, an
effective theory for the traced Polyakov loop $\ell(\bx)$ (or one can
take higher-dimensional representations into
account~\cite{Pisarski:2000eq}) is useful.  The effective theory can
be generally expanded as
\begin{equation}
 \Gamma^{\text{(glue)}} = \int d^d x\, \Bigl[ (\bnabla \ell)^2
  + c_2 |\ell|^2 + c_4 |\ell|^4 + c_{\Nc} \text{Re}(\ell^{\Nc})
  + \cdots \Bigr]
\label{eq:Seff}
\end{equation}
with $T$-dependent coefficients $c_i(T)$.  The fourth term is implied
by $\mathrm{Z}_{\Nc}$ center symmetry.  The effect of dynamical quarks
can be implemented as discussed in Sec.~\ref{sec:NJL}.  The above
effective action is supposed to capture the deconfinement phase
transition of the ($d$+1)-dimensional pure Yang-Mills theory at finite
temperature with $d$ being the number of spatial dimensions.

\begin{figure}
 \begin{center}
 \includegraphics[width=0.5\textwidth]{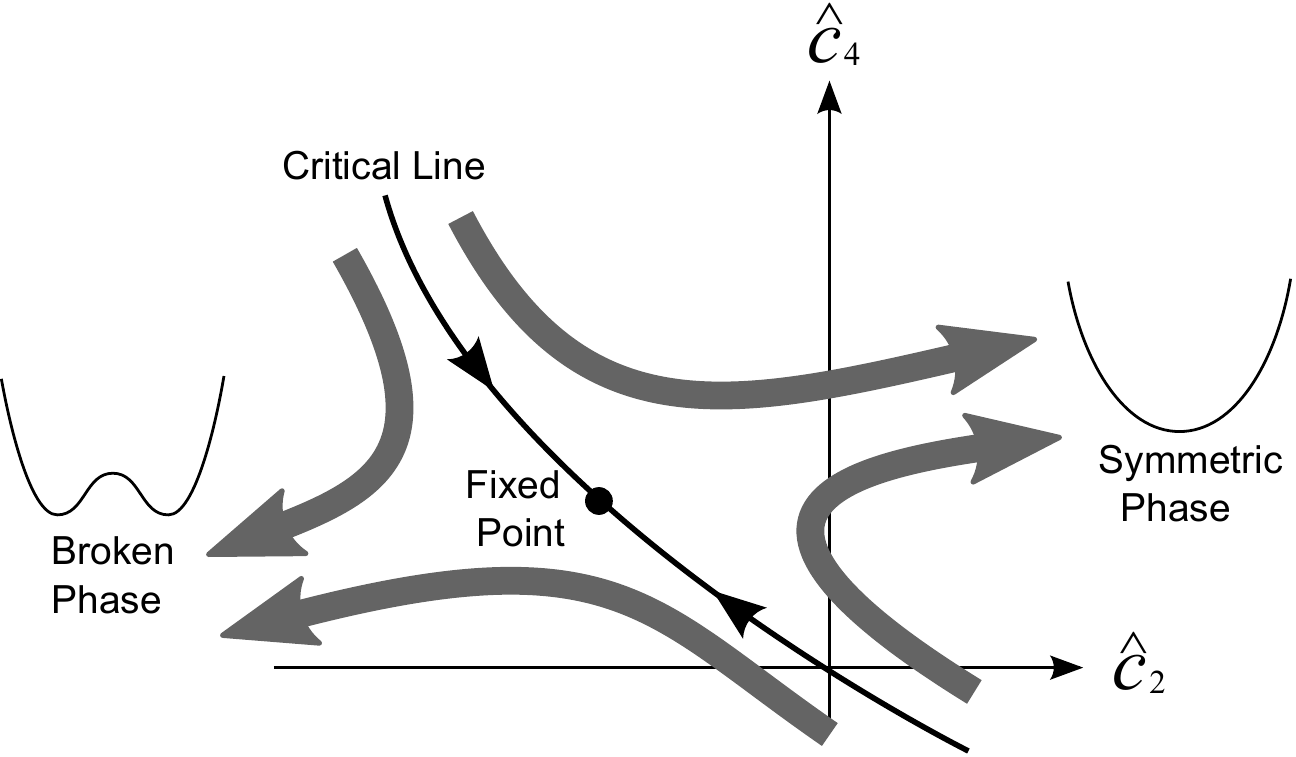}
 \end{center}
 \caption{Schematic diagram of the RG flow in the dimensionless
   parameter space that characterizes the shape of the effective
   potential.  The critical line defines the phase transition, on
   which the flow runs into the Wilson-Fisher fixed point.}
 \label{fig:flow}
\end{figure}

The order of the phase transition and the critical phenomena are
characterized by the behavior of the effective action or the flow of
the coefficients $c_i$ in Eq.~\eqref{eq:Seff}.  Let us consider a
coarse-grained action $\Gamma_k^{\text{glue}}$ at some momentum scale
$k$, which is defined by an effective action with the quantum and
thermal fluctuations above $k$ integrated out (see also
Sec.~\ref{sec:quark-meson} for a theoretical framework).  The full
effective action is therefore retrieved in the $k\to0$ limit.  Then,
naturally, the coefficients are dependent on $k$ as well as $T$, that
is, $c_i(T;k)$.  To take account of the rescaling process in the
renormalization group (RG) equation, we make these coefficients
dimensionless using $k$ and denote them as $\hat{c}_i$.

If the phase transition is of second order, with decreasing $k$, these
dimensionless $\hat{c}_i$'s flow toward an infrared (IR) fixed point
along the critical line (or hypersurface in general), as schematically
illustrated in Fig.~\ref{fig:flow}.  As $T$ changes, the initial point
of the RG flow moves, and when it crosses the critical hypersurface,
the destination of the flow drastically changes from the broken phase
(with $\hat{c}_2\to-\infty$) to the symmetric phase (with
$\hat{c}_2\to+\infty$).  If the initial point sits exactly on the
critical hypersurface, the flow heads for the IR (Wilson-Fisher) fixed
point that describes the second-order critical point, and the critical
phenomena are uniquely determined by the flow pattern near the fixed
point.  That is, according to the general theory of critical
phenomena, the critical behavior associated with the second-order
phase transition (approaching $\Tc$ from the ordered phase) is
characterized by
\begin{equation}
 \Phi \propto \langle\ell(\bx)\rangle \propto t^\beta \;\;,\quad
 \chi \propto t^{-\gamma} \;,\quad
 C \propto t^{-\alpha} \;,\quad
 \xi \propto t^{-\nu} \;,\quad
 \langle\ell(0)\ell(\bx)\rangle \propto \frac{1}{|\bx|^{d-2+\eta}}
\end{equation}
as a function of the reduced temperature, $t=|T-\Tc|/\Tc$.  The
critical behavior of the order parameter is specified by $\beta$, the
susceptibility $\chi$ by $\gamma$, the specific heat $C$ by $\alpha$,
and the correlation length $\xi$ by $\nu$.  The two-point spatial
correlation function $\langle\ell(0)\ell(\bx)\rangle$ has an anomalous
dimension $\eta$ at $T=\Tc$.  These critical exponents are not all
independent but they satisfy the scaling relations;
\begin{equation}
 \begin{tabular}{lp{1em}l}
  (Rushbrooke scaling) && $\alpha+2\beta+\gamma=2$ \;,\\
  (Fisher scaling)     && $\gamma=(2-\eta)\nu$ \;,\\
  (Josephson scaling)  && $d \nu = 2-\alpha$ \;.
 \end{tabular}
\end{equation}
It has been established that the properties near the IR fixed point
can be classified with the global symmetry of the theory and the
spatial dimensions $d$ and this idea of the classification is commonly
referred to as the \textit{universality}.  In scalar theories in
particular, $d=4$ is known as the critical dimension.  The IR fixed
point moves to the origin when $d\to 4$, and eventually merges with
a ultraviolet (UV) fixed point right at $d=4$.  When $d>4$, the fixed
point goes across the origin and is hidden in the unstable and
unphysical region with $\hat{c}_4<0$.  Then, the flow goes into the
Gaussian fixed point at the origin that describes a free theory, so
that the critical exponents take the classical (mean-field) values
(see Tab.~\ref{tab:SY}).

According to the universality argument, the deconfinement phase
transition in ($d$+1) dimensions can be analyzed by a scalar theory
sharing the same global symmetry (i.e.\ $\mathrm{Z}_{\Nc}$ center
symmetry) in $d$ dimensions~\cite{Svetitsky:1982gs,Svetitsky:1985ye}.
The critical exponents can be deduced in a simpler theory than the
Yang-Mills theory \textit{if} the phase transition is of second order
described by the same IR fixed point
(\textit{Svetitsky-Yaffe conjecture}).  The key point is that this
argument gives us the critical exponents, but it does not guarantee
anything about the order of the phase transition.  Table~\ref{tab:SY}
summarizes the Svetitsky-Yaffe prediction for the critical exponents
and the results of the numerical tests.

\begin{table}[t]
 \begin{tabular}{lllll}
  Spatial dimension & Gauge group
   & \multicolumn{2}{l}{Svetitsky-Yaffe conjecture}
   & Numerical test
   \\ \hline \\[-8pt]
  $d=2$ & \begin{minipage}{7em}$\mathrm{U(1)}$\\
                                $\mathrm{Z}(N>4)$\\
                                $\mathrm{SU}(N>4)$ \end{minipage}
   & \multicolumn{2}{l}{(KT)~~ $\eta=1/4$} \\ \\[-8pt] \cline{2-5} \\[-8pt]
   & \begin{minipage}{7em}$\mathrm{Z(2),\; Z(4)}$\\
                           $\mathrm{SU(2)}$\\ \end{minipage}
   & \begin{minipage}{5em}$\beta=0.125$\\
                          $\gamma=1.75$\\
                          $\alpha=0$ \end{minipage}
   & \begin{minipage}{5em}$\nu=1$\\
                          $\eta=0.25$ \end{minipage}
   & (Universal Second Order)~\cite{Engels:1996dz} \\ \\[-8pt]
                                          \cline{2-5} \\[-8pt]
   & $\mathrm{SU(4)}$ &
     \multicolumn{2}{l}{(Varying Exponents)}
   & (Second Order)~\cite{deForcrand:2003wa} \\ \\[-8pt]
                                          \cline{2-5} \\[-8pt]
   & \begin{minipage}{7em}$\mathrm{Z(3)}$\\
                           $\mathrm{SU(3)}$ \end{minipage}
   & \begin{minipage}{5em}$\beta=0.11$\\
                          $\gamma=1.44$\\
                          $\alpha=0.33$ \end{minipage}
   & \begin{minipage}{5em}$\nu=0.83$\\
                          $\eta=0.27$ \end{minipage}
   & (Universal Second Order)~\cite{Christensen:1991rx} \\ \\[-8pt]
                                                 \hline \\[-8pt]
  $d=3$ & \begin{minipage}{7em}$\mathrm{U(1)}$\\
                                $\mathrm{Z}(N\ge 4)$\\
                                $\mathrm{SU}(N\ge 4)$ \end{minipage}
   & \begin{minipage}{5em}$\beta=0.35$\\
                          $\gamma=1.32$\\
                          $\alpha=-0.01$ \end{minipage}
   & \begin{minipage}{5em}$\nu=0.67$\\
                          $\eta=0.03$ \end{minipage}
   & (First Order)~\cite{Lucini:2003zr} \\ \\[-8pt] \cline{2-5} \\[-8pt]
   & \begin{minipage}{7em}$\mathrm{Z(2)}$\\
                           $\mathrm{SU(2)}$ \end{minipage}
   & \begin{minipage}{5em}$\beta=0.33$\\
                          $\gamma=1.24$\\
                          $\alpha=0.11$ \end{minipage}
   & \begin{minipage}{5em}$\nu=0.63$\\
                          $\eta=0.03$ \end{minipage}
   & (Universal Second Order)~\cite{Engels:1989fz} \\ \\[-8pt]
                                          \cline{2-5} \\[-8pt]
   & \begin{minipage}{7em}$\mathrm{Z(3)}$\\
                           $\mathrm{SU(3)}$ \end{minipage}
   & \multicolumn{2}{l}{(First Order)}
   & (First Order)~\cite{Alves:1990yq} \\ \\[-8pt] \hline \\[-8pt]
  $d\ge4$ & \begin{minipage}{7em}$\mathrm{U(1)}$\\
                                  $\mathrm{Z}(N\neq 3)$\\
                                  $\mathrm{SU}(N\neq 3)$ \end{minipage}
   & \begin{minipage}{5em}$\beta=1/2$\\
                          $\gamma=1$\\
                          $\alpha=0$ \end{minipage}
   & \begin{minipage}{5em}$\nu=1/2$\\
                          $\eta=0$ \end{minipage} \\ \\[-8pt]
                                         \cline{2-5} \\[-8pt]
   & \begin{minipage}{7em}$\mathrm{Z(3)}$\\
                           $\mathrm{SU(3)}$ \end{minipage}
   & \multicolumn{2}{l}{(First Order)} \\ \\[-8pt] \hline
 \end{tabular}
 \caption{Svetitsky-Yaffe conjecture for various gauge theories and the
   numerical test in the lattice simulation.  KT in $d=2$ represents
   the Kosterlitz-Thouless transition.  Table adapted from
   Ref.~\cite{Svetitsky:1985ye}.}
 \label{tab:SY}
\end{table}

The situation relevant to QCD of our interest is the
$\mathrm{SU}(\Nc)$ case at $d=3$.  The universal behavior of the
second-order phase transition has been confirmed in the numerical
simulation of the $\mathrm{SU(2)}$ Yang-Mills theory.  In the $\Nc=3$
case, the effective scalar theory does not have a stable IR fixed
point, which implies that the phase transition is possibly of
first order.  Indeed the $\mathrm{SU(3)}$ Yang-Mills theory has turned
out to exhibit a first-order phase transition in the numerical
simulation.  The situation is subtle for $\Nc\ge4$.  One may think that
the corresponding scalar theory with $\mathrm{Z}(\Nc)$ symmetry may
have a first-order phase transition, but this $\mathrm{Z}(\Nc)$
symmetry is dynamically enhanced to $\mathrm{U}(1)$ symmetry at the
critical point, and then a second-order phase transition is
theoretically possible as listed in the Svetitsky-Yaffe conjecture.
It has been, however, established that the order of the phase
transition is not second but first for the $\mathrm{SU}(6)$ and
$\mathrm{SU}(8)$ Yang-Mills theories.  Hence, the gauge theories at
$\Nc\ge4$ seem not to fall into the Svetitsky-Yaffe universality class.
References for the lattice simulations are listed in
Tab.~\ref{tab:SY}.

\subsection{Flavor symmetry}
\label{sec:flavor}

The quark part in the QCD Lagrangian density~\eqref{eq:QCD_Lag} has
global symmetries in flavor space.  Because the gauge part is not
sensitive to the flavor structure, we can safely concentrate on the
quark part in what follows.

A quark field $\psi$ is decomposed into two chiral sectors,
\begin{equation}
 \psiL \equiv \frac{1-\gamma_5}{2}\psi \;, \qquad
 \psiR \equiv \frac{1+\gamma_5}{2}\psi \;.
\end{equation}
The quark Lagrangian density is then expressed in terms of
$\psi_{\text{L,R}}$, that reads,
\begin{equation}
 \bar{\psi} ( i\gamma^\mu D_\mu - \boldm) \psi
  = \bpsiL i\gamma^\mu D_\mu \psiL + \bpsiR i\gamma^\mu D_\mu \psiR
    - \bpsiL \boldm \psiR - \bpsiR \boldm \psiL \;,
\end{equation}
where we suppressed the color and the flavor indices to simplify the
notation.  The covariant derivative is defined as usual as
$(D_\mu)_{ij}=\partial_\mu\delta_{ij}-igA_\mu^a T^a_{ij}$ and the mass
matrix takes the form of $\boldm=\text{diag}(\mmu,\mmd,\mms,\dots)$ in
flavor space.

Here let us assume the chiral limit in which there are $\Nf$ massless
flavors ($\mmu=\mmd=\mms=\dots=0$).  In the chiral limit the left-
and right-handed sectors are totally disconnected from each other and
any mixing between $\psiL$ and $\psiR$ is forbidden on the Lagrangian
level.  Therefore, the Lagrangian density is invariant under two
independent chiral rotations by
\begin{equation}
 \begin{array}{lp{1em}l}
    \psiL \;\to\; \psiL' = L\,\psiL \;,
 && \psiR \;\to\; \psiR' = R\,\psiR \;,\\
    L = \exp\bigl(i\theta_{\text{L}}^a t^a\bigr)
    \:\in\; \mathrm{U}(\Nf)_{\text{L}} \;,
 && R = \exp\bigl(i\theta_{\text{R}}^a t^a\bigr)
    \:\in\; \mathrm{U}(\Nf)_{\text{R}} \;,
 \end{array}
\label{eq:chiral_rot}
\end{equation}
where $t^a$'s denote the $\mathrm{su}(\Nf)$ algebras and the flavor
index runs over $a=1, \cdots, \Nf^2-1$.  This means that the QCD
Lagrangian density has the following symmetry,
\begin{equation}
 \mathrm{U}(\Nf)_{\text{L}}\times\mathrm{U}(\Nf)_{\text{R}}
 \;\;\simeq\;\; \mathrm{SU}(\Nf)_{\text{L}}\times
   \mathrm{SU}(\Nf)_{\text{R}} \times\mathrm{U}(1)_{\text{V}}\times
   \mathrm{U}(1)_{\text{A}} \;,
\label{eq:unun}
\end{equation}
apart from the discrete symmetry.  (The center elements of
$\mathrm{SU}(\Nf)$ also belong to $\mathrm{U}(1)$, and so
$\mathrm{Z}(\Nf)$ is redundant in the right-hand side above;  see
Ref.~\cite{Fukushima:2010bq}.)

Realization of the global symmetry always leads to the existence of
the conserved N\"{o}ther current.  Under the infinitesimal (coordinate
dependent) transformations
$\psiL'\simeq (1+i\delta\theta_{\text{L}}^a(x) t^a)\psiL$ and
$\psiR'\simeq (1+i\delta\theta_{\text{R}}^a(x) t^a)\psiR$, the change
in the Lagrangian density should be at most the surface term,
$\delta\calL=\theta_{\text{L}}^a(x)\partial_\mu j_{\text{L}}^{\mu\,a}(x)
+\theta_{\text{R}}^a(x)\partial_\mu j_{\text{R}}^{\mu\,a}(x)$, where
\begin{equation}
 j_{\text{L}}^{\mu\,a} = \bar{\psi}_{\text{L}} \gamma^\mu t^a
  \psi_{\text{L}} \;,\qquad
 j_{\text{R}}^{\mu\,a} = \bar{\psi}_{\text{R}} \gamma^\mu t^a
  \psi_{\text{R}} \;.
\end{equation}
Then we can make appropriate combinations corresponding to the vector
and the axial-vector transformations, i.e.
\begin{equation}
 Q_{\text{V}}^a = Q_{\text{R}}^a + Q_{\text{L}}^a\;,
 \qquad
 Q_{\text{A}}^a = Q_{\text{R}}^a - Q_{\text{L}}^a\;,
\end{equation}
where we defined
$Q_{\text{L,R}}^a = \int d^3x\, \psi_{\text{L,R}}^\dagger t^a
\psi_{\text{L,R}}$.

\subsubsection{Chiral symmetry}
\label{sec:chiralsym}

Among the global symmetry~\eqref{eq:unun} the first part of
$\mathrm{SU}(\Nf)_{\text{L}}\times\mathrm{SU}(\Nf)_{\text{R}}$ is
called chiral symmetry specifically.  We postpone the discussions on
the remaining part of
$\mathrm{U}(1)_{\text{L}}\times\mathrm{U}(1)_{\text{R}}$.  Here we
give a concise primer of spontaneous chiral symmetry breaking and
associated low-energy theorems.

So far, we have seen what symmetries the QCD Lagrangian density
possesses.  However, the chiral invariance is not manifest in the
low-lying hadron spectra where any degenerate patterns between parity
partners are absent.  The resolution is that chiral symmetry is
dynamically broken due to the strong interaction.  The generator of
the axial transformation, $Q_{\text{A}}^a$, does not annihilate the
ground state then.  In a precise expression, some operator $\calO(x)$
exists so that
\begin{equation}
 \langle \Omega| \bigl[ i Q_{\text{A}}^a, \calO(x) \bigr]
  |\Omega \rangle \neq 0 \;,
\label{eq:ssb}
\end{equation}
where $|\Omega\rangle$ is the physical state.  The symmetry generated
by $Q_{\text{A}}^a$ is spontaneously broken in this state
$|\Omega\rangle$ if Eq.~\eqref{eq:ssb} holds, and the above non-zero
expectation value is nothing but the order parameter.  The simplest
choice would be $\calO=\bar{\psi}t^a\psi$, leading to the chiral
condensate as an order parameter;
\begin{equation}
 \langle\bar{\psi}\psi\rangle
 = \langle \bpsiL\psiR + \bpsiR\psiL \rangle \neq 0 \;.
\end{equation}
Since $\bar{\psi}\psi$ is unchanged under
$\mathrm{SU}(\Nf)_{\text{V}}$ transformation generated by
$Q_{\text{V}}^a$, the physical state is invariant and
$Q_{\text{V}}^a |\Omega\rangle = 0$, which obeys the Vafa-Witten
theorem~\cite{Vafa:1984xg}.  In summary, chiral symmetry is
spontaneously broken by $\langle\bar{\psi}\psi\rangle\neq0$ as
\begin{equation}
 \mathrm{SU}(\Nf)_{\text{L}} \times \mathrm{SU}(\Nf)_{\text{R}}
 \;\;\to\;\; \mathrm{SU}(\Nf)_{\text{V}} \;.
\label{eq:chiral_broken}
\end{equation}

The physical manifestation of such spontaneous symmetry breaking is
the presence of massless scalar particles, namely, the Nambu-Goldstone
(NG) bosons.  In QCD the pions are identified as the approximate NG
bosons.  We note that the original theory contains only fermions with
spin $1/2$ and gauge bosons with spin $1$ and, therefore, the NG
bosons must be composite states generated through non-perturbative
dynamics.  The number of NG bosons is specified by the dimension of
the $G/H$ manifold, where
$G=\mathrm{SU}(\Nf)_{\text{L}}\times\mathrm{SU}(\Nf)_{\text{R}}$ is
the original chiral symmetry whereas
$H=\mathrm{SU}(\Nf)_{\text{V}}$ the unbroken vectorial symmetry;
there appear as many NG bosons as $\text{dim}(G/H) = \Nf^2-1$.

The axial current, $J_{\text{A}}^{\mu\,a}$, has a direct coupling to
the NG boson $\pi^b$, the strength of which can be parametrized by the
decay constant $f_\pi$ as
\begin{equation}
 \langle\Omega| J_{\text{A}}^{\mu\,a}(0) |\pi^b(q) \rangle
  = i f_\pi q^\mu\, \delta^{ab} \;.
\end{equation}
In the experimental point of view $f_\pi$ is a useful measure, and the
empirical value is $f_\pi=(92.4\pm0.2)\MeV$.  In a medium $f_\pi$
serves as an order parameter for (partial) chiral symmetry
restoration.  The divergence of the axial current is non-vanishing due
to an explicit symmetry breaking by the presence of $\mq$, i.e.
\begin{equation}
 \partial_\mu J_{\text{A}}^{\mu\,a} = f_\pi m_\pi^2 \pi^a \;,
\label{eq:PCAC}
\end{equation}
which is known as the partially conserved axial-vector current (PCAC).
If the explicit breaking is turned off, the pion is precisely massless
as dictated by the NG theorem and thus the axial current is
conserved.  Utilizing the PCAC and the soft pion theorems, one can
derive low-energy theorems.  The most well-known is the
Gell-Mann--Oakes--Renner (GOR) relation,
\begin{equation}
 f_\pi^2 m_\pi^2 = -\mq\langle\bar{\psi}\psi\rangle \;,
\label{eq:GOR}
\end{equation}
and another important low-energy theorem is the Goldberger--Treiman
relation,
$f_\pi\, g_{\pi NN} = m_N\, g_A$, where $g_{\pi NN}$ is the
pion-nucleon coupling, $m_N$ is the nucleon mass, and $g_A$ is the
nucleon axial charge.

\subsubsection{$\UA$ symmetry and the quantum anomaly}
\label{sec:instanton}

The QCD Lagrangian density~\eqref{eq:QCD_Lag} has not only chiral
symmetry
$\mathrm{SU}(\Nf)_{\text{L}}\times\mathrm{SU}(\Nf)_{\text{R}}$ but
also $\mathrm{U}(1)_{\text{V}}\times\mathrm{U}(1)_{\text{A}}$.  The
$\mathrm{U}(1)_{\text{V}}$ symmetry corresponds to the baryon number
conservation and should not be broken except in
color-superconductivity at asymptotically high density.  Then,
according to the NG theorem, there should be one more massless boson
associated with the $\UA$ symmetry breaking.
There is, however, no such light particle in the iso-singlet
pseudo-scalar channel.  Instead of the NG boson, the lightest meson in
this channel is $\eta'$ whose mass is $m_{\eta'}=958\MeV$, that is too
heavy to be the NG boson.  The problem of missing $\UA$ NG boson is
called the $\UA$ problem.

Now it is a textbook knowledge that the $\UA$ symmetry in the
classical Lagrangian is \textit{explicitly} broken via quantum
effects.  Such quantum anomaly appears ubiquitously in the gauge field
theory.  If we choose to adhere the vector gauge symmetry, the quantum
anomaly arises in the axial current generally as
\begin{equation}
 \partial_\mu j_{\text{A}}^{\mu\,a} = -\frac{g^2}{16\pi^2}
  \epsilon^{\mu\nu\rho\sigma} \tr\bigl[ t^a\, F_{\mu\nu} F_{\rho\sigma}
  \bigr] \;,
\label{eq:div_j5}
\end{equation}
which is non-vanishing only for $a=0$ (iso-singlet) because $t^a$'s
are traceless for $a=1,\dots,\Nf^2-1$.  Therefore the chiral symmetry
breaking pattern is truly the one given by
Eq.~\eqref{eq:chiral_broken} and $\eta'$ cannot be the NG boson.  We
make a remark here that it is not such easy to disprove the existence
of the massless pole because $J_{\text{A}}^{\mu\,0}$ can be modified
into a conserved but gauge-variant form.  Interested readers can
consult Ref.~\cite{Kogut:1974kt}.

From the microscopic point of view $\UA$ symmetry is broken
through the instanton-induced interactions.  Let us take a look at the
derivation with an assumption that an instanton with the size $\rho$
is placed at $z$~\cite{'tHooft:1976up}.  Then, the partition function
can be expressed in a form of the functional integration over the
collective coordinates as
\begin{equation}
 Z = \int d^4 z\, d\rho\, \rho^{-1} (\det D[\calA])^{-1/2}
  (\det G[\calA])\, e^{-8\pi^2/g^2} \;,
\label{eq:Z_inst}
\end{equation}
after the one-loop integration around the instanton background
$\calA(x)$.  Here $D[\calA]$ and $G[\calA]$ are the gluon and the
(massless) quark propagator inverse, respectively, in the presence of
the background $\calA(x)$.

It is important to note that a fermionic zero-mode is accompanied by
the instanton gauge configuration, which has definite chirality in
accord with the sign of the winding number $Q_{\text{W}}$;  for
$\calA(x)$ with $Q_{\text{W}}=1$ for instance, there exists a
wave-function that satisfies $G[\calA]\psi_{\text{R}0}=0$ (apart from
the mass term).  Therefore, the partition function $Z$ should be zero
in this case.  Now we deform the theory slightly so that $Z$ stays
finite, that is, we insert a source term, $\bpsiL J(x)\psiR$, in the
fermionic part $G$.  Then, at low energies, we can approximate the
Dirac determinant as
\begin{equation}
 \det[G+J] \;\simeq\; \det\int d^4 x\,\bar{\psi}_{\text{L}0} J(x)
  \psi_{\text{R}0} \;.
\label{eq:inst_inserted}
\end{equation}
Here det in the right-hand side is taken in flavor space.  Now we
shall consider an effective theory of QCD with all gluonic fields
integrated out.  Then this effective theory should be written in
terms of quark fields $\bar{\psi}(x)$ and $\psi(x)$.  The question is
how to incorporate the instanton effect with the quark degrees of
freedom only, or how to find an effective interaction vertex induced
by instanton.  We can find such an instanton-induced interaction to
reproduce Eqs.~\eqref{eq:Z_inst} and \eqref{eq:inst_inserted};  the
following vertex can mimic Eq.~\eqref{eq:inst_inserted} and thus
embody the instanton effect;
\begin{equation}
 \calL_{\UA} \sim n(\rho,T=0)\, \bigl[ \det\bar{\psi}
  (1-\gamma^5)\psi + \det\bar{\psi} (1+\gamma^5)\psi \bigr] \;,
\label{eq:inst_induce}
\end{equation}
including the latter term coming from an anti-instanton.  As expected,
the interaction~\eqref{eq:inst_induce} breaks the $\UA$ symmetry
explicitly, while chiral
$\mathrm{SU}(\Nf)_{\text{L}}\times\mathrm{SU}(\Nf)_{\text{R}}$
symmetry is kept unbroken.  This determinant form had been postulated
earlier from the symmetry consideration~\cite{Kobayashi:1970ji} and
the above interaction~\eqref{eq:inst_induce} is now called the
Kobayashi-Maskawa-'t~Hooft (KMT) interaction.

In the above derivation it is clear that the interaction strength is
proportional to the instanton density that is given in the one-loop
perturbation theory by
\begin{equation}
 n(\rho,T=0) = \frac{8\pi^2}{g^2}\,\rho^{-5} e^{-8\pi^2/g^2} \;,
\end{equation}
originating from $\det D[\calA]$.  Thus, the strength of the $\UA$
breaking interaction depends on how many instantons the system would
accommodate, though the anomaly itself is never diminished and
Eq.~\eqref{eq:div_j5} is not altered by the finite-$T$ effect.  This
fact opens an interesting possibility.  That is, at finite $T$, the
one-loop instanton density is modified as~\cite{Pisarski:1980md},
\begin{equation}
 n(\rho,T) = \biggl(\frac{8\pi^2}{g^2}\, \rho^{-5} e^{-8\pi^2/g(\rho)^2}
  \biggr)\, \exp\biggl[ -\pi^2\rho^2 T^2 \biggl(
  \frac{2\Nc}{3}+\frac{\Nf}{3} \biggr) \biggr] \;.
\end{equation}
The latter exponential factor suppresses the instanton excitation
significantly.  Similarly, at finite density, the exponential
suppression factor has been found~\cite{Schafer:1999fe}, and after
integrating over the instanton size, the suppression factor should be
$T^{-14}$ or $\mu^{-14}$.  In any case these expressions are
perturbative ones, and any quantitative estimate is not really
trustworthy near $\Tc$.  Nevertheless, it is quite suggestive that
$n(T\sim\Tc)$ in the above expression is already a half of $n(T=0)$.
Then, one may well speculate that the $\UA$ symmetry is
\textit{effectively} restored before the critical
temperature/density is reached~\cite{Shuryak:1993ee}, and if so, as we
will see in the next subsection, the universality class and the
critical properties would be significantly affected.

\subsubsection{Chiral phase transition and critical phenomena}
\label{sec:chiral}

The order of the chiral phase transition was systematically
investigated first by the RG analysis~\cite{Pisarski:1983ms}.  As we
explained, if the phase transition is of second order, the fixed-point
structure yields the critical exponents not relying on the microscopic
dynamics.  One can also say that, reversely, the absence of the stable
fixed point suggests that the theory should exhibit a first-order
phase transition.

To clarify the nature of the phase transition we need not to solve QCD
at finite temperature but we can make use of an effective description
with chiral symmetry.  The simplest model is the linear sigma model
composed from the meson field matrices,
$M_{ij}\sim \bar{\psi}_{\text{R}j} \psi_{\text{L}i}$ and its conjugate
$M^\dagger$ that transform as
\begin{equation}
 M \;\to\; L^\dagger M R \;, \qquad
 M^\dagger \;\to\; R^\dagger M^\dagger L \;.
\end{equation}
The general Lagrangian density consistent with chiral symmetry is
decomposed into $\calL=\calL_\sigma+\calL_{\UA}+\calL'$, where
\begin{equation}
 \calL_\sigma = \frac{1}{2}\tr\bigl(\partial_\mu M \partial^\mu
  M^\dagger \bigr) + \frac{1}{2}\mu^2 \tr\bigl( M M^\dagger \bigr)
  - \frac{\pi^2 g_1}{3} \bigl[ \tr\bigl( M M^\dagger \bigr) \bigr]^2
  - \frac{\pi^2 g_2}{3} \tr\bigl( M M^\dagger M M^\dagger \bigr) \;,
\label{eq:lsm}
\end{equation}
and the instanton-induced KMT interaction
$\calL_{\UA}= c \bigl( \det M + \det M^\dagger \bigr)$ which is
reminiscent of Eq.~\eqref{eq:inst_induce}, and the explicit breaking
part $\calL'=f_0\,\tr(t^0 M)+f_3\,\tr(t^3 M)+f_8\,\tr (t^8 M)+\cdots$
ranging over all the Cartan subalgebras.  Here, in $\calL_\sigma$, the
first is the kinetic term, the second is the mass term, and the third
and the fourth are potential terms that results in the spontaneous
symmetry breaking, respectively.  From the symmetry property $M$ can
be decomposed as
\begin{equation}
 M = \Sigma + i \Pi = \sum_{a=0}^{\Nf^2-1} t^a (\sigma_a + i \pi_a)
\label{eq:chiral_field}
\end{equation}
with the scalar fields,
$\Sigma = t^a \sigma_a = \tfrac{1}{2}(M+M^\dagger)$, and the
pseudo-scalar fields,
$\Pi = t^a \pi_a = \tfrac{1}{2i}(M-M^\dagger)$.  (Note that $M$
changes to $M^\dagger$ under the parity transformation.)   Because $M$
transforms as $V^\dagger M V$ under a vectorial rotation ($V=L=R$),
one can classify the matrix components in terms of the hadron
language.  In the $\Nc=3$ case, for example,
\begin{equation}
 \begin{split}
 &\Sigma = \begin{pmatrix}
  \frac{1}{\sqrt{2}}(\sigma + a_0^0) & a_0^+ & K^{\ast +} \\
  a_0^- & \frac{1}{\sqrt{2}}(\sigma - a_0^0) & K^{\ast 0} \\
  K^{\ast -} & \bar{K}^{\ast 0} & \zeta
\end{pmatrix} \;, \\
 &\Pi = \begin{pmatrix}
 \frac{1}{\sqrt{2}}\pi^0 + \frac{1}{\sqrt{6}}\eta_8
  + \frac{1}{\sqrt{3}}\eta_0 & \pi^+ & K^+ \\
 \pi^- & -\frac{1}{\sqrt{2}}\pi^0 + \frac{1}{\sqrt{6}}\eta_8
  + \frac{1}{\sqrt{3}}\eta_0 & K^0 \\
 K^- & \bar{K}^0 & -\frac{2}{\sqrt{6}}\eta_8
  + \frac{1}{\sqrt{3}}\eta_0
\end{pmatrix}
 \end{split}
\label{eq:matrix_lsm}
\end{equation}
with two independent condensates $\sigma$ and $\zeta$ (isospin
symmetry is assumed) corresponding to two chiral condensates,
$\langle\bar{u}u\rangle=\langle\bar{d}d\rangle$ and
$\langle\bar{s}s\rangle$.  In this case $\eta_0$ is anomalous, while
there are eight NG bosons, $\pi^0$, $\pi^\pm$, $K^0$, $\bar{K}^0$,
$K^\pm$, and $\eta_8$, in the chiral limit.  In reality
$\mathrm{SU}(3)_{\text{V}}$ symmetry is substantially broken by the
strange quark mass $\mms$, so that the masses in the strange sector,
$K^0$, $\bar{K}^0$, $K^\pm$, $\eta_8$, are lifted up by $\mms$.
Besides, $\eta_8$ and $\eta_0$ can mix together by $\mms$, which
amounts to physical states of $\eta$ and $\eta'$.

One could put this linear sigma model directly in a finite-$T$
environment and investigate the phase transition of chiral
restoration.  For the present purpose to study the order of the phase
transition and the critical properties, however, the universality
argument is extremely useful.  It is sufficient to begin with a
dimensionally reduced description after the integration over
temperature fluctuations, as already discussed in
Sec.~\ref{sec:deconf}, and then we can simply focus on the
3-dimensional linear sigma model with $T$-dependent couplings
$g_1(T;k)$, $g_2(T;k)$, $c(T;k)$, etc, at the RG scale $k$.  Since we
are interested in the critical phenomena only, we shall work in the
chiral limit, $f_0=f_3=f_8=\cdots=0$.

\begin{table}[t]
\begin{center}
 \begin{tabular}{lp{1em}lp{1em}ll}
  Symmetry breaking pattern
   && Flavors
   && \multicolumn{2}{l}{Pisarski-Wilczek conjecture}
   \\ \\[-8pt] \hline \\[-8pt]
     $\mathrm{SU}(\Nf)_{\text{L}}\times\mathrm{SU}(\Nf)_{\text{R}}
      \to \mathrm{SU}(\Nf)_{\text{V}}$
   && $\Nf=2$
   && \begin{minipage}{5.5em}
     $\beta=0.38$\\
     $\gamma=1.44$\\
     $\alpha=-0.19$ \end{minipage}
   & \begin{minipage}{5.5em}
     $\nu=0.73$\\
     $\eta=0.03$ \end{minipage} \\ \\[-8pt] \cline{3-6} \\[-8pt]
   && $\Nf\ge3$
   && \multicolumn{2}{l}{(First Order)} \\ \\[-8pt] \hline \\[-8pt]
     $\mathrm{SU}(\Nf)_{\text{L}}\times\mathrm{SU}(\Nf)_{\text{R}}
      \times\mathrm{U}(1)_{\text{A}} \to
      \mathrm{SU}(\Nf)_{\text{V}}$
   && $\Nf=1$
   && \begin{minipage}{5.5em}
     $\beta=0.35$\\
     $\gamma=1.32$\\
     $\alpha=-0.015$ \end{minipage}
   & \begin{minipage}{5.5em}
     $\nu=0.67$\\
     $\eta=0.038$ \end{minipage} \\ \\[-8pt] \cline{3-6} \\[-8pt]
   && $\Nf\ge2$
   && \multicolumn{2}{l}{(First Order)} \\ \\[-8pt]
  \hline
 \end{tabular}
\end{center}
 \caption{Pisarski-Wilczek conjecture for the chiral phase transition
   in $d=3$ space.}
 \label{tab:chiral}
\end{table}

\paragraph{Without the axial anomaly}
Let us first consider the case without the $\UA$-breaking term
(i.e.\ $c=0$).  This is possible if the effective restoration of $\UA$
symmetry occurs slightly below $\Tc$ due to instanton suppression as
argued in Sec.~\ref{sec:instanton}.  In this case the symmetry
breaking pattern is
\begin{equation}
 \mathrm{U}(\Nf)_{\text{L}}\times \mathrm{U}(\Nf)_{\text{R}}
  \;\;\to\;\; \mathrm{U}(\Nf)_{\text{V}} \;,
\end{equation}
and there are $\Nf^2$ massless NG bosons.  The critical phenomena can
be analyzed by the $\epsilon$ expansion, in which the spatial
dimension is taken as $d=4-\epsilon$ and $\epsilon$ is assumed to be a
small number.  Because $d=4$ is the critical dimension, the coupling
constant around the fixed point is $\sim\calO(\epsilon)$, which
justifies the perturbation theory.

One can then carry out the one-loop calculation using the
Lagrangian~\eqref{eq:lsm} to find the following $\beta$ functions;
\begin{equation}
 \begin{split}
 & \beta_1 \equiv k\frac{d \hat{g}_1}{d k}
   = -\epsilon \hat{g}_1 + \frac{\Nf^2+4}{3} \hat{g}_1^2
      + \frac{4\Nf}{3}\hat{g}_1 \hat{g}_2 + \hat{g}_2^2 \;, \\
 & \beta_2 \equiv k\frac{d \hat{g}_2}{d k} = -\epsilon \hat{g}_2
      + 2\hat{g}_1 \hat{g}_2 + \frac{2\Nf}{3}\hat{g}_2^2
 \end{split}
\label{eq:beta}
\end{equation}
for general $\Nf$.  We introduced a notation for the dimensionless
couplings, $\hat{g}_1=g_1 k^{-\epsilon}$ and
$\hat{g}_2=g_2 k^{-\epsilon}$.  The first term in the right-hand
side of Eq.~\eqref{eq:beta} involving $\epsilon$ appears trivially
from $k^{-\epsilon}$ in $\hat{g}_1$ and $\hat{g}_2$.  One can solve
this set of differential equations to draw the RG flow pattern.  The
fixed points can be found readily from the conditions,
$\beta_1=\beta_2=0$.  Two solutions always exist, one at
$(\hat{g}_1^\ast,\hat{g}_2^\ast)=(0,0)$, and the other at
\begin{equation}
 (\hat{g}_1^\ast, \hat{g}_2^\ast)
  = \Bigl(\frac{3\epsilon}{\Nf^2+4},0\Bigr)\;,
\label{eq:fixed1}
\end{equation}
and two more fixed-points can appear for $\Nf\le \sqrt{3}$ at
\begin{equation}
 (\hat{g}_1^\ast, \hat{g}_2^\ast)=\frac{3\epsilon}{2(27-8\Nf^2+\Nf^4)}
  \biggl( 9 - \Nf^2 \mp \Nf \sqrt{2(3-\Nf^2)}\;,\;
         -5\Nf + \Nf^3 \pm 3\sqrt{2(3-\Nf^2)} \biggr) \;.
\label{eq:fixed2}
\end{equation}
From explicit analysis it turns out that Eq.~\eqref{eq:fixed1} is an
IR-stable fixed point for $0\le\Nf\le\sqrt{2}$, and then this fixed
point would represent the critical phenomena of the $\mathrm{O}(2\Nf)$
universality class (which is rather close to the situation
\textit{with} the axial anomaly).  Therefore, when $\Nf=1$, the phase
transition can be of second order, and if so, it is characterized by
the $\mathrm{O}(2)$ critical exponents.  For $\Nf>\sqrt{3}$ this
fixed-point becomes unstable in the $\hat{g}_2$ direction and there is
no IR-stable fixed-point at all.  Thus, when $\Nf\ge 2$, it is
impossible to have a second-order phase transition and it should be of
fluctuation-induced first order.

One might have had an impression that the leading-order $\epsilon$
expansion is insufficient to conclude whether the phase transition is
of first order.  It should be mentioned that this conclusion of the
fluctuation-induced first-order transition has been confirmed
non-perturbatively in a more sophisticated framework of the functional
RG method~\cite{Berges:1996ib,Fukushima:2010ji}.

\paragraph{With the axial anomaly}
If the instanton-induced $\UA$-breaking interaction remains
non-vanishing, as is a more conventional scenario, the critical
properties are drastically changed especially in the case with
$\Nf=2$.  Because, in the $\Nf=2$ case, the original group is
$\mathrm{SU}(2)_{\text{L}}\times \mathrm{SU}(2)_{\text{R}} \simeq
\mathrm{O}(4)$, the critical phenomena should belong to the
$\mathrm{O}(4)$ universality class.  This can be analyzed in the
$\epsilon$ expansion in the same way as previously.  The instanton
term $\det M+\det M^\dagger$ yields substantial mass terms for the
scalar particles only, and $\Sigma$ decouples from the low-lying
dynamics.  Then, with only $\Pi$, two terms involving $g_1$ and $g_2$
in Eq.~\eqref{eq:lsm} are no longer independent, and the $\beta$
function is reduced to $\beta_1$ in Eq.~\eqref{eq:beta} with
$\hat{g}_2=0$, which has a stable fixed-point unlike the previous case
without the anomaly.

For $\Nf\ge3$ no IR-stable fixed point is found, and it is likely that
the phase transition is of first order.  In fact, even at the tree
level, the KMT interaction at $\Nf=3$ involves terms like
$\sigma^3$ in the chiral limit where $\zeta=\sigma$.  The presence of
such cubic term strongly implies that the phase transition should be
of first order.

To summarize the expected critical properties of the chiral phase
transition in $d=3$ space, we shall make a table based on this
Pisarski-Wilczek conjecture~\cite{Pisarski:1983ms} as in
Tab.~\ref{tab:chiral}.

\subsection{Highlights of the lattice-QCD results}
\label{sec:highlights}

In this subsection let us look over the numerical results from the
Monte-Carlo simulation of QCD on the lattice.  Although there are some
data available for spatially 2-dimensional case, we will limit our
discussions here to the realistic situation with $d=3$ in space.  For
details on the lattice-QCD simulation, readers may consult
Refs.~\cite{Fodor:2009ax,Philipsen:2012nu}.

\begin{figure}
 \includegraphics[width=0.46\textwidth]{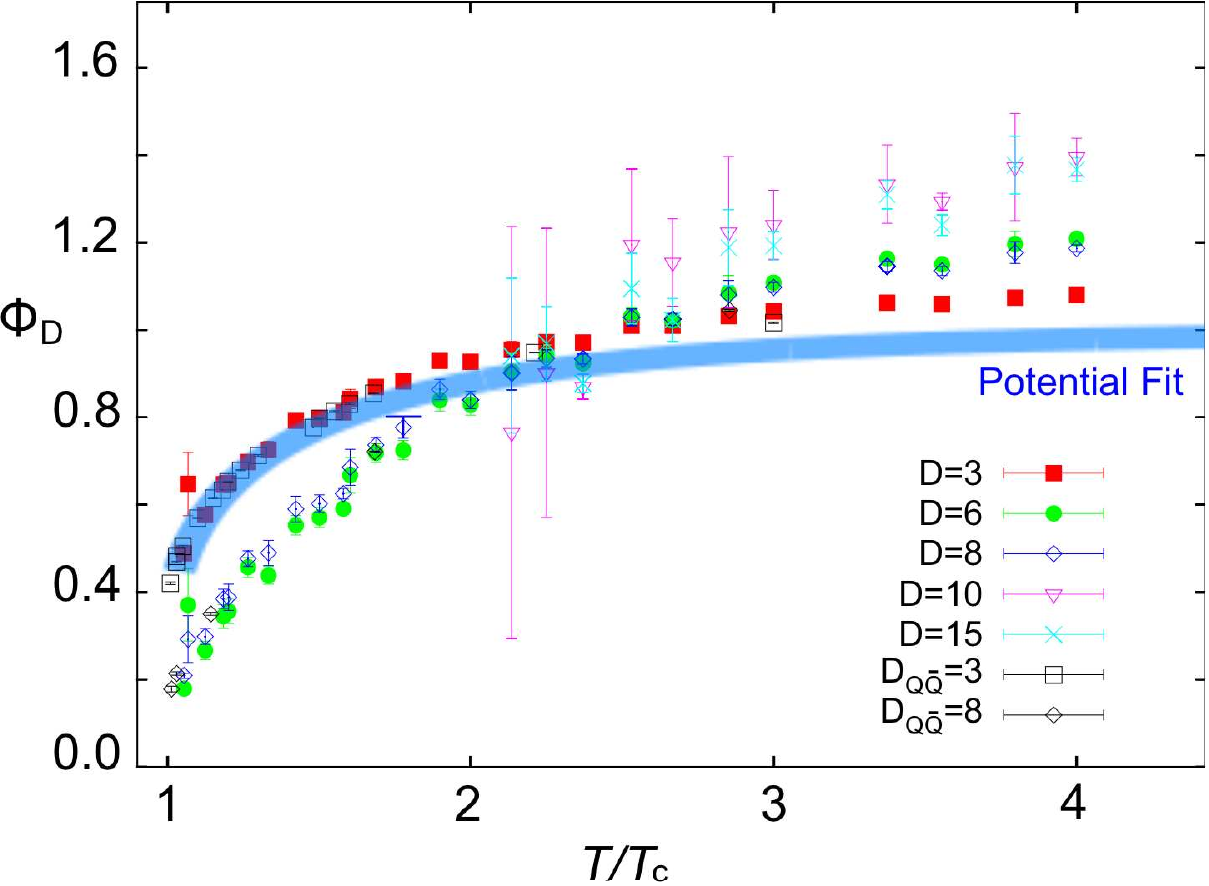}\hspace{2em}
 \includegraphics[width=0.46\textwidth]{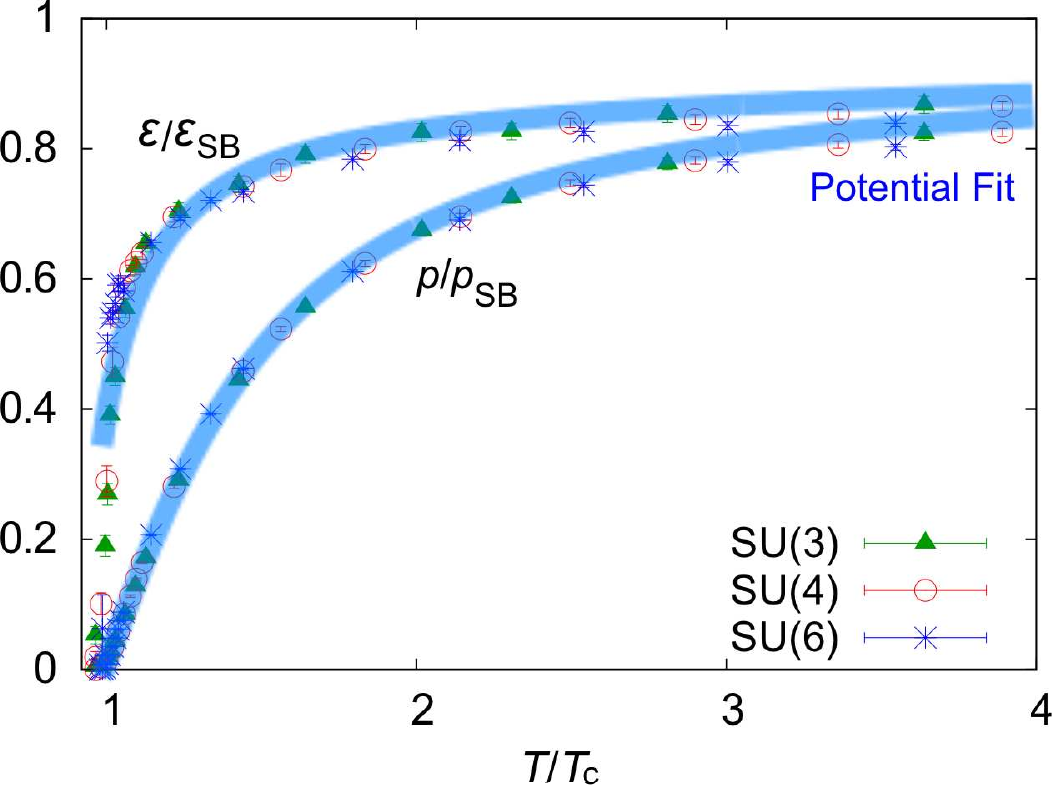}
 \caption{(Left) Renormalized Polyakov loop in various representations
   (with the dimension $D$) in the SU(3) Yang-Mills theory.  The blue
   curve represents the potential fit from
   $\Omega_{\text{glue}}[\Phi]$ discussed in Sec.~\ref{sec:NJL}.
   Figure adapted from Ref.~\cite{Gupta:2007ax}.  (Right) Energy
   density and pressure in the $\mathrm{SU}(\Nc)$ Yang-Mills theories
   normalized by the Stefan-Boltzmann value.  The blue curves
   represent the potential fit from $\Omega_{\text{glue}}[\Phi]$
   again.  Figure adapted from Ref.~\cite{Datta:2010sq}.}
 \label{fig:purelattice}
\end{figure}

\paragraph{Pure Yang-Mills theories}
The left panel of Fig.~\ref{fig:purelattice} shows a figure adapted
from Ref.~\cite{Gupta:2007ax} for the renormalized Polyakov loop in
various representations in color space.  It is the renormalization
effect that renders $\Phi$ exceed the unity.  We see that the order
parameter clearly shows a discontinuous jump at $T=\Tc$.  (We should
note that $\Phi=0$ below $\Tc$ where center symmetry is unbroken which
is omitted in Fig.~\ref{fig:purelattice}.)  In the pure Yang-Mills
theory the physical scale is set with the string tension $\sigma$, and
the critical temperature is found as
$\Tc=(0.630\pm0.005)\sqrt{\sigma}\simeq
280\MeV$~\cite{Beinlich:1997ia}.  In the right panel of
Fig.~\ref{fig:purelattice} a figure adapted from
Ref.~\cite{Datta:2010sq} presents thermodynamic quantities such as the
internal energy density $\varepsilon$ and the pressure $p$ for the
pure Yang-Mills theories with various gauge groups.  From this figure
we can understand that the thermodynamics has only weak dependence on
the number of colors once properly normalized by the Stefan-Boltzmann
value (for more general large-$\Nc$ studies, see
Refs.~\cite{Panero:2009tv,Mykkanen:2012ri}).  It is also evident in
the behavior of $\varepsilon(T)$ that the physical degrees of freedom
rapidly change at $T=\Tc$ corresponding to the liberation of colored
excitations.  On the other hand, the pressure $p$ is always continuous
regardless of the order of the phase transition.  The state-of-the-art
lattice data at high precision is available in
Ref.~\cite{Borsanyi:2012ve}.  The blue curves in
Fig.~\ref{fig:purelattice} represent the potential fit results for
later convenience, which we will discuss in Sec.~\ref{sec:NJL}.

\begin{figure}
\begin{center}
 \includegraphics[width=0.36\textwidth]{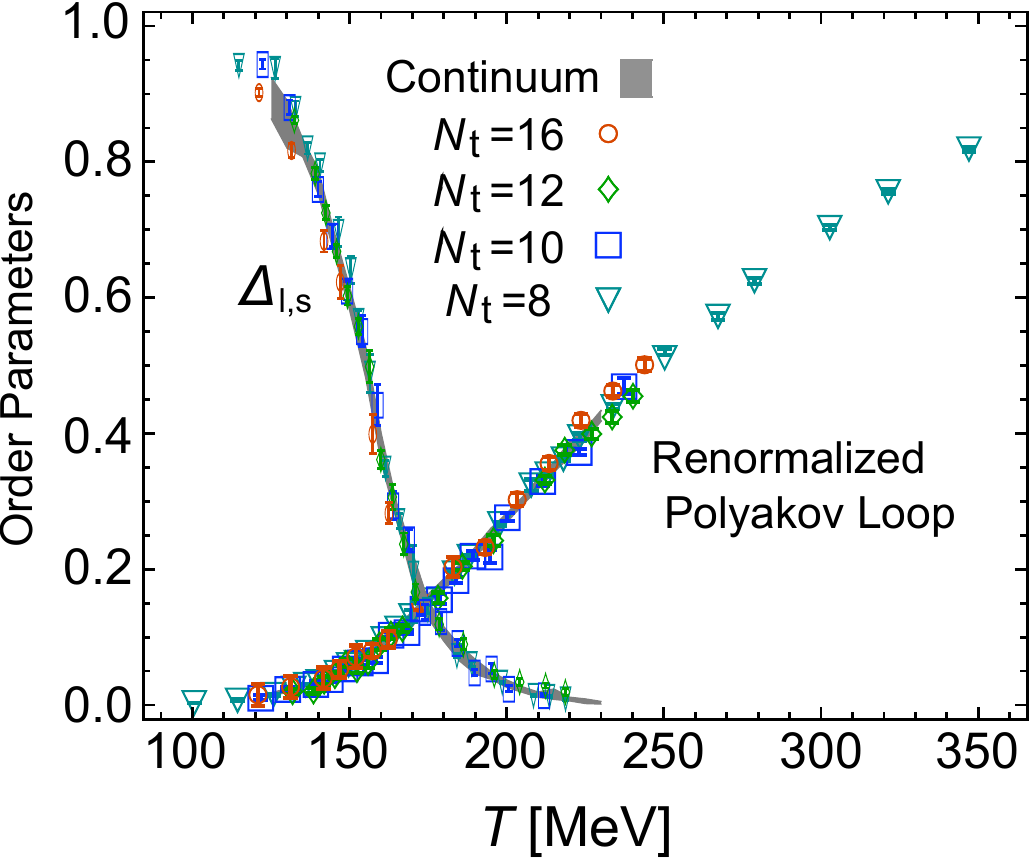}\hspace{1.8em}
 \includegraphics[width=0.5\textwidth]{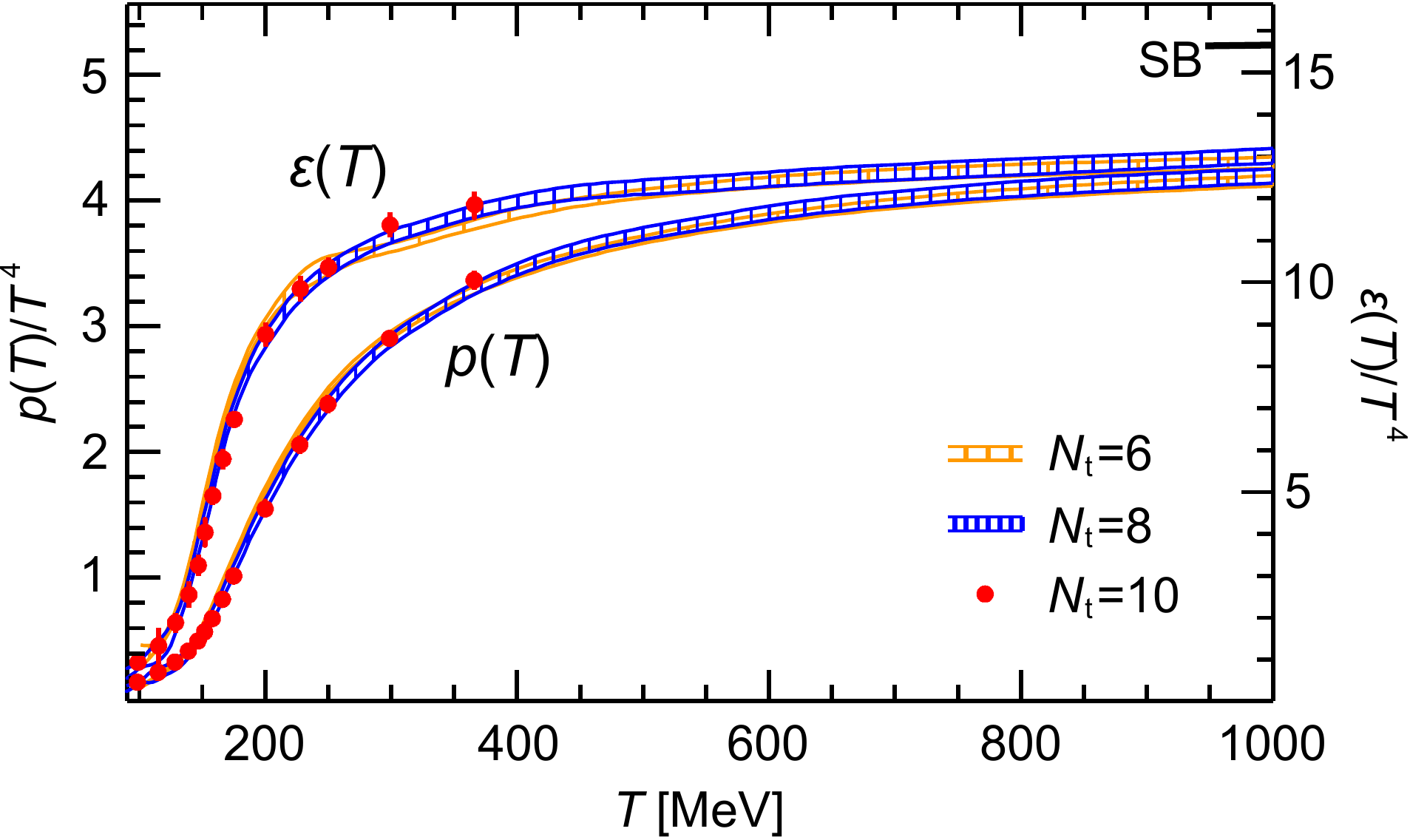}
\end{center}
 \caption{(Left) Renormalized Polyakov loop and the subtracted and
   normalized chiral condensate in the (2+1)-flavor QCD simulation.
   The definition of $\Delta_{\text{l},\text{s}}$ is explained in
   Eq.~\eqref{eq:subtracted}.  Figure adapted from
   Ref.~\cite{Borsanyi:2010bp} and originally two separate plots are
   superimposed with the horizontal axis properly adjusted.  (Right)
   Energy density and pressure in the (2+1)-flavor QCD simulation.
   Figure adapted from Ref.~\cite{Borsanyi:2010cj} and originally two
   separate plots are superimposed with the vertical axis adjusted by
   the Stefan-Boltzmann limit.}  
 \label{fig:fulllattice}
\end{figure}

\paragraph{Full QCD}
Figure~\ref{fig:fulllattice} is the counterpart of
Fig.~\ref{fig:purelattice} for the full QCD simulation with dynamical
quarks.  In this case not only the Polyakov loop $\Phi$ but also the
chiral condensate $\langle\bar{\psi}\psi\rangle$ is an interesting
quantity and they both show smooth crossover as seen in the left panel
of Fig.~\ref{fig:fulllattice}.  To conclude that they are crossover in
a rigorous way, the finite-volume scaling is
necessary~\cite{Aoki:2006we}.  For the practical purpose, it is more
appropriate to use a subtracted order parameter at $T$ defined
by~\cite{Cheng:2007jq},
\begin{equation}
 \Delta_{\text{l},\text{s}} \equiv \frac{M(T)}{M(0)}\;,\qquad
 M(T) \equiv \hat{m}_{\text{s}} \Bigl(
  \langle\bar{\psi}\psi\rangle_{\text{l}}- \frac{m_{\text{l}}}{\mms}
   \langle\bar{\psi}\psi\rangle_{\text{s}}\Bigr) N_\tau^4 \;,
\label{eq:subtracted}
\end{equation}
rather than the na\"{i}ve chiral condensate, where
$\hat{m}_{\text{s}}$ is the strange quark mass in the lattice unit and
$N_\tau$ is the site number along the temporal direction.  This
combination is chosen to be free from an additive divergence
$\sim\mq\Lambda^2$ inherent in the chiral condensate.  This subtracted
order parameter, $\Delta_{\text{l},\text{s}}$, is plotted in the left
panel of Fig.~\ref{fig:fulllattice}.

Interestingly, as seen from the left panel of
Fig.~\ref{fig:fulllattice}, two crossovers of deconfinement and
chiral restoration happen almost simultaneously;  the renormalized
Polyakov loop and $\Delta_{\text{l},\text{s}}$ start increasing and
decreasing, respectively, in the same temperature range.  One may
define the pseudo-critical temperature (see Eq.~\eqref{eq:suscept})
but it is not meaningful to ask whether two pseudo-critical
temperatures coincide or not quantitatively.  As we will discuss
later, the chiral crossover is pretty close to the second-order phase
transition and critical scaling properties are expected, while the
deconfinement crossover spreads over a wide temperature range.  In
this sense, it is often emphasized that only the chiral restoration is
critical and there is no phase transition associated with
deconfinement.  This statement is indeed true, but we would insist
that it is still a highly non-trivial problem how these two crossovers
can be such correlated (see Ref.~\cite{Hatta:2003ga} for a possible
scenario).  In view of the internal energy density, $\varepsilon(T)$,
in the right panel of Fig.~\ref{fig:fulllattice}, deconfinement
crossover occurs in a narrower window of the temperature than
suggested from $\Phi$ in the left panel.  We see that the physical
degrees of freedom rapidly increase within the temperature range
$\lesssim100\MeV$ and the phenomenon of color deconfinement makes
reasonable sense though it does not come along with critical scaling.
In the same way as the pure gluonic case, the pressure $p(T)$ is a
continuous function in the whole $T$ range.

Sometimes, to quantify the crossover location, the pseudo-critical
temperature is defined.  The pseudo-critical temperature should
coincide with the critical temperature when crossover turns to a
second-order phase transition in the chiral limit.  One should keep in
mind that the pseudo-critical temperature is a quantity only for
convenience and it does not add any physical insight.  There are
several different prescriptions which lead to different
pseudo-critical temperatures.

A frequently used prescription for the pseudo-critical temperature is
to see the peak position of the chiral susceptibility,
\begin{equation}
 \chi_{\bar{\psi}\psi} = \frac{T}{V}\frac{\partial^2 \ln Z}
  {\partial \mq^2} \;,
\label{eq:suscept}
\end{equation}
which gives $\Tc=151(3)(3)\MeV$ for chiral restoration, where the
first error comes from the $T\neq0$ and the second from $T=0$
analyses~\cite{Aoki:2006br}.  We note that this estimate gives the
systematic error but does not have uncertainty from the choice of
the prescription.  In view of the order parameter behavior in
Fig.~\ref{fig:fulllattice}, we should consider that the
pseudo-critical temperatures from various definitions could not avoid
substantial uncertainty more than $\Delta\Tc\sim \pm20\MeV$.

\begin{figure}
\begin{center}
 \includegraphics[width=0.44\textwidth]{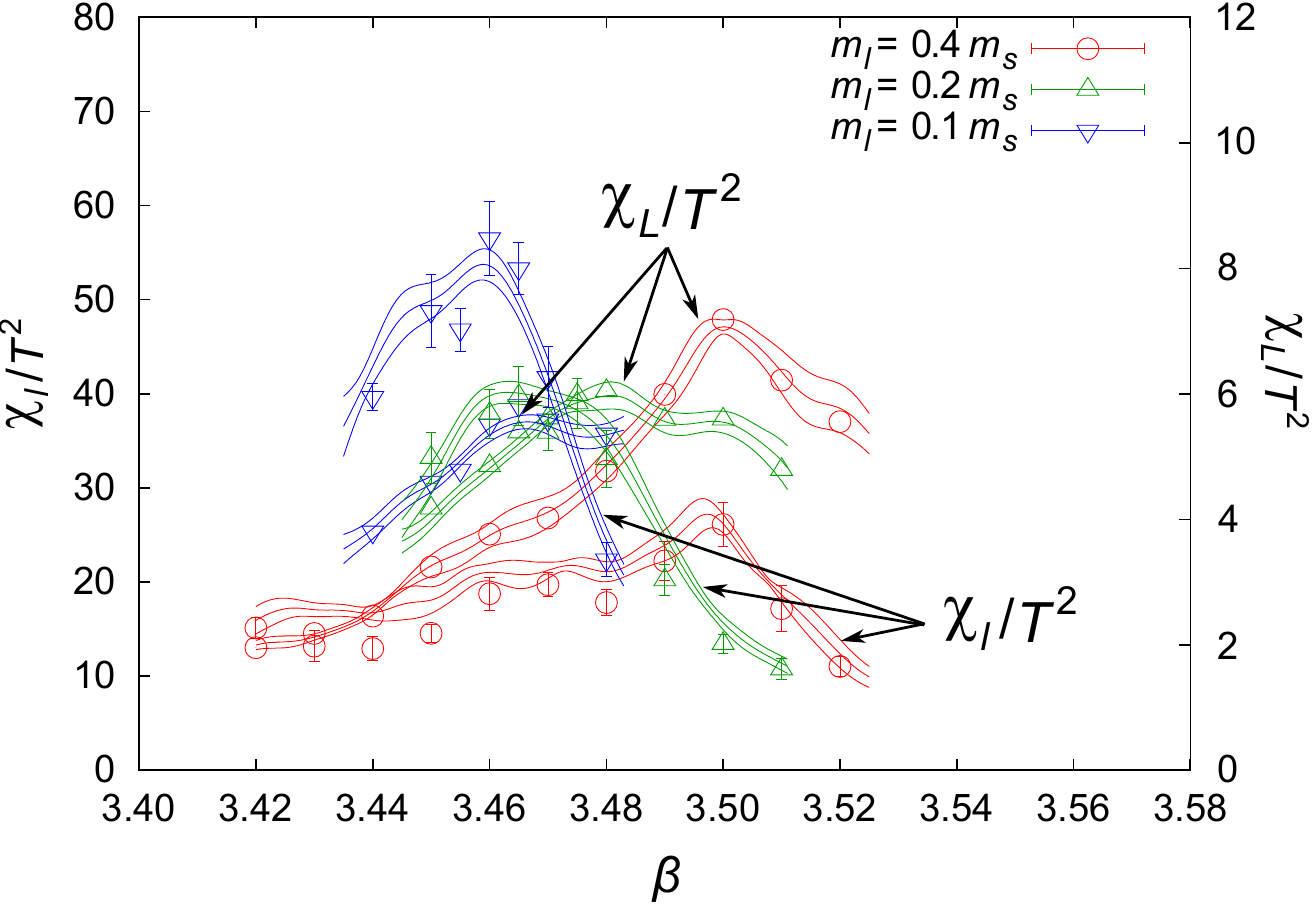}\hspace{1.8em}
 \includegraphics[width=0.36\textwidth]{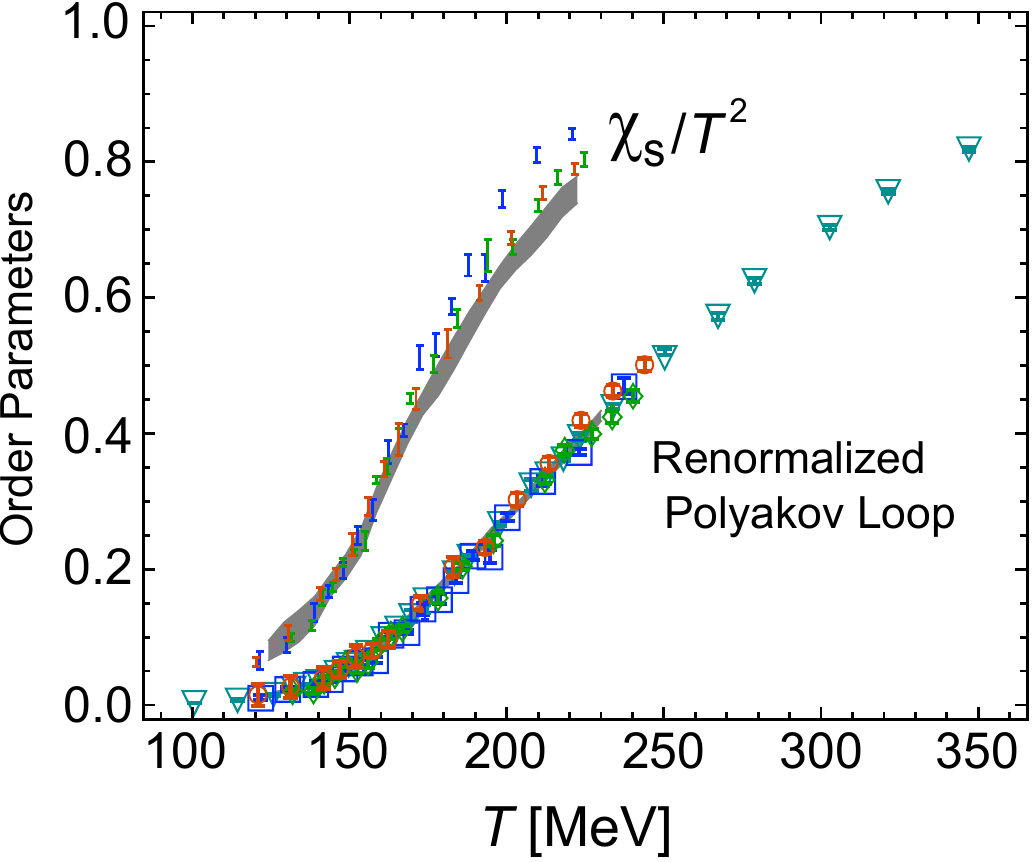}
\end{center}
 \caption{(Left) Peaks in the Polyakov loop susceptibility $\chi_L$
   and the chiral susceptibility $\chi_l$, from which the
   pseudo-critical temperatures are defined for various quark masses.
   Figure adapted from Ref.~\cite{Cheng:2006qk} and originally three
   separate plots are superimposed.  (Right) Strange quark number
   susceptibility in the (2+1)-flavor lattice simulation together with
   the renormalized Polyakov loop.  The pseudo-critical temperature
   for deconfinement is defined from the inflexion point of the
   susceptibility with respect to the temperature.  Figure adapted
   from Ref.~\cite{Aoki:2006br} and originally two separate plots are
   superimposed.}
 \label{fig:suscept}
\end{figure}

It is obvious at first glance from Fig.~\ref{fig:fulllattice} that
there is no prominent phase transition at all associate with
deconfinement in full QCD data, while $\Phi$ serves as a good order
parameter in the pure Yang-Mills theories.  Thus, the
pseudo-critical temperature for deconfinement is less meaningful than
that for the chiral phase transition.  One way to locate the
deconfinement pseudo-critical temperature is to look for the inflexion
point of $\Phi(T)$ or the peak position of the temperature derivative
of the Polyakov loop, $d\Phi(T)/dT$.  This prescription gives an
estimate $\Tc=176(3)(4)\MeV$~\cite{Aoki:2006br}.  Needless to say, the
real uncertainty width should be much larger than the errors, that is,
$\Delta\Tc\sim \pm 50\MeV$ at least from the curve in the left panel
of Fig.~\ref{fig:fulllattice}.  Another possible way to define the
pseudo-critical point is to use the peak position of the Polyakov loop
susceptibility in the same manner as for chiral restoration.  The left
panel of Fig.~\ref{fig:suscept} shows an example at various light
quark masses.  We can notice a general trend that two pseudo-critical
temperatures stick to each other, which is clearer for larger quark
mass ($m_{\text{l}}=0.4\mms$) and becomes vague for small (physical)
quark mass ($m_{\text{l}}=0.1\mms$).  The latest lattice data can be
found in Ref.~\cite{Bazavov:2011nk}.

Instead of the Polyakov loop, it is also possible to utilize the
strange quark number susceptibility to define the pseudo-critical
temperature;
\begin{equation}
 \chi_{\text{s}} = \frac{T}{V}\frac{\partial^2 \ln Z}
  {\partial \mu_{\text{s}}^2} \;,
\end{equation}
where $\mu_{\text{s}}$ is the chemical potential for strange quarks.
This quantity, $\chi_{\text{s}}$, behaves in a similar way to the
Polyakov loop (see the right panel of Fig.~\ref{fig:suscept}), which
can be easily understood in a model as commented below
Eq.~\eqref{eq:l_coupling}.  The inflexion point of $\chi_{\text{s}}$
gives the pseudo-critical point consistent with the definition by
means of the Polyakov loop,
i.e.\ $\Tc=175(2)(4)\MeV$~\cite{Aoki:2006br}.

In the application of the lattice-QCD data to the heavy-ion collision
experiment it is crucially important to check how close to criticality
the QCD phase transition is.  If $u$- and $d$-quarks are massless, the
chiral phase transition should be of second order that belongs to the
O(4) universality class as argued in Sec.~\ref{sec:chiral}.  If the
physical quark mass is situated in the critical region, the subtracted
order parameter $M$ in Eq.~\eqref{eq:subtracted} should obey the
critical scaling, which can be parametrized in terms of the
universality scaling function and scaling variables.  The lattice data
as reported in Ref.~\cite{Ejiri:2009ac,Kaczmarek:2011zz} is consistent
with the O(4) or O(2) universality class. (The O(2) universality class
is also expected for the coarse staggered fermion.)  Therefore, the
pseudo-critical temperature for chiral restoration conveys a definite
meaning if it is fixed such that the magnetic equation of state is
fitted.

\paragraph{Finite density}
Because the chiral pseudo-critical temperature is relatively
well-defined, it makes sense to draw a phase boundary on the QCD phase
diagram using this pseudo-critical temperature.  The Monte-Carlo
simulation is not applicable so far at finite density, but it is
feasible to include the density effect by means of the Taylor
expansion in terms of $\muq/T$.  In this way one can estimate the
$\muq$-dependence of the pseudo-critical temperature or the curvature
of the phase boundary with respect to $\muq$.  That is, the
pseudo-critical temperature is expanded as
\begin{equation}
 \frac{\Tc(\muq)}{\Tc} = 1 - \kappa_{\text{q}}\Bigl(\frac{\muq}{T}
  \Bigr)^2 + \calO\Bigl( \Bigl(\frac{\muq}{T}\Bigr)^4
  \Bigr) \;,
\end{equation}
and recent lattice-QCD results yield the estimate,
$\kappa_{\text{q}}=0.059$ based on $\chi_{\bar{\psi}\psi}$ in
Ref.~\cite{Kaczmarek:2011zz}, $\kappa_{\text{q}}=0.059$ based on
$\chi_{\bar{\psi}\psi}$ and $\kappa_{\text{q}}=0.080$ based on
$\chi_{\text{s}}$ in Ref.~\cite{Endrodi:2011gv} (see also
Ref.~\cite{Philipsen:2008gf} for former results).  These values are
smaller than the curvature expected from the freeze-out line of the
heavy-ion collision experiment, but more than twice larger values are
reported based on the EoS~\cite{Borsanyi:2012cr}.  Also, there are
suggestive data on the QCD critical point, which will be introduced
later in Sec.~\ref{sec:QCDcp}.

\paragraph{Effective $\UA$ symmetry restoration}
Apart from thermodynamics, there are useful data available from the
lattice simulation, and among many, let us mention on suggestive data
for effective $\UA$ symmetry restoration.  It is not quite feasible to
measure the strength of the KMT interaction or the instanton density
directly, and instead, the (pure) topological susceptibility
$\chi_{\rm top}$ can detect the instanton fluctuations and should be
connected to the KMT interaction.  The lattice simulation at finite
$T$ has resulted in a rapid drop of $\chi_{\rm top}$ around
$T\sim\Tc$~\cite{Vicari:2008jw}, which is a circumstantial evidence
for the $\UA$ restoration above $\Tc$.  More direct analyses on the
$\UA$ symmetry encoded in the Dirac spectra are
ongoing at finite $T$ but zero $\muq$~\cite{:2012jaa}.  It might be
even possible to probe the in-medium $\eta'$ mass
experimentally~\cite{Csorgo:2009pa}, and thus quantitative outputs
from the lattice-QCD simulation should be more and more important in
the future studies.

\section{Nuclear Matter}
\label{sec:nuclear}

In this section we make a review over fundamental properties of
nuclear matter and raise some theoretical issues at high baryon
density before quark matter emerges.  Among various observables
in nuclear matter, the saturation density and the binding energy are
the most basic physical quantities.  The binding energy is inferred
from the semi-empirical mass formula, i.e.\ the Bethe-Weizs\"{a}cker
mass formula, $M(Z,N) = Z m_p + N m_n - B(Z,N)$, with the binding
energy parametrized as
\begin{equation}
 B(Z,N) = a_V A - a_S A^{2/3} - a_C \frac{Z^2}{A^{1/3}}
  - a_{\text{sym}} \frac{(N-Z)^2}{A} + \delta_A \;,
\end{equation}
where $A$ is the mass number.  The first, the second, the third, and
the fourth terms represent the volume energy, the surface energy, the
Coulomb energy, and the symmetry energy, respectively.  The last term
is the pairing energy.  For nuclear matter where $A$ and the volume
$V$ are infinitely large with $\rho=A/V$ fixed, the surface term and
the pairing term are irrelevant.  For symmetric nuclear matter ($Z=N$)
the symmetry energy also drops off.  (In this article our concerns are
limited to symmetric nuclear matter only.)  Then, the empirical value
for the binding energy per nucleon in symmetric nuclear matter is
$B/A\simeq 16.3\MeV$, and this is realized at the saturation density,
$\rho_0\simeq 0.17\;\text{nucleons}/\fm^3$.  The Fermi momentum
corresponding to $\rho_0$ is $\kF\simeq 260\MeV$ in a free Fermi gas,
which is less than one third of the nucleon mass $\mn\simeq 939\MeV$.
Therefore, in this sense of $\kF/\mn<1$, normal nuclear matter can be
regarded as a \textit{dilute} system of nucleons which allows for an
expansion in terms of $\kF$.  This fact is essential to distinguish it
from quarkyonic matter as we will address in details in
Sec~\ref{sec:quarkyonic}.  From this observation it should be
legitimate to approximate the binding energy in the following expanded
form;
\begin{equation}
 \bar{B}(\kF) = \frac{B}{A} = -\frac{3\kF^2}{10\mn} + \alpha
  \frac{\kF^3}{\mn^2} - \beta\frac{\kF^4}{\mn^3} \;,
\end{equation}
where the first term is the kinetic energy of a free Fermi gas and the
cubic and the quartic terms represent the interaction effects with
dimensionless parameters $\alpha$ and $\beta$.  In
Ref.~\cite{Kaiser:2001jx} it is argued that $\alpha=5.27$ and
$\beta=12.22$ can reproduce the binding energy and the saturation
point, and also the nuclear matter compressibility,
$K=\kF^2(\partial^2\bar{B}/\partial\kF^2)=236\MeV$, within a
reasonable range.  This simple parametrization works fine to describe
the ground state properties of nuclear matter, but it does not suffice
for the finite-$T$ research on the phase transitions.  It is, thus,
indispensable to develop systematic instruments to evaluate $\alpha$
and $\beta$ from microscopic dynamics.

\subsection{Relativistic mean-field model}
\label{sec:Walecka}

In the Walecka model or the $\sigma$-$\omega$
model~\cite{Walecka:1974qa}, nuclear matter is approximated by a Fermi
gas of in-medium nucleons that feel mean-fields in the scalar channel
(denoted by $\sigma$) and the vector channel (denoted by $\omega$).
This is an old and simple model but is still useful to test
theoretical ideas and derive an EoS even in contemporary contexts (see
Ref.~\cite{Floerchinger:2012xd} for example for the application to the
relativistic heavy-ion collision).  Moreover, the calculation
techniques for the mean-field treatment in the Walecka model are in
complete parallel to those in quark matter.  This analogue is quite
useful not only on the technical level but also for the physical
interpretation of the first-order phase transition and the associated
critical point at the terminal of the first-order phase boundary.

\subsubsection{Saturation properties}
\label{sec:w_calc}

The bare nucleon-nucleon ($NN$) potential has an origin in the pion
exchange and also heavier mesons for a shorter distance.  In the
mean-field approximation it is assumed that this interaction effects
can be renormalized in a one-body potential represented by the
mean-field variables, so that we can drop the pionic degrees of
freedom until they are thermally excited at high enough temperature.
Then, the Walecka model is defined by the Lagrangian density,
\begin{equation}
 \calL = \bar{\psi}\bigl( i\gamma_\mu\partial^\mu + \muB \gamma_0
  - g_\omega \gamma_\mu \omega^\mu - \mn + g_\sigma \sigma \bigr)\psi
  +\frac{1}{2}\bigl( \partial_\mu\sigma\partial^\mu\sigma
  - m_\sigma^2\sigma^2 \bigr)
  -\frac{1}{4}\omega_{\mu\nu}\omega^{\mu\nu}
  +\frac{1}{2}m_\omega^2\omega_\mu\omega^\mu
\label{eq:Lag_w}
\end{equation}
with the following model parameters; the bare nucleon mass $\mn$,
the $\sigma$-meson mass $m_\sigma$, the $\omega$-meson mass
$m_\omega$, the coupling constants in the scalar channel $g_\sigma$,
and in the vector channel $g_\omega$, which are determined to fit the
nuclear matter properties.  We note that $\psi$ above represents the
Dirac spinor for nucleon as if it were a point particle.  Since the
nucleon is a composite state having a complicated internal structure,
the nucleon mass at high density cannot avoid ambiguity, which will be
the central subject of Sec.~\ref{sec:nucleon_mass}.

In the mean-field approximation, both $\sigma$ and $\omega_0$ can take
a finite value induced by the non-zero chiral condensate
$\langle\bar{\psi}\psi\rangle$ and the baryon density
$\langle\psi^\dagger\psi\rangle$.  If we are interested in the
isospin non-symmetric situation such as neutron matter, we have to
include the $\rho$ meson as well.  In the above we choose a convention
in which $\sigma$ is negative (i.e.\ the same sign as the chiral
condensate).  In this approximation the mean-field Lagrangian density
reads,
\begin{equation}
 \calL_{\text{MF}} = \bar{\psi}\bigl[ i\gamma_\mu\partial^\mu
  + (\muB-g_\omega \omega^0)\gamma_0 - (\mn - g_\sigma \sigma)
  \bigr]\psi
  - \frac{1}{2}m_\sigma^2 \sigma^2 + \frac{1}{2}m_\omega^2 (\omega^0)^2 \;.
\end{equation}
From this we can immediately identify the renormalized or in-medium
nucleon mass and the effective baryon chemical potential as
\begin{equation}
 \mn^\ast = \mn - g_\sigma \sigma \;, \qquad
 \muB^\ast = \muB - g_\omega \omega^0 \;.
\label{eq:gap_eqs}
\end{equation}
Then, the thermodynamic potential divided by the volume $V$ (denoted
simply by $\Omega$ throughout this review) for a free quasi-nucleon
gas is expressed as
\begin{equation}
 \Omega = -2\cdot 2 T\! \int\frac{d^3 p}{(2\pi)^3} \!\biggl\{
  \ln\bigl[1+e^{-\beta(\omega-\muB^\ast)}\bigr]
  + \ln\bigl[1+e^{-\beta(\omega+\muB^\ast)}\bigr]
  \biggr\} + \frac{m_\sigma^2 (\mn^\ast-\mn)^2}{2g_\sigma^2} -
  \frac{m_\omega^2 (\muB^\ast-\muB)^2}{2g_\omega^2} \;,
\label{eq:Omega_w}
\end{equation}
where $2$ comes from the isospin degeneracy and another $2$ from the
spin degeneracy.  We note that the zero-point oscillation energy is
dropped, which corresponds to the approximation in the nuclear
hydrodynamics~\cite{Buballa:1996tm}.  It should be mentioned that one
should be careful of the treatment of the zero-point oscillation
energy if the fermion mass is small, as explained in
Sec.~\ref{sec:quark-meson}.  The mean-field variables, $\mn^\ast$ and
$\muB^\ast$, are to be fixed from the stationary conditions;
$\partial\Omega/\partial\mn^\ast=\partial\Omega/\partial\muB^\ast=0$,
leading to the gap equations,
\begin{align}
 \mn^\ast &= \mn - 4\frac{g_\sigma^2}{m_\sigma^2}\int\frac{d^3 p}{(2\pi)^3}
  \frac{\mn}{\omega_p}\bigl[ \nF(\omega_p-\muB^\ast)
  + \nF(\omega_p+\muB^\ast) \bigr] \;,\\
 \muB^\ast &= \muB - 4\frac{g_\omega^2}{m_\omega^2}\int\frac{d^3 p}{(2\pi)^3}
  \bigl[ \nF(\omega_p-\muB^\ast) - \nF(\omega_p+\muB^\ast) \bigr] \;.
\end{align}
With the solutions of these equations substituted to $\Omega$ in
Eq.~\eqref{eq:Omega_w}, we can get the thermodynamic
potential as a function of $T$ and $\muB$, from which all
thermodynamic quantities are calculable.  The baryon density is
expressed by $\rho = -\partial\Omega/\partial\muB$ and the internal
energy density by $\varepsilon = \Omega + T s + \muB \rho$.

\begin{figure}
 \includegraphics[width=0.4\textwidth]{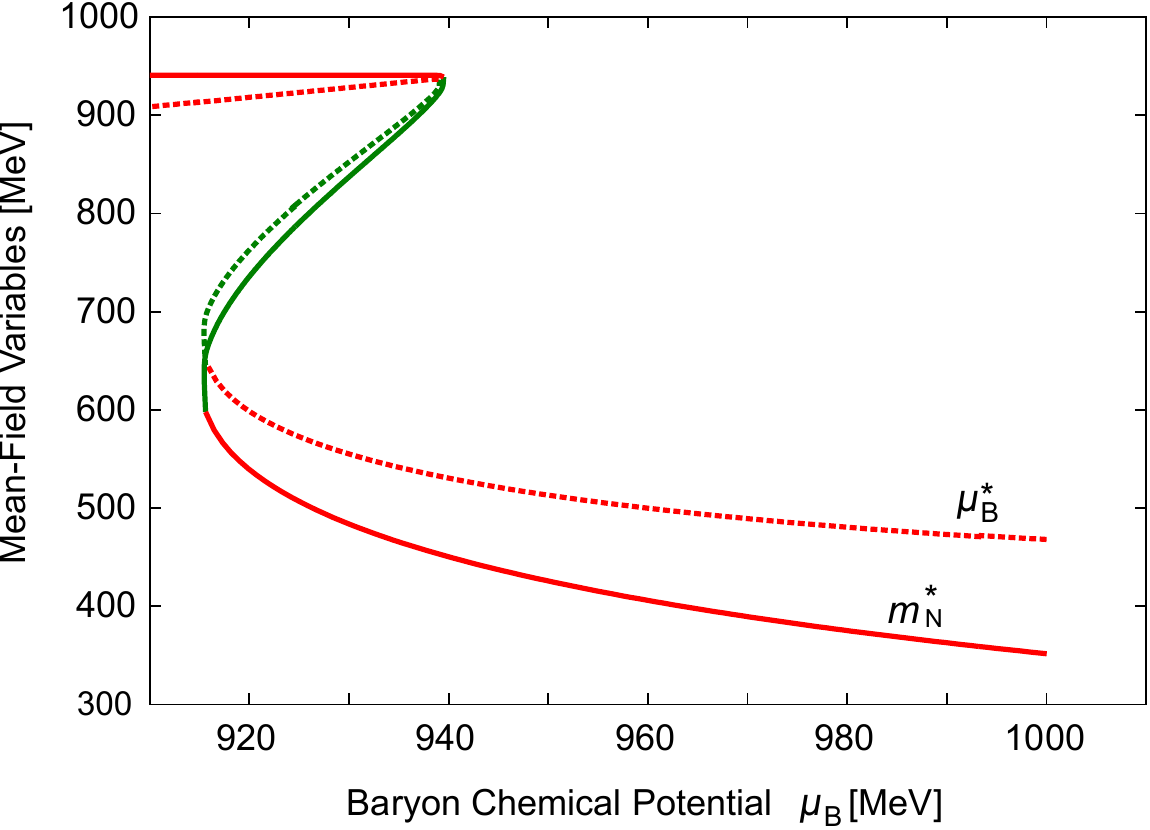}\hspace{4em}
 \includegraphics[width=0.41\textwidth]{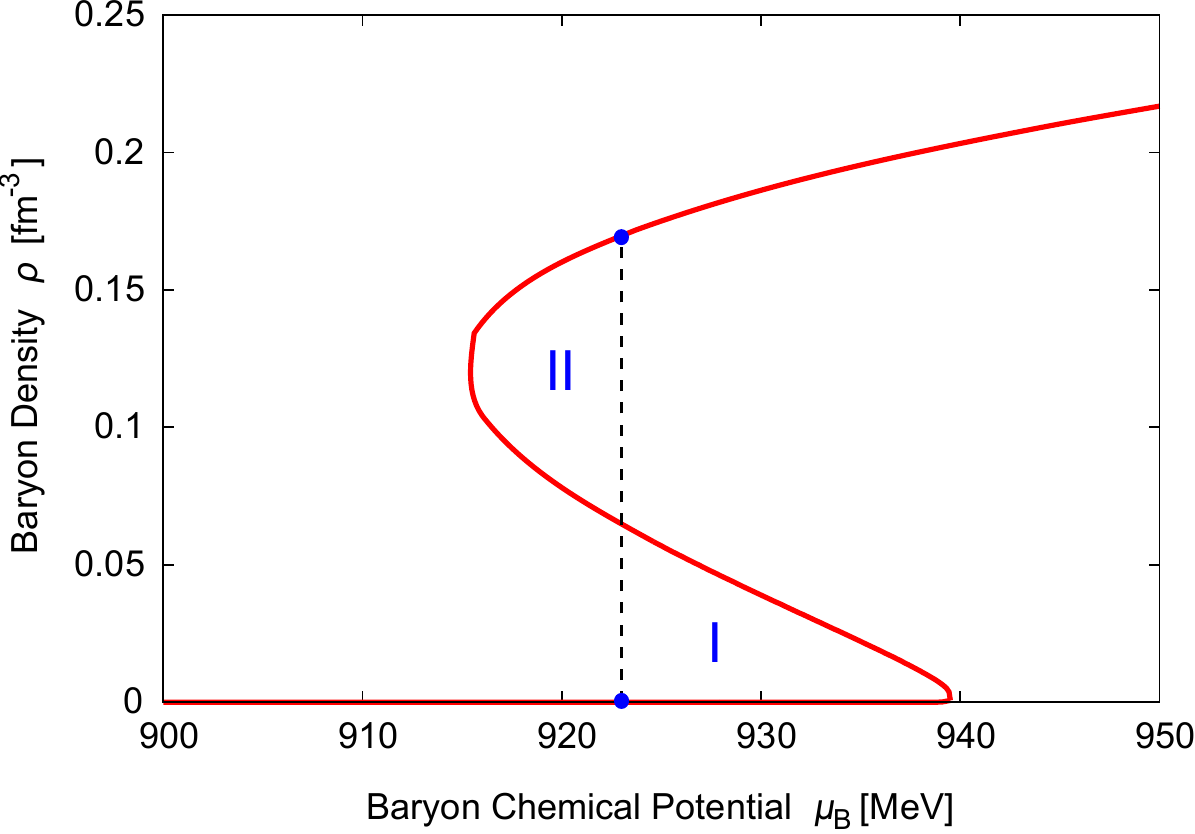}
 \caption{(Left) Mean-field variables as functions of $\muB$.  There
   is a first-order phase transition from the low-density state in
   which the nucleon mass is the bare one to the high-density state in
   which the in-medium nucleon mass drops to almost a half.  The green
   curves represent the unstable solutions of the gap equation.
   (Right) Baryon density as a function of the chemical potential.
   The first-order phase transition occurs so that the area in the
   region~I is equal to that in the region~II.}
 \label{fig:walecka}
\end{figure}

Two coupling constants are chosen so as to satisfy the following
saturation properties of nuclear matter, that is,
\begin{equation}
 \frac{\varepsilon}{\rho}\biggr|_{\rho=\rho_0} = \mn - 16.3\MeV\;,
 \qquad
 \frac{d (\varepsilon/\rho)}{d\rho} \biggr|_{\rho=\rho_0} = 0\;.
\label{eq:saturation}
\end{equation}
We make a remark that the latter is understood as a stability
condition for nuclear droplet.  In fact, this density derivative
at $T=0$ leads to $d(\varepsilon/\rho)/d\rho
=\muB/\rho - \varepsilon/\rho^2 = p/\rho^2 = 0$.  We note that this
$p$ is a pressure difference between the nuclear droplet and the empty
vacuum.  When we discuss the liquid-gas phase transition of quark
matter, we will come back to this point.  A set of parameters
consistent with the nuclear matter properties
is~\cite{Buballa:1996tm},
\begin{equation}
 \mn = 939\MeV\;,\quad
 m_\sigma = 550\MeV\;,\quad
 m_\omega = 783\MeV\;,\quad
 g_s = 10.3\;,\quad
 g_\omega = 12.7\;.
\end{equation}

The behavior of the in-medium mass $\mn^\ast$ is plotted in the left
panel of Fig.~\ref{fig:walecka} together with the effective chemical
potential $\muB^\ast$.  It is clear from Fig.~\ref{fig:walecka} that
there is a first-order liquid-gas phase transition of nuclear matter
(or the liquid-vacuum phase transition, strictly speaking).  The
nuclear matter should exhibit a first-order phase transition at
$\muB=\mn-B/A\simeq 923\MeV$.  In thermodynamics, equivalently, we can
find this critical value of $\muB$ from the condition that the
thermodynamic potential $\Omega$ becomes balanced in two states.  In
view of the curve of $\rho(\muB)$ in the right panel of
Fig.~\ref{fig:walecka} the phase transition point is determined so
that the area in the region~I is equal to that in the
region~II.\ \ This can be understood from
$\Delta\Omega({\muB}_c) = \Omega({\muB}_c)-\Omega({\muB}_c)
=-\oint_{{\muB}_c}^{{\muB}_c}\rho(\mu)d\mu=0$ along the path from the
lower dot to the upper dot in the right panel of
Fig.~\ref{fig:walecka}.  Such a way to find ${\muB}_c$ is reminiscent
of the Maxwell construction for $p$ and $V$.

The left panel of Fig.~\ref{fig:walecka-t} is the saturation curve of
the energy per nucleon as a function of the baryon density $\rho$.
Obviously the saturation point and the binding energy agree with the
values in Eq.~\eqref{eq:saturation} as a result of the fit.  In this
simplest setup, however, the compressibility cannot be reproduced
well, and a potential term $V(\sigma,\omega)$ should be added in
Eq.~\eqref{eq:Lag_w}~\cite{Boguta:1977xi}, which we will not discuss,
for we are mostly interested in the qualitative aspects of the phase
structure.

\subsubsection{Liquid-gas phase transition of symmetric nuclear matter}

\begin{figure}
\begin{center}
 \includegraphics[width=0.39\textwidth]{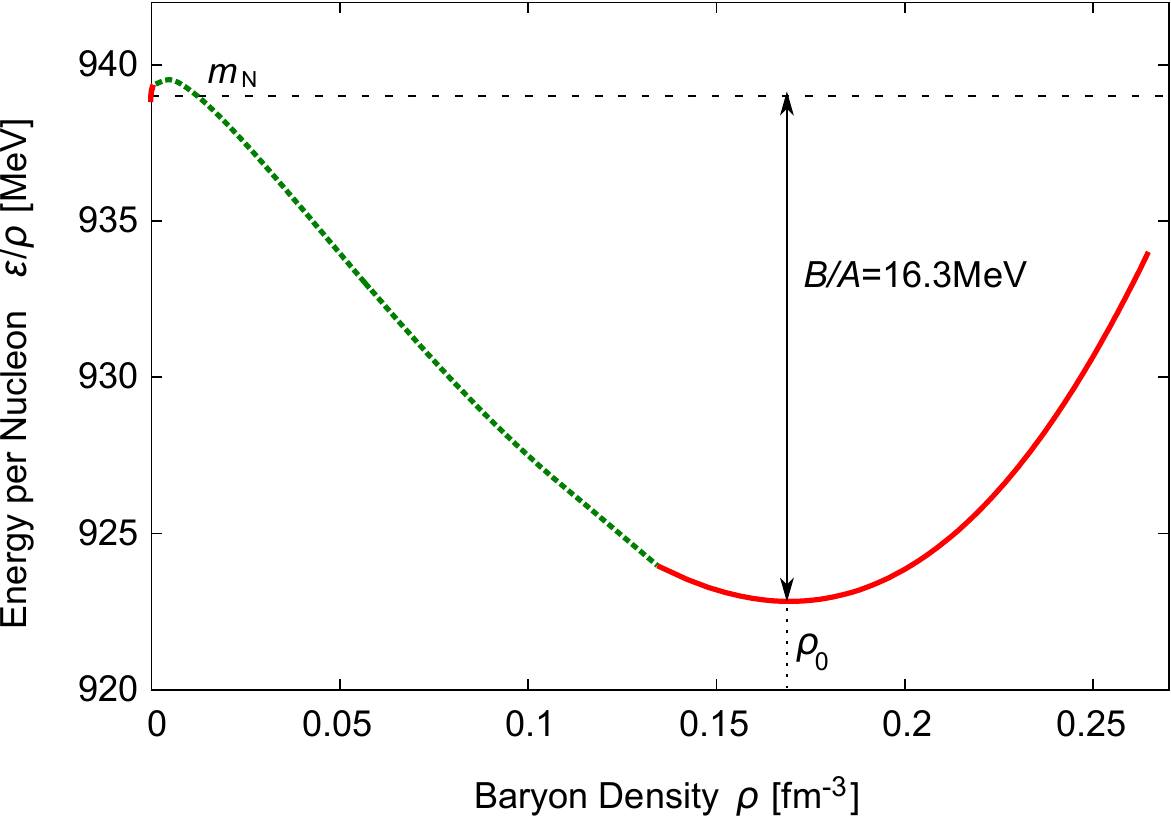} \hspace{4em}
 \includegraphics[width=0.39\textwidth]{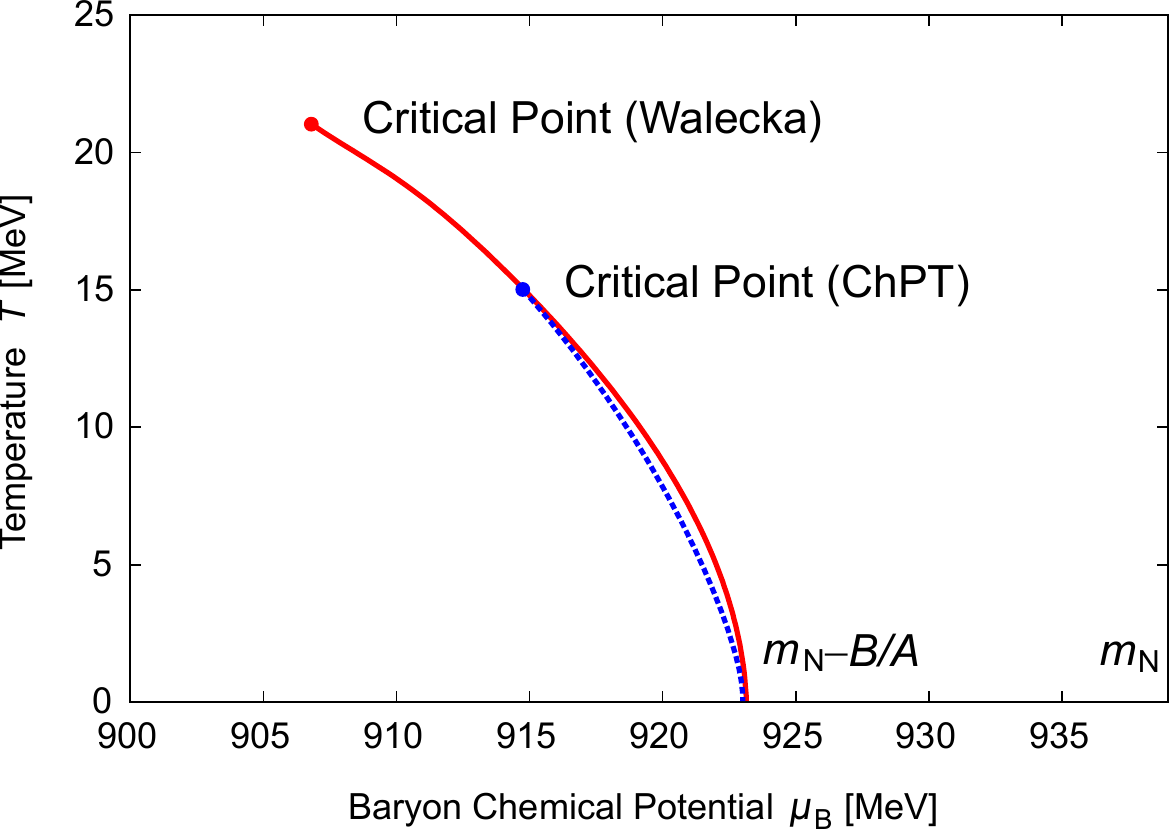}
\end{center}
 \caption{(Left) The saturation curve of the energy per nucleon as a
   function of the baryon density $\rho$.  The global minimum is
   located at the saturation density.  The dotted curve represents the
   unstable solution of the gap equation.  (Right) First-order phase
   boundary and the critical point in the Walecka model and in the
   chiral perturbation theory in the three-loop
   order~\cite{Fiorilla:2011sr}.}
 \label{fig:walecka-t}
\end{figure}

The existence of the first-order phase transition is an inevitable
consequence from the saturation properties of symmetric nuclear
matter.  This world as it is, in other words, embodies the mixed
state, and in general, a self-bound system of fermionic particles
should have a first-order phase transition.

Differently than the second-order phase transition governed by the
global symmetry and its spontaneous breaking, the first-order phase
transition in nuclear matter does not change the symmetry pattern at
all.  Because two states before and after the transition have a
different density as seen clearly in the right panel of
Fig.~\ref{fig:walecka}, one can regard this type of the transition as
associated with the density.  Then, typically, the dilute phase
corresponds to a gaseous state and the dense phase corresponds to a
liquid state, and nuclear matter hence results in the liquid-gas phase
transition.  Right at the first-order phase transition a mixed state
should be formed, during which the gas and the liquid parts can
coexist in space.  The most optimal spatial shape should emerge from
the competition between the Coulomb energy and the surface energy (see
Ref.~\cite{Tatsumi:2011tt} and references therein), and the resulting
peculiar shapes are referred to as various ``pasta'' phases.

To fit the saturation properties the Walecka model was solved at
$T=0$, and it is straightforward to extend such model calculations to
the finite-$T$ situation.  We have already written the gap
equations~\eqref{eq:gap_eqs} down with the matter terms.  Then,
solving them numerically, we can make a prediction for the first-order
boundary elongated toward $T\neq0$.  In the Walecka model meson
fluctuations are totally neglected, and thus, we should not apply this
model to the analysis at high $T$ where abundant pions are thermally
excited.  As noticed in the numerical results in
Fig.~\ref{fig:walecka-t}, the terminal point of the first-order phase
boundary (i.e.\ the critical point) lies around $T\sim 20\MeV$, which
is small enough to justify our neglecting pion fluctuations.

The phase boundary tends to bend toward smaller chemical potential
with increasing temperature, which can be understood from the
Clausius-Clapeyron relation~\cite{Halasz:1998qr}.  On the
first-order phase transition the thermodynamic potential should be
balanced between two phases.  If we move along the phase boundary from
$T$ and $\muB$ by infinitesimal $dT$ and $d\muB$, the thermodynamic
potentials, $\Omega_{\text{dilute}}$ in a dilute state and
$\Omega_{\text{dense}}$ in a dense state should remain equal, which
immediately leads to
\begin{equation}
 \frac{\partial\Omega_{\text{dilute}}}{\partial T}dT
  + \frac{\partial\Omega_{\text{dilute}}}{\partial\muB}d\muB =
 \frac{\partial\Omega_{\text{dense}}}{\partial T}dT
  + \frac{\partial\Omega_{\text{dense}}}{\partial\muB}d\muB
 \quad \Rightarrow \quad
 \frac{dT}{d\muB} = -\frac{\Delta\rho}{\Delta s} \;,
\end{equation}
where $\Delta\rho=\rho_{\text{dense}}-\rho_{\text{dilute}}$ is the
density difference and $\Delta s=s_{\text{dense}}-s_{\text{dilute}}$
is the entropy difference.  The slope $dT/d\muB$ is negative in
Fig.~\ref{fig:walecka-t} and this implies that $\Delta\rho$ and
$\Delta s$ should have the same sign.  In most cases, indeed, the
entropy density becomes greater in the dense state than in the dilute
state, which means $\Delta\rho/\Delta s>0$.

From the theoretical point of view the underlying mechanism in quark
matter to produce a first-order phase transition is exactly the same
as that in the Walecka model we elucidated here.  In the case of quark
matter, however, there is no strong support for the existence of the
first-order phase transition.

\subsection{Chiral perturbation theory}

One of the most advanced methods to deal with nuclear matter in a
relativistic framework is the chiral perturbation theory.  We have
already introduced a Lagrangian~\eqref{eq:lsm} that describes
low-energy chiral dynamics.  It was necessary to utilize the linear
sigma model because $\sigma$ can be a soft mode at chiral restoration.
At low temperatures, however, chiral symmetry is significantly broken,
and the low-energy dynamics is completely ruled by the light NG
bosons.  Hence, it should be the most appropriate to formulate an
effective theory in terms of the NG bosons only, which is obtainable
by integrating $\sigma$ out from the linear sigma model.  In such a
way, one can come by a non-linear representation of the low-energy
effective theory.  Actually, thanks to the low-energy theorems, one
can adopt any representation only to get the same physics results
(Haag theorem)~\cite{Haag:1958vt}, which provides us with a systematic
and model-independent tool to disclose the properties of hadronic
matter.

\subsubsection{Effective Lagrangian and the power counting}

In the vacuum we do not have to consider the scalar particles because
they are much heavier than the NG bosons.  Then, as seen in
Eq.~\eqref{eq:chiral_field}, the theory is defined as a function of
$\btau\cdot\bpi$ where $\btau$ is the Pauli matrices in isospin space
in the $\Nf=2$ case.  It is the most useful choice to employ a
representation in terms of unitary matrix,
\begin{equation}
 U = e^{i\btau\cdot\bpi/f_\pi} \;, \qquad
 U \;\to\; L^\dagger\, U\, R \;,
\end{equation}
having the same transformation property as $M$ with the $\sigma$ meson
dropped off.  The chiral-symmetric Lagrangian density is to be
expanded in terms of $U$ and derivatives;  the lowest-order is
\begin{equation}
 \calL_\pi^{(2)} = \frac{f_\pi^2}{4}\tr(D_\mu U D^\mu U^\dagger)
  +\frac{f_\pi^2}{4}\tr(\chi U^\dagger + U\chi^\dagger) \;,
\label{eq:LagE2}
\end{equation}
with two derivatives and and one symmetry breaking term.  The
parameter in the latter term is $\chi = 2B_0 (s+ip)$ where $s$ is the
quark mass matrix.  Using $B_0$ one can express the chiral condensate
as $\langle\bar{q}q\rangle=-f_\pi^2 B_0$ and the pion mass as
$m_\pi^2 = (m_u+m_d)B_0$ which is consistent with the
Gell-Mann-Oakes-Renner relation~\eqref{eq:GOR}.  The covariant
derivative is defined as
$D_\mu U = \partial_\mu U + i\ell_\mu U - i U r_\mu$ that
transforms as $D_\mu U\;\to\;L^\dagger(x)(D_\mu U)R(x)$ with properly
defined $\ell_\mu$ and $r_\mu$ for local chiral rotations $L(x)$ and
$R(x)$ as in the gauge theory.

The chiral effective Lagrangian~\eqref{eq:LagE2} can be regarded as
the lowest-order $\calO(p^2)$ contribution in the expansion with
respect to pion momentum scale $p$ as compared to the typical chiral
scale $\Lambda_\chi=4\pi f_\pi\sim 1\GeV$.  Then, the pion mass
$m_\pi$ is $\calO(p)$ and then $\chi\propto m_\pi^2$ is $\calO(p^2)$.
We can perform the theoretical calculation for physical observables up
to $\calO(p^2)$ using this Lagrangian~\eqref{eq:LagE2} at the
tree-level.  To go to the next order of $\calO(p^4)$, one should
include one-loop diagrams from $\calL_\pi^{(2)}$ together with the
tree-level contributions from $\calL_\pi^{(4)}$ at $\calO(p^4)$.  In
the $\mathrm{SU}(\Nf=3)$ case there are 10 independent terms
(including $\chi$, $\ell_\mu$, $r_\mu$) and 10 low-energy constants
are required at $\calO(p^4)$.  In the $\mathrm{SU}(2)$ case there are
less parameters.  These constants are fixed to fit with the
experimental data of the meson masses, $\pi$-$\pi$ scattering, rare
pion decays, $K_{\ell 4}$ decay, $f_K/f_\pi$, etc, and then one can
carry theoretical predictions out.  This systematic program of the
low-energy expansion of the effective theory of QCD in terms of the NG
bosons is called the chiral perturbation theory (ChPT).

Let us consider how to apply the ChPT to figure out the behavior of
the chiral condensate at finite temperature and density.  Because the
operator $\bar{\psi}\psi$ and the quark mass $\mq$ are conjugate to
each other, the chiral condensate $\langle\bar{\psi}\psi\rangle$ is
derived from the derivative of the thermodynamic potential $\Omega$
with respect to $\mq$, that is,
$\langle\bar{\psi}\psi\rangle=-\partial\Omega/\partial\mq$.  Thanks
to the GOR relation~\eqref{eq:GOR}, it is a straightforward procedure
to convert the $\mq$-derivative to that in terms of the pion mass
squared $m_\pi^2$.  Then, in a medium at finite $T$ and density
$\rho$, the deviation of the in-medium chiral condensate from the
vacuum value $\langle\bar{\psi}\psi\rangle_0$ can be expressed as
\begin{equation}
 \frac{\langle\bar{\psi}\psi\rangle_{T,\rho}}
  {\langle\bar{\psi}\psi\rangle_0}
  = 1 - f_\pi^{-2}\biggl( \frac{\partial\Omega}
  {\partial m_\pi^2} + \frac{\sigmaN}{m_\pi^2}
  \frac{\partial\Omega}{\partial\mn} \biggr) \;,
\end{equation}
where the second contribution has the nuclear sigma term that is
defined and estimated as~\cite{Gasser:1990ce,Alarcon:2011zs},
\begin{equation}
 \sigmaN \equiv \mq\frac{\partial\mn}{\partial\mq} \simeq 45\MeV \;.
\end{equation}

\begin{figure}
\begin{center}
 \includegraphics[width=0.39\textwidth]{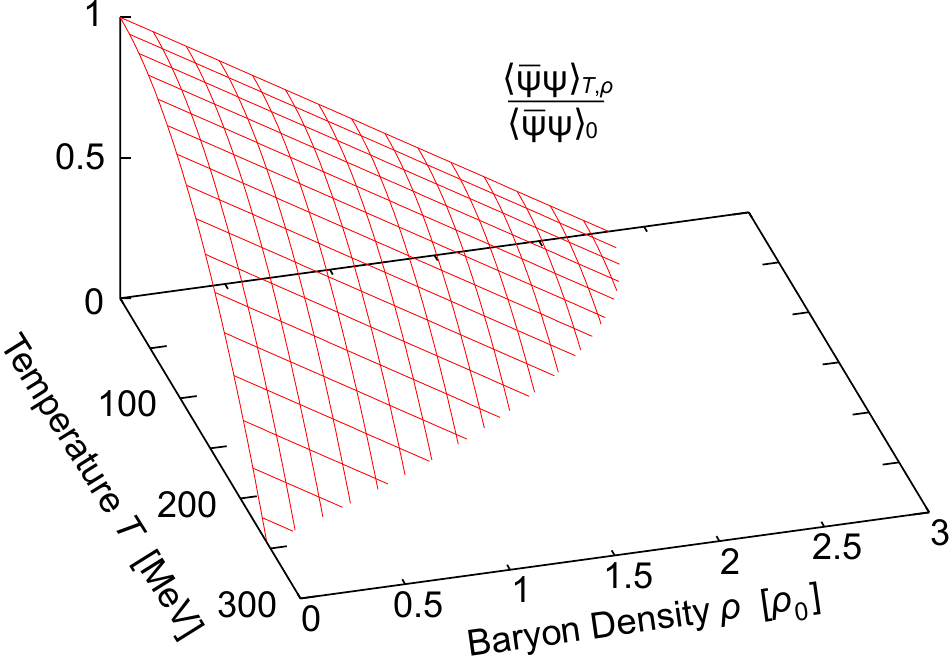} \hspace{4em}
 \includegraphics[width=0.38\textwidth]{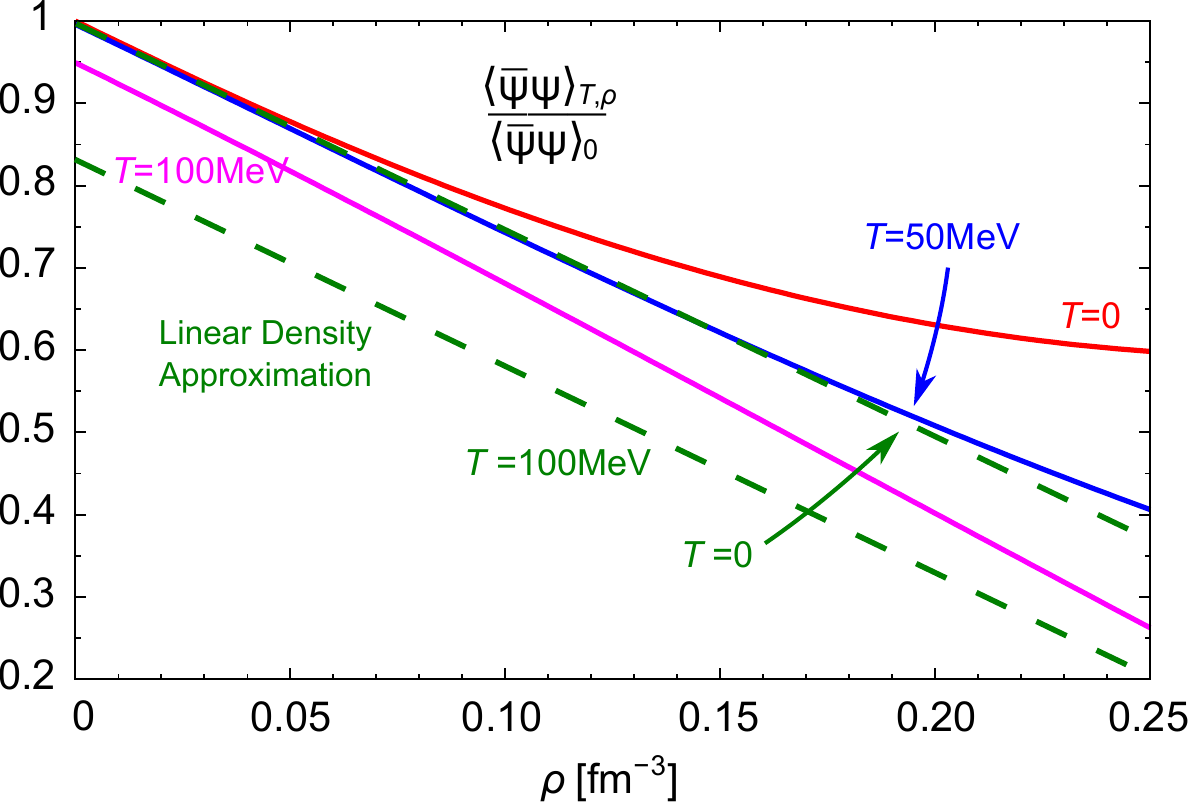}
\end{center}
 \caption{(Left) Chiral condensate at finite $T$ and $\rho$ in the
   limit of the free pion gas and the linear density approximation
   given in Eq.~\eqref{eq:chiralex}.  (Right) Chiral condensate
   calculated in the NLO three-loop order.  Figure adapted from
   Ref.~\cite{Fiorilla:2011qr} and the linear density approximation
   results (dashed lines) are overlaid for reference.}
 \label{fig:ChPTLO}
\end{figure}

The lowest-order correction originates from contributions induced by
$T$ from the free pion gas and $\rho$ from the free nucleon gas.  If
the pion mass can be negligible (i.e.\ $m_\pi\simeq 0$) and $\rho$ is
sufficiently small enough to approximate the scalar density
$\rho_{\text{s}}=-\partial\Omega/\partial\mn$ with the density
$\rho$, the analytical integration is possible, leading
to~\cite{Gasser:1986vb,Gerber:1988tt},
\begin{equation}
 \frac{\langle\bar{\psi}\psi\rangle_{T,\rho}}
  {\langle\bar{\psi}\psi\rangle_0} = 1
  - \frac{T^2}{8 f_\pi^2} - \frac{\sigmaN}{m_\pi^2 f_\pi^2}\,\rho
  + \calO(T^4,\rho^2) \;,
\label{eq:chiralex}
\end{equation}
at the lowest order in the linear density approximation.  The
finite-$m_\pi$ correction is not really negligible, in fact, and the
expansion breaks down around $T\sim 150\MeV$ due to the neglected
excitations of other mesonic states.  Because the $\sigma$ meson
decouples from the dynamics, one has no chance to describe the chiral
phase transition using the non-linear representation.  Nevertheless,
it would be suggestive to see explicitly where the chiral condensate
vanishes with increasing $T$ and $\rho$ according to the lowest-order
expression~\eqref{eq:chiralex}.  The results are depicted in the left
panel of Fig.~\ref{fig:ChPTLO}.  It seems that the chiral condensate
disappears to draw a sort of the phase boundary on the $T$-$\rho$
plane, though the approximation in Eq.~\eqref{eq:chiralex} breaks down
there.  Even though we cannot reach the chiral phase transition in the
ChPT, such in-medium reduction of the chiral condensate or $f_\pi$
can be an experimental measure for the partial restoration of chiral
symmetry~\cite{Suzuki:2002ae} (see also Eq.~\eqref{eq:mediumf}).

\subsubsection{Application to nuclear matter}

It is necessary to take account of the heavy baryonic degrees of
freedom into the dynamics at high density.  The fermion part is
expressed by the Lagrangian,
$\calL = \bar{\psi}\bigl[ i\feyn{\partial}
-g(\sigma-i\btau\cdot\bpi) \bigr]\psi$, which is converted into the
non-linear representation after dropping the scalar field.  The usual
procedure to write the baryon theory down is to perform the field
redefinition as
\begin{equation}
 \xi = e^{i\btau\cdot\bpi/2f_\pi}\;, \qquad
 N_{\text{L}} = \xi^\dagger\psiL \;, \qquad
 N_{\text{R}} = \xi\psiR \;, \qquad
 U = \xi \xi \;.
\end{equation}
Then the fermionic Lagrangian is translated into the form similar to
the gauge theory, $\calL=\bar{N}(i\feyn{D}-\feyn{A}\gamma_5-\boldm)N$,
with $D_\mu=\partial_\mu+iV_\mu$ and the vector and the axial-vector
fields,
\begin{equation}
 V_\mu = -\frac{i}{2}(\xi^\dagger\partial_\mu\xi
  + \xi\partial_\mu\xi^\dagger)\;,\qquad
 A_\mu = -\frac{i}{2}(\xi^\dagger\partial_\mu\xi
  - \xi\partial_\mu\xi^\dagger)\;.
\end{equation}
Including the renormalization constants that change the axial-vector
coupling from unity to $g_A\simeq 1.26$, the nucleon effective
Lagrangian is expanded in terms of the pion fields up to the quadratic
order as
\begin{equation}
 \calL_{\text{N}} = \bar{N}(i\feyn{\partial} - \mn)N
  + \frac{g_A}{f_\pi}\bar{N}\gamma^\mu\gamma_5 \frac{\btau}{2} N \cdot
  \partial_\mu\bpi - \frac{1}{4f_\pi^2}\bar{N}\gamma^\mu \btau\cdot
  \bpi\times\partial_\mu\bpi N + \frac{\sigmaN}{2f_\pi^2}
  \bpi^2\bar{N}N
\end{equation}
with the pseudo-vector $\pi NN$-vertex and the Tomozawa-Weinberg
$\pi\pi NN$-contact vertex.  With these interaction vertices the
thermodynamics and the chiral condensate are derived;  the
next-to-leading-order (NLO) results are calculated and shown in
the right panel of Fig.~\ref{fig:ChPTLO} adapted from
Ref.~\cite{Fiorilla:2011qr}.  The linear density approximation comes
to work fine for $T\gtrsim 50\MeV$, while the $T$-dependence at NLO
shows a significant deviation from Eq.~\eqref{eq:chiralex}.

From the point of view of the \textit{systematic} expansion with
respect to the energy scale, it is not straightforward to formulate
perturbative expansions involving baryons with the consistent power
counting.   A new scale, i.e.\ the nucleon mass $\mn$, is not small as
compared to the chiral symmetry breaking scale $\Lambda_\chi$ and
appears in higher loops that contribute to lower
order.  Thus, the apparent one-to-one correspondence between the loop
and the derivative expansions is, at first glance, lost.  Yet, within
the fully relativistic framework one can retrieve the consistent power
counting for multi-loop diagrams using appropriate renormalization
conditions~\cite{Becher:1999he,Fuchs:2003qc}.  Another resolution to
avoid this drawback is to adopt a heavy baryon reduction yielding the
non-relativistic limit of the original theory, that is, the heavy
baryon chiral perturbation theory (HBChPT) where the power counting is
restored~\cite{Jenkins:1990jv}.  We write the nucleon momentum as
\begin{equation}
 p^\mu = \mn v^\mu + k^\mu \;,
\end{equation}
where $v^\mu$ is the four-velocity with $v^2=1$ and $k^\mu$ is the
residual momentum of order $\LQCD$, so that one can perform the
derivative expansion systematically in energy range below
$\Lambda_\chi$.  The Lagrangian density is written in terms of the
nucleon field $B$ carrying $v^\mu$ which is related to the original
field $N$ via
\begin{equation}
 B(x) = \exp\bigl( i \mn v\cdot x \bigr)\, N(x) \;.
\end{equation}
With this the nucleon part of the Lagrangian density becomes
\begin{equation}
 \calL_{\text{N}} = \bar{N}\bigl(i\gamma^\mu\partial_\mu - \mn\bigr)N
 = \bar{B}\, iv^\mu \partial_\mu B \;.
\end{equation}
Therefore no hard scale appears in the theory.  Higher-order terms are
arranged in powers of $k^\mu/\Lambda_\chi$ and $k^\mu/\mn$.  Loop
calculations in the HBChPT framework are simpler as compared to the
relativistic treatment and the HBChPT has been applied to a broad
range of the nuclear physics problems.  For reviews, see
Refs.~\cite{Bernard:1992qa,Bernard:1995dp,Park:1993jf}.

In the nuclear medium the methodology of chiral effective field theory
can also be formulated.  The in-medium ChPT Lagrangian is derived from
the generating functional after integrating the nucleon fields out but
keeping the particle-hole part of the fermion
determinant.  The in-medium pion decay constants,
\begin{equation}
 \langle N| J_{\text{A}}^0 (x)|\pi(q)\rangle = iq^0\, f_\pi^t\, e^{iqx}\;,
  \qquad
 \langle N| J_{\text{A}}^i (x)|\pi(q)\rangle = iq^i\, f_\pi^s\, e^{iqx}\;,
\end{equation}
are found at one-loop order in symmetric nuclear matter
as~\cite{Meissner:2001gz}
\begin{equation}
 f_\pi^t(\rho)
  = f_\pi\left[ 1 - \frac{\rho}{\rho_0}(0.26 \pm 0.04)\right]\;,
  \qquad
 f_\pi^s(\rho)
  = f_\pi\left[ 1 - \frac{\rho}{\rho_0}(1.23 \pm 0.07)\right]\;.
\label{eq:mediumf}
\end{equation}
A model-independent analysis valid at low density leads to a scaling
relation between $f_\pi^t$ and the in-medium quark
condensate~\cite{Jido:2008bk}, with which one sees that the condensate
decreases in nuclear matter, indicating partial restoration of chiral
symmetry (see Ref.~\cite{Hayano:2008vn} for a review).

\subsection{Mesons, baryons, and exotica}

From the ChPT one can calculate how the chiral condensate decreases.
If the chiral condensate significantly drops, then, it should have
sizable influences on the hadron spectra such as the nucleon mass.
Indeed we have already elucidated that the in-medium effective mass in
the Walecka model is different largely from the bare value.  The
question is, then, how chiral symmetry becomes manifest for the
hadronic degrees of freedom if (partial) chiral restoration occurs in
dense matter.

\subsubsection{Mended symmetry}
\label{sec:mended}

In general a large variety of representations are possible for
composite states made from quarks that belong to the fundamental
representation.  When chiral symmetry $G$ were not spontaneously
broken down into $H$, the physical states would be classified by
irreducible representations of $G$.  With broken chiral symmetry, the
physical states belong no longer to those irreducible representations,
but instead are superposition of all the possible representations.
The physical state $|\alpha\rangle$ is expressed in a series of a
complete set of states $|J\rangle$ characterized by good quantum
numbers $J$;
\begin{equation}
 |\alpha\rangle = \sum_J c_J |J\rangle \;.
\end{equation}
The coefficients $c_J$ contain information about the broken chiral
symmetry.  The state $|\alpha\rangle$ can further be constrained by
applying the large-$\Nc$ approximation (see Sec.~\ref{sec:largeNc}
also) as well as current algebraic
relations~\cite{Weinberg:1969hw,Weinberg:1990xn}.  The states for the
low-lying excitations with zero helicity are classified by
\textit{a limited number} of the irreducible representations and form
a closed algebra which is referred to as \textit{mended symmetries}.

As shown in Ref.~\cite{Weinberg:1969hw}, assuming that the
Adler-Weisberger sum rules for the scattering process,
$\pi_a + \alpha(\lambda) \to \pi_b + \beta(\lambda^\prime)$ for
helicities $\lambda$ and $\lambda^\prime$, can be saturated with
narrow one-particle states, the set of the sum rules is put into the
Lie-algebraic form,
\begin{equation}
 \left[ X_a, X_b \right] = if_{abc}T_c \;,
\label{eq:lie1}
\end{equation}
where $f_{abc}$ is the structure constant, $T_c$ is the generator of
the $\mathrm{SU}(\Nf)_{\text{V}}$ group that satisfies
\begin{equation}
 \left[ T_a, T_b \right] = if_{abc}T_c \;,
\label{eq:lie2}
\end{equation} 
and $X_a$ is the axial-vector coupling matrix defined from the matrix
elements at zero invariant momentum transfer of the axial-vector
current between states with collinear momenta via
\begin{equation}
 \langle \vec{q} \lambda^\prime \beta | (J_{\text{A}a}^0+J_{\text{A}a}^3)
  | \vec{p} \lambda \alpha \rangle = (2\pi)^3 2p^+\delta^3(\vec{p}-\vec{q})
  \delta_{\lambda^\prime\lambda}
  \left[ X_a(\lambda)\right]_{\beta\alpha} \;.
\end{equation}
Thus, from this definition, some non-trivial calculations lead us to
the following commutation relation,
\begin{equation}
 \left[ T_a, X_b \right] = if_{abc}X_c \;.
\label{eq:lie3}
\end{equation}
Putting together, Eqs.~\eqref{eq:lie1}, \eqref{eq:lie2}, and
\eqref{eq:lie3} close the algebra and one sees that the $X_a$ together
with the unbroken $\mathrm{SU}(\Nf)_{\text{V}}$ generators $T_c$
satisfies the commutation relations of the Lie algebra of the broken
symmetry group $G$.

In the broken phase the $X_a$ is not symmetry generators.  In fact, it
does not commute with mass-squared matrix $m^2$.  We note that $m^2$
here refers to the general hadron masses and is not necessarily
vanishing in the chiral symmetric phase.  The non-commutativity which
follows from Eq.~\eqref{eq:lie1} along with the Jacobi identity
defines $m_4^2$, that is, the chiral-symmetry breaking part of $m^2$ as
\begin{equation}
 \left[ X_b, \left[ m^2, X_a \right] \right] = -m_4^2\delta_{ab} \;.
\end{equation}
In the $\Nf=2$ case, for example, $m_4$ transforms as the fourth
component of a chiral four-vector (like the $\sigma$ meson).  This
tells us that the mass matrix is a sum of a chiral-singlet $m_0^2$
that satisfies $[X_a,m_0^2]=0$ and a non-singlet $m_4^2$, i.e.
\begin{equation}
 m^2 = m_0^2 + m_4^2 \;.
\label{eq:mass}
\end{equation}
When $m_4^2$ vanishes, i.e.\ chiral symmetry is restored, $X_a$
becomes true symmetry generators, so that $X_a$ commutes with $m^2$
and hadrons form degenerate multiplets filling out a complete set of
representations of $G$.

In the chiral broken phase the algebraic representations of $X_a$ do
not always coincide with the mass eigenstates as accounted for by
$\left[ X_a, m^2 \right] \neq 0$.  In general, there, physical particle states are expressed as a sum of all
possible elements of various representations of the Lie algebra
composed of $X_a$ and $T_c$.  The $X_a$ is therefore entirely
determined by the mixing angles which define the coefficients of the
representations in the sum.  Let the physical states of the low-lying
$\lambda=0$ mesons, the scalar $\sigma$, pseudo-scalar $\pi$, vector
$\rho$ and axial-vector $a_1$ mesons, be in the following admixture of
the chiral representations~\cite{Weinberg:1969hw,Gilman:1967qs},
\begin{equation}
 \begin{split}
 |\pi\rangle &=
  |\left(\Nf, \Nf^\ast\right) \oplus \left(\Nf^\ast, \Nf\right)\rangle 
  \sin\theta
  {}+ |\left(\Nf^2\!-\!1, 1\right) \oplus \left(1, \Nf^2\!-\!1\right)\rangle 
  \cos\theta \;,\\
 |a_1(\lambda=0)\rangle &=
  |\left(\Nf, \Nf^\ast\right) \oplus \left(\Nf^\ast, \Nf\right)\rangle 
  \cos\theta
  {}- |\left(\Nf^2\!-\!1, 1\right) \oplus \left(1, \Nf^2\!-\!1\right)\rangle 
  \sin\theta \;,\\
 |\sigma\rangle &=
  |\left(\Nf, \Nf^\ast\right) \oplus \left(\Nf^\ast, \Nf\right)\rangle\;,\\
 |\rho(\lambda=0)\rangle &=
  |\left(\Nf^2\!-\!1, 1\right) \oplus \left(1, \Nf^2\!-\!1\right)\rangle \;.
 \end{split}
\label{eq:repmixing}
\end{equation}
This quartet structure based on the chiral algebraic sum rules is
known as a notion of mended symmetry~\cite{Weinberg:1990xn}.  This
leads to the decay widths as functions of the mixing angle $\theta$
and masses as well as $f_\pi$.  The experimental rate for
$\rho \to \pi+\pi$ and the pion decay constant $f_\pi$ give
approximately $\theta \approx 45^\circ$ and one then finds
\begin{equation}
 m_\rho \approx m_\sigma\;,
  \qquad
 m_{a_1}^2 \approx 2m_\rho^2\;,
\end{equation}
indicating that the above algebraic consideration indeed yields the
reasonable order of magnitude for the masses.

Apparently, those helicity-zero mesons as well as the pion all become
massless at a second-order phase transition where $m_\sigma$ becomes
vanishing~\cite{Weinberg:1990xn}.  When an external parameter, such as
$T$ and $\rho$, drives a system from broken to unbroken phase, the
mixing angle is in general supposed to
have some intrinsic dependence on the external parameter.  Such
medium evolution of the mixing angle needs to be pinned down so that
some signals of the chiral phase transition in hadronic quantities
would be revealed.

\subsubsection{The problem of mass}
\label{sec:nucleon_mass}

In introducing baryon fields, there are two alternative ways of
assigning chirality to the nucleons.  Let us consider a pair of chiral
partners, $\psi_1$ and $\psi_2$, that realizes chiral
$\mathrm{SU}(2)_{\text{L}}\times\mathrm{SU}(2)_{\text{R}}$.  One is,
\begin{equation}
 \begin{split}
 \text{(Standard Assignment)}\qquad
  & \psi_{1\text{L}} \to L\psi_{1\text{L}} \;,
    \qquad \psi_{1\text{R}} \to R\psi_{1\text{R}} \;, \\
  & \psi_{2\text{L}} \to L\psi_{2\text{L}} \;,
    \qquad \psi_{2\text{R}} \to R\psi_{2\text{R}} \;,
 \end{split}
\end{equation}
which is anchored on the standard chiral symmetry structure where the
entire constituent quark or nucleon mass (in the chiral limit) is
generated via spontaneous chiral symmetry breaking.  The other
is~\cite{Detar:1988kn}, 
\begin{equation}
 \begin{split}
 \text{(Mirror Assignment)}\qquad
  & \psi_{1\text{L}} \to L\psi_{1\text{L}} \;,
    \qquad \psi_{1\text{R}} \to R\psi_{1\text{R}} \;, \\
  & \psi_{2\text{L}} \to R\psi_{2\text{L}} \;,
    \qquad \psi_{2\text{R}} \to L\psi_{2\text{R}} \;,
  \end{split}
\end{equation}
which allows for a chiral invariant mass term,
\begin{equation}
 \calL_m = m_0 \bigl( \bar{\psi}_{2\text{L}}\psi_{1\text{R}}
  - \bar{\psi}_{2\text{R}}\psi_{1\text{L}}
  - \bar{\psi}_{1\text{L}}\psi_{2\text{R}}
  + \bar{\psi}_{1\text{R}}\psi_{2\text{L}} \bigr) \;,
\end{equation} 
with which $m_0$ remains non-zero at chiral restoration.  One should
notice that the left- and right-handedness of a fermion is the
representation of the Lorentz group, whereas chiral symmetry is
entirely associated with the internal symmetry.  Therefore those must
be in general dealt with independently, which enables us to adopt the
above non-standard chiral rotation.  In other words, the mirror
assignment is possible since the nucleon is not an elementary particle
but a composite state made from quarks.

A part of the nucleon mass, $m_0$, must arise from a mechanism that is
not associated with spontaneous chiral symmetry breaking.  Various
observables both in the vacuum such as baryon
decays~\cite{Nemoto:1998um} and pion-nucleon
scattering~\cite{Gallas:2009qp,Paeng:2011hy} and in a medium such as
nuclear matter
properties~\cite{Hatsuda:1988mv,Zschiesche:2006zj,Sasaki:2010bp,Gallas:2011qp}
and the thermal gluon condensate~\cite{Sasaki:2011ff} have been
explored based on chiral Lagrangian approaches with mirror baryons.
The analysis tells us that an $m_0$ of a few hundred MeV cannot be
ruled out.  In the non-linear realization of chiral symmetry only the
properties coming from the vectorial transformation are
remained~\cite{Weinberg:1968de}, and therefore the standard and mirror
assignments cannot be distinguished.  Even if indistinguishable in
matter-free space, they are expected to start showing their
differences due to the remaining $m_0$ as one approaches the chiral
restoration point.

What is the origin of such a mass $m_0$?  Recall that the emergence of
$\LQCD$ is signaling the breaking of conformal invariance, i.e.\ the
trace anomaly, of QCD in the chiral limit where the theory has no
dimensionful parameters.  Spontaneous chiral symmetry breaking, which
gives rise to a nucleon mass, and the trace anomaly are intimately
linked to each other~\cite{Bardeen:1985sm} and dynamical scales in
hadronic systems are then considered to originate from them.  Hence,
the origin of $m_0$ can be attributed to the non-vanishing gluon
condensate in chiral symmetric phase~\cite{Sasaki:2011ff} (see also
discussions in Sec.~\ref{sec:dilaton}).

This is reminiscent of Weinberg's chiral-singlet mass in
Eq.~\eqref{eq:mass}.  The chiral representation
mixing~\eqref{eq:repmixing} would be expected to be resolved when
chiral symmetry is restored in such a way that the $\pi$ and $\sigma$
mesons form a degenerate multiplet as chiral partners, as presumed from
the argument of the $\mathrm{O}(4)$-universality class in $\Nf=2$ QCD.\ \
This is in fact realized when the mixing goes like $\cos\theta \to 0$
and with which the $\pi$ and $\sigma$ belong to the same
representation $|(\Nf,\Nf^\ast)\oplus(\Nf^\ast,\Nf)\rangle$ and being
massless protected by the NG theorem, whereas this is not the case for
the $\rho$ and $a_1$ mesons being degenerate in the same
representation at chiral restoration and in general having a
non-vanishing common mass.  It is also possible that those
lowest-lying mesons form a quartet being all the mesons massless when
the mixing remains unresolved as demonstrated using the mended
symmetry notion in Ref.~\cite{Weinberg:1990xn} where $m_\rho=m_\sigma$
is predicted.  Or even with a resolving mixing, if there are some
mechanisms that the $\rho$ and $a_1$ mesons become massless, the
massless quartet can be well accommodated to the requirement of chiral
restoration, such as in the context of the Brown-Rho (BR)
scaling~\cite{Brown:1991kk}, an analysis of RG flows within a class of
chiral effective field theory~\cite{Harada:2005br,Hidaka:2005ac} and
of the non-linear sigma model implementing the trace
anomaly~\cite{Sasaki:2011ff}.

Yet, massive vector bosons with no mixing at chiral restoration are
not prohibited by symmetries.  The mixing angle can be in general a
function of temperature and density and therefore would develop in a
hot/dense medium.  Even if the mixing is preferred being
$m_\rho=m_\sigma$ in matter-free space, it might evolve so that
$\cos\theta \to 0$, not optionized in Ref.~\cite{Weinberg:1990xn},
namely $m_\pi=m_\sigma = 0$ and $m_\rho = m_{a_1} \neq 0$ can be allowed.

\subsubsection{Thermodynamics of dilatons}
\label{sec:dilaton}

The trace anomaly is implemented in a chiral Lagrangian by introducing
a dilaton (or scalar glueball) field $\chi$ representing the gluon
condensate
$\langle G_{\mu\nu}G^{\mu\nu} \rangle$~\cite{Schechter:1980ak,Migdal:1982jp}. 
Requiring that the dilatation current associated with scale
transformation is saturated by a condensed dilation field leads to the
potential,
\begin{equation}
 V_\chi = \frac{B}{4}\biggl(\frac{\chi}{\chi_0}\biggr)^4
  \biggl[ \ln\biggl(\frac{\chi}{\chi_0}\biggr)^4 - 1 \biggr]\;.
\end{equation}
The two parameters, i.e.\ the bag constant $B$ and a scale parameter
$\chi_0$, are fixed from the vacuum energy density
$\mathcal{E} = \frac{1}{4}B$ and the glueball mass
$M_\chi = 2\sqrt{B}/\chi_0$.  Effective Lagrangians are written so
that they possess chiral and scale invariance, mimicking dynamical
breaking of those symmetries of QCD in low energies.

In relativistic nuclear physics, a scalar boson plays an essential
role as a mediator of the medium-range $NN$ interaction as we
discussed using the Walecka model in Sec.~\ref{sec:Walecka}.  The
important lesson from the model studies is that this scalar degree of
freedom \textit{cannot be identified with the scalar meson that
  appears as the chiral partner of the NG boson in a linear sigma
  model} since otherwise no stable nuclear ground state is achieved
and one cannot reproduce the basic saturation
properties~\cite{Kerman:1974yk,Furnstahl:1995zb}.  This problem would
be resolved in a non-linear representation as in the ChPT, or
different chiral approaches have been introduced with modified
interaction potentials among the chiral fields and the
dilaton~\cite{Furnstahl:1995zb,Heide:1993yz,Papazoglou:1997uw}.  A
non-linear chiral Lagrangian with a
\textit{soft dilaton}~\cite{Miransky:1989qc} is derived when a heavy
component of the dilaton, responsible for the pure glueball dynamics,
is integrated out~\cite{Furnstahl:1995by}.

Besides those approaches implementing the dilaton, a parity doublet
model based on the mirror assignment can also be accommodated to the
nuclear ground state~\cite{Zschiesche:2006zj}.  The presence of a
chiral invariant mass $m_0$ of the nucleons is essential there and a
relatively large $m_0 \sim 800\MeV$ is favored in order to have the
known nuclear matter properties.  (This number becomes reduced
substantially when a tetra-quark state is introduced and $m_0$ in
nuclear matter results in $\sim 500\MeV$~\cite{Gallas:2011qp}.)

It appears that $m_0$ plays a similar role to the dilaton in nuclear
matter and thus a non-vanishing condensate of the dilaton field is a
conceivable origin of $m_0$~\cite{Sasaki:2011ff,Paeng:2011hy}.  The
scalar meson as the chiral partner of the pion generates only the mass
difference of the nucleon parity doublers.

\subsubsection{Four-quark condensate in a medium}

Four-quark states were suggested as the lightest scalar 
mesons~\cite{Jaffe:1976ig,Jaffe:1976ih} and this explains naturally
the degenerate $a_0(980)$ and $f_0(980)$ states.  How does the
four-quark condensate evolve with temperature or density?  The NG
theorem by itself does not dictate a particular operator to fulfill a
non-trivial commutation relation with the broken charge
$Q_{\text{A}}$.  Any operators $\calO$ yielding non-commutativity to
$Q_{\text{A}}$ guarantee spontaneous chiral symmetry breaking as
discussed in Sec.~\ref{sec:chiralsym}.  Besides
$\calO\sim\bar{\psi}t^a\psi$, some operators with higher dimensions
such as a four-quark are possible.

In matter-free space the four-quark condensates are often evaluated
assuming factorization,
\begin{equation}
 \langle \bar{\psi}\bar{\psi}\psi\psi \rangle
 \simeq \langle \bar{\psi}\psi \rangle^2 \;.
\label{eq:factor}
\end{equation}
This indicates that the four-quark condensate seems to vanish at the
chiral phase transition as $\langle\bar{\psi}\psi\rangle$.  However,
the factorized form~\eqref{eq:factor} is justified in the large-$\Nc$
limit only and not quite accurate in the real QCD case with $\Nc=3$.
Its validity in a hot/dense medium is even less justifiable.

Therefore, the issue is that when $\langle\bar{\psi}\psi\rangle$
vanishes, some four-quark operators (apart from the chiral singlet
part) can condense so that it breaks chiral symmetry dynamically.
Such a possibility has been considered at zero temperature and zero
density in Refs.~\cite{Knecht:1994zb,Stern:1998dy}, where a
non-standard pattern of dynamical symmetry breaking was suggested.
This pattern keeps the center of chiral group unbroken, i.e.
\begin{equation}
 \mathrm{SU}(\Nf)_{\text{L}} \times \mathrm{SU}(\Nf)_{\text{R}} \to
  \mathrm{SU}(\Nf)_{\text{V}} \times \mathrm{Z}_{\Nf} \;.
\label{eq:breaking}
\end{equation}
The discrete $\mathrm{Z}_{\Nf}$ symmetry protects a theory from
condensate of quark bilinears.  Spontaneous symmetry breaking is
driven by quartic condensates which are invariant under both
$\mathrm{SU}(\Nf)_{\text{V}}$ and $\mathrm{Z}_{\Nf}$ transformation.
However, this pattern is strictly ruled out in QCD both at $T=0$ and
$T\neq0$ but $\muq=0$ since the axial-vector correlator becomes
greater than the pseudo-scalar correlator then, which violates the QCD
inequality~\cite{Kogan:1998zc}.  This argument does not exclude the
above unorthodox pattern in the presence of a dense medium because the
QCD inequalities do not hold then.  Several studies have been made,
such as a similar dynamical breaking in an O(2) scalar
model~\cite{Watanabe:2003xt}, constructing a Ginzburg-Landau free
energy consistent with Eq.~\eqref{eq:breaking}, and the exploration of
its thermodynamic consequences~\cite{Harada:2009nq}.

\subsection{Hidden local symmetries}
\label{sec:hls}

The low-energy pion dynamics is well described by the non-linear sigma
model on the manifold $G/H$ which is an effective Lagrangian in terms 
only of NG bosons transforming non-linearly under chiral 
rotation. 
The approach using the hidden local symmetry
(HLS)~\cite{Bando:1984ej,Bando:1987br} is based on the fact that the
non-linear sigma model is gauge equivalent to a model possessing
$G_{\rm global}\times H_{\rm local}$ symmetry 
(see Ref.~\cite{Harada:2003jx} for a review and references therein).
As demonstrated explicitly in Refs.~\cite{Kugo:1985jc,Kugo:1981fe},
the dynamical generation of gauge bosons is a common phenomenon which
appears in a variety of systems not restricted to particular models.
Furthermore, one-loop quantization of the HLS theory based on the
derivative expansion as made in the ordinary ChPT is
established~\cite{Harada:2003jx}.  This enables us to evaluate loop
effects systematically and thus the low-energy quantities with good
precision come out combined with the RG analysis.  Below, we briefly
introduce the HLS approach and discuss several issues in strong
coupling gauge theories.

\subsubsection{$G_{\text{global}} \times H_{\text{local}}$ model}

The Lagrangian is based on a $G_{\text{global}} \times H_{\text{local}}$ 
symmetry, where $G_{\text{global}}=[\mathrm{SU}(\Nf)_{\text{L}}
 \times \mathrm{SU}(\Nf)_{\text{R}}]_{\text{global}}$ is chiral
symmetry and $H_{\text{local}}=[\mathrm{SU}(\Nf)_{\text{V}}]_{\text{local}}$
is the HLS.\ \ The entire symmetry
$G_{\text{global}}\times H_{\text{local}}$ is spontaneously broken to
its diagonal subgroup $\mathrm{SU}(2)_{\text{V}}$.  The basic
quantities are the HLS gauge boson, $V_\mu$, and two matrix valued
variables $\xi_{\text{L}}$, $\xi_{\text{R}}$, which are combined in an
$\Nf \times \Nf$ special-unitary matrix
$U = \xi_{\text{L}}^\dagger \xi_{\text{R}}$.  There exists an
ambiguity in the decomposition and this redundancy is identified with
a local gauge transformation $H_{\text{local}}$.  These variables are
conventionally parameterized as
\begin{equation}
 \xi_{\text{L,R}}(x) = e^{i\sigma (x)/{f_\sigma}}e^{\mp i\pi (x)/{f_\pi}}\,,
\end{equation}
where $\pi = \pi^a T_a$ denotes the pseudo-scalar NG bosons associated
with the spontaneous breaking of chiral symmetry, and
$\sigma = \sigma^a T_a$ the NG bosons associated with the spontaneous
breaking of $H_{\text{local}}$.  The $\sigma$ is the longitudinal part
of the vector meson and absorbed into the HLS gauge boson through the
Higgs mechanism, so that the gauge boson acquires its mass.
$f_\pi$ and $f_\sigma$ are the decay constants of the associated
particles.

The main building blocks are $\hat{\alpha}_{\perp,\parallel}^\mu$
defined from the Maurer-Cartan 1-forms as
\begin{equation}
 \hat{\alpha}_{\perp, \parallel}^{\mu}
 = \frac{1}{2i}\left( D^\mu\xi_{\text{R}} \cdot \xi_{\text{R}}^\dagger
 {}\mp D^\mu\xi_{\text{L}} \cdot \xi_{\text{L}}^\dagger \right)\;,
\end{equation}
where the covariant derivatives of $\xi_{\text{L,R}}$ are given by
\begin{equation}
 \begin{split}
 D_\mu \xi_\text{\text{L}}
  &= \partial_\mu\xi_{\text{L}} - iV_\mu\xi_{\text{L}}
  + i\xi_{\text{L}}{\mathcal{L}}_\mu \;, \\
 D_\mu \xi_{\text{R}}
  &= \partial_\mu\xi_{\text{R}} - iV_\mu\xi_{\text{R}}
  + i\xi_{\text{R}}{\mathcal{R}}_\mu
 \end{split}
\end{equation}
with ${\mathcal{L}}_\mu$ and ${\mathcal{R}}_\mu$ being the external
gauge fields (weak bosons and photons) introduced by gauging
$G_{\text{global}}$.  The most general Lagrangian with lowest
derivatives is given by~\cite{Bando:1984ej}
\begin{equation}
 \calL
 = f_\pi^2\tr\bigl(\hat{\alpha}_{\perp\mu}\hat{\alpha}_{\perp}^\mu \bigr)
 + f_\sigma^2\tr\bigl( \hat{\alpha}_{\parallel\mu}
   \hat{\alpha}_{\parallel}^\mu \bigr)
 - \frac{1}{2g^2}\tr\bigl( V_{\mu\nu}V^{\mu\nu} \bigr)\;,
\end{equation}
where $g$ is the HLS gauge coupling and the field strength is
$V_{\mu\nu}=\partial_\mu V_\nu-\partial_\nu V_\mu-i[V_\mu,V_\nu]$.
When the vector-meson kinetic term is ignored in low energy, the
second term of the Lagrangian vanishes with the equation of motion of
$V_\mu$ and the well-known non-linear sigma model on the coset space
$G/H$ is derived.  The leading-order Lagrangian contains two arbitrary
parameters, $a = f_\sigma^2/f_\pi^2$ and $g$.  For a particular choice
$a=2$, the known phenomenological facts are successfully reproduced;
universality of the vector meson coupling, the
Kawarabayashi-Suzuki-Riazuddin-Fayyazudin (KSRF) relation, and vector
meson dominance of the pion electromagnetic form
factor~\cite{Bando:1984ej}.

There are other models of vector mesons introduced as the matter 
field, 
the anti-symmetric tensor field 
or in the massive Yang-Mills method.
It is shown that they are equivalent to the gauge-fixed form of the
HLS Lagrangian.  For details, see
e.g.\ Refs.~\cite{Bando:1987br,Meissner:1987ge,Birse:1996hd,Harada:2003jx}
and references therein.

\subsubsection{Modeling strongly-coupled gauge theories}

In principle one can enlarge the gauge group with more redundant
variables.  With such a generalized HLS
(GHLS)~\cite{Bando:1987br,Bando:1987ym}, the lowest-lying axial-vector
mesons are embedded in a $G_{\text{global}}\times G_{\text{local}}$
model.  Other higher-lying vector modes can also be incorporated
systematically by introducing enlarged gauge symmetries.  This idea is
employed in so-called dimensional deconstruction which resembles the
nearest neighbor interactions in condensed matter physics,
constructed on a lattice~\cite{ArkaniHamed:2001ca,ArkaniHamed:2001nc,
ArkaniHamed:2002sp}.  The theory possesses gauge group which is a
product of many numbers of the same gauge group.  The interaction to
matter fields is characterized by
``theory space locality''~\cite{ArkaniHamed:2001ed} where the mixing
of left and right chirality is generated only through gauge boson
exchanges.  Taking the continuum limit generates an extra dimension.
Inspired by those studies, a dimensionally-deconstructed QCD was
proposed~\cite{Son:2003et}.  The resultant five-dimensional Yang-Mills
action is shown to reproduce some of known low-energy hadron
properties.

The identical action implemented with the Chern-Simons action is
obtained from the string theory based on the
gauge/gravity duality (see Sec.~\ref{sec:SSM} for the Sakai-Sugimoto
model)~\cite{Sakai:2004cn,Sakai:2005yt}.  Performing the Kaluza-Klein
mode expansion yields a four-dimensional action which contains the NG
bosons and an infinite tower of vector and axial-vector mesons.  When
all the vector modes except the lowest-lying vector meson are
integrated out, the $\calO(p^4)$ terms as well as the $\calO(p^2)$
terms of the HLS Lagrangian are generated~\cite{Harada:2010cn}.  The
higher vector mesons are now accommodated in the $\calO(p^4)$ part.

Another type of duality of interest is the electric/magnetic duality
in $\calN=1$ supersymmetric Yang-Mills theory (i.e.\ Seiberg
duality)~\cite{Seiberg:1994pq}.  For $\frac{3}{2}\Nc < \Nf < 3\Nc$
(conformal window), there exists a magnetic theory possessing the
$\mathrm{SU}(\Nf-\Nc)$ gauge symmetry which is dual to the original
$\mathrm{SU}(\Nc)$ theory regarded as the electric theory.  The two
theories flow to the same IR fixed point and thus exhibit the
identical low-energy dynamics.  When $\Nf$ decreases to $\Nf=\Nc$, the
magnetic gauge symmetry induced at the composite level is
spontaneously broken, which corresponds to the system with confinement
and spontaneously broken chiral symmetry.  It should be emphasized
that the induced gauge symmetry has nothing to do with the original
$\mathrm{SU}(\Nc)$ gauge symmetry, but it is rather analogous to the
HLS~\cite{Komargodski:2010mc}.

Is the Seiberg duality realized in non-supersymmetric QCD under
consideration?  The answer is yes~\cite{Harada:1999zj,Kitano:2011zk}:
The HLS theory is shown to exhibit chiral symmetry restoration by its
own dynamics for large $\Nf \sim 5\,(\Nc/3)$, as in the $CP^{N-1}$
non-linear sigma model~\cite{Bando:1987br}.  Therefore the
$\mathrm{SU}(\Nf)$ HLS theory is considered to be the Higgsed magnetic
theory dual to the electric theory, i.e.\ $\mathrm{SU}(\Nc)$ QCD.\ \
Some consequences of this ``QCD/HLS duality'' in hadron physics are
discussed e.g.\ in Refs.~\cite{Kitano:2011zk,Abel:2012un}.

Here we stress that the HLS theory appears as a potentially practical
approach for various gauge theories in the strong coupling region.
Obviously low-energy hadron physics, not only in matter-free space but
also in a hot and/or dense medium, is a case of great interest.  The
HLS approach has been applied to QCD matter in a wide range of density
and temperature.  For reviews, see
Refs.~\cite{Brown:1995qt,Furnstahl:1996wv,Serot:1997xg,%
Casalbuoni:1999wu,Brown:2001nh,Brown:2009az} and references therein.

\subsubsection{Chiral perturbation theory with HLS}

Beyond the tree-level HLS results, loop effects are systematically
evaluated at a given order of the derivative expansion as in the
ordinary ChPT~\cite{Harada:1992np,Tanabashi:1993np}.  The essential
concept is the HLS gauge bosons treated as \textit{light}.  The chiral
counting of the HLS gauge coupling is assigned of order $p$, which is
equivalent to the order assignment by
$m_V \sim \calO(p)$~\cite{Georgi:1989gp,Georgi:1989xy}.  The pionic
coupling is derivative-type typically in non-linear realization.  With
the above counting, the vector mesons and their interaction can
naturally be put in a perturbative treatment on the same footing with
the NG bosons, yielding consistent counting rules.  In real QCD the
vector meson mass is not very small compared to the chiral symmetry
breaking scale $\Lambda_\chi \sim 1\GeV$.  However, there exists a
limit of large-$\Nc$ in which the above expansion is certainly
justified since
\begin{equation}
 \frac{m_V^2}{\Lambda_\chi^2} \sim \frac{1}{\Nc}\;.
\end{equation}
Thus the entire procedure for the loop calculations should be regarded
as an extrapolation from large-$\Nc$ to $\Nc=3$ QCD.\ \ Indeed,
combined with RG equations (RGEs), this works quite well and one sees
the low-energy constants and the relevant quantities in the vector
meson sector in good agreement with
measurements~\cite{Harada:2003jx}.

Quadratic divergences that arise from the pion loops drive a system
to chiral symmetry restoration (Wigner-Weyl realization).  In the HLS
theory this happens with the following conditions dictated by matching
of the current correlation functions to those in the operator product
expansion (OPE)~\cite{Harada:2000kb}:
\begin{equation}
 a \to 1\;, \qquad  m_V\to 0\;.
\label{eq:vm}
\end{equation}
The vector meson thus becomes massless toward chiral symmetry
restoration where $f_\pi = 0$, and falls into the same chiral
multiplet together with the pion.  Such somewhat non-standard
realization called the vector manifestation (VM) is similar to the
vector realization.  The essential difference is that in the vector
realization the NG boson pole residue $f_\pi$ remains non-vanishing
whereas chiral symmetry is ``unbroken'', i.e.\
$f_\pi \neq 0$, $\langle\bar{\psi}\psi\rangle = 0$, therefore
\textit{the system is not in the Wigner-Weyl phase} (see
Refs.~\cite{Georgi:1989gp,Georgi:1989xy} for more details).  Here one
should be cautious of the fact that although the two realizations are
formulated in the HLS framework, the theory contains a limited number
of low-lying mesons, pseudo-scalar, and vector mesons, and in
particular lack of a scalar particle near the restoration point is
unsatisfactory from the point of view of building effective theories
that describe the chiral phase transition.  In fact, the two
conditions in Eq.~\eqref{eq:vm} correspond to the fixed points of the
RGEs \textit{separately}.  Thus higher loops or other resonances would
modify the RGEs and consequently parametric conditions of symmetry
restoration.  Nevertheless, the power of this effective field theory
is that one can examine several scenarios allowed as phenomenological
options, including the $m_V \neq 0$ case, within the single
framework~\cite{Harada:2008hj}.  How the scalar mode emerges from
non-linear sigma models is an issue~\cite{Beane:1994ds,Sasaki:2011ff}
and a na\"{i}ve introduction of an elementary scalar field is known to
cause the \textit{naturalness problem} at loop level as in the Higgs
sector of the Standard Model.

An interesting extension of the phase structure study was carried out
based on the GHLS at one loop~\cite{Harada:2005br,Hidaka:2005ac}.  The
chiral representation mixing between the pseudo-scalar and
axial-vector meson states~\eqref{eq:repmixing} is present in the
theory and its RG evolution indicates how the mixing is resolved
toward chiral symmetry restoration driven by external parameters such
as $\Nf$, temperature, and density.  Three associated fixed points are
found and their characteristic features are summarized below with the
vector and axial-vector meson masses $m_{V,A}$:
\begin{center}
 \begin{tabular}{lll}
 \textit{Ginzburg-Landau type} &
  $\cos\theta \to 0$ \text{ thus } $(m_V/m_A)^2\to 1$ &
  VD fulfilled \\
 \textit{Vector Manifestation type} &
  $\sin\theta \to 0$ \text{ thus } $(m_V/m_A)^2\to 0$ &
  VD violated \\
 \textit{Hybrid type} &
  $\sin\theta \to \sqrt{1/3}$ \text{ thus } $(m_V/m_A)^2\to 1/3$ &
  VD violated
 \end{tabular}
\end{center}
where VD represents the vector dominance.  Which realization is
eventually chosen is a question of the actual QCD dynamics.  Violation
of the VD in the latter two scenarios is rather strong, in particular
near the chiral restoration point; violated by 50\% in the Vector
Manifestation type and 33\% in the Hybrid type.  Its impact on QCD
phenomenology is obvious.  One example is dilepton yields from
in-medium $\rho$ mesons which are suppressed when the VD is violated,
whereas the VD is often \textit{assumed to be valid at any temperature
  and density} in many calculations.  This will be discussed more in
Sec.~\ref{sec:dilepton}.

Another important finding in the GHLS approach is the spectral
function sum rules that are locked to theory space locality.  The
extended condition for locality in the GHLS is shown to be stable
against the one-loop corrections~\cite{Harada:2005br} and this
immediately leads to the celebrated Weinberg sum rules in the
pole-saturated forms~\cite{Weinberg:1967kj},
\begin{equation}
 f_\pi^2 + f_A^2 = f_V^2 \;, \qquad
 f_A^2 m_A^2 = f_V^2 m_V^2 \;,
\end{equation}
with the masses and decay constants of the corresponding particles.
In other words, the sum rules invariant under the RG evolution are
guaranteed by theory space locality.

\subsection{Approaches in the large-$\Nc$ limit}
\label{sec:largeNc}

We here give detailed explanations on a non-perturbative approach
based on the limit of the infinite number of colors (see
Ref.~\cite{Lucini:2012gg} for a recent review).  The large-$\Nc$
limit was first considered on the diagrammatic
level~\cite{'tHooft:1973jz,Witten:1979kh}, and such diagrammatic
considerations are still useful for some kind of duality in
supersymmetric and non-supersymmetric Yang-Mills theories
(i.e.\ planar equivalence; see Ref.~\cite{Hanada:2011ju} for example).
For non-perturbative applications we will go specially into the Skyrme
model~\cite{Skyrme:1962vh} and the Sakai-Sugimoto
model~\cite{Sakai:2004cn} below.

\subsubsection{Counting rule}
\label{sec:Nccounting}

If the number of colors $\Nc$ is chosen to be infinitely large, the
theoretical treatment becomes simplified, and it is even possible to
extract some model-independent results.  The na\"{i}ve limit of
$\Nc\to\infty$ makes, however, all physical observables blow up to
infinity.  First, let us consider the structure of the leading
diagrams, and the physically sensible way to take the the large-$\Nc$
limit.

\begin{figure}
\begin{center}
\begin{minipage}{0.35\textwidth}
 \includegraphics[width=\textwidth]{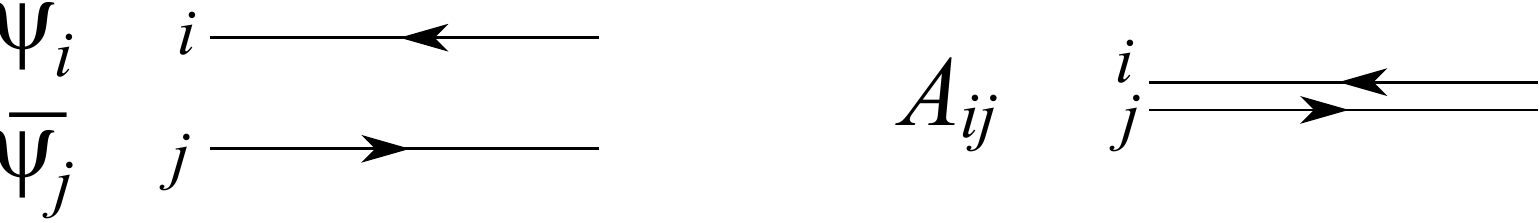}
\end{minipage} \hspace{5em}
\begin{minipage}{0.4\textwidth}
 \includegraphics[width=\textwidth]{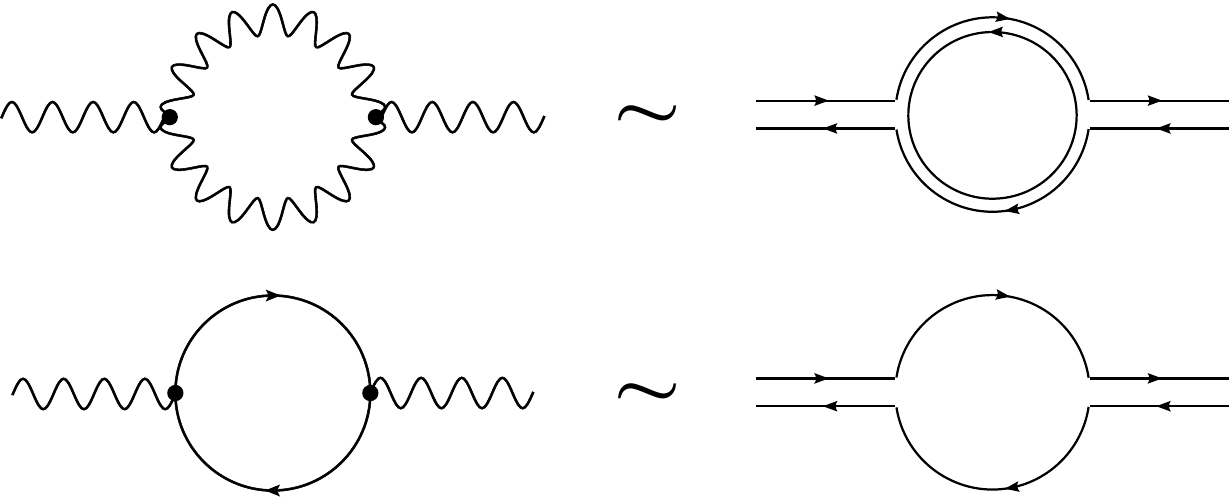}
\end{minipage}
 \caption{(Left) Double-line notation:  a quark and anti-quark are
   represented by a single-line, while the gauge fields involve two
   lines as a combination of the fundamental and anti-fundamental
   indices.  (Right) Quark loops are suppressed by $1/\Nc$ as compared
   to gluon loops in the large-$\Nc$ limit.}
 \label{fig:double-line}
\end{center}
\end{figure}

It is a common procedure to introduce the double-line notation in
which each line refers to the index in the fundamental
representation.  Thus, a quark $\psi_i$ and an anti-quark
$\bar{\psi}_j$ are represented by a single-line.  Because the adjoint
representation is composed from the non-singlet part with the
fundamental and anti-fundamental indices, the gauge field
$A_{ij}=A^a T^a_{ij}$ is represented by a double-line, as illustrated
in the left of Fig.~\ref{fig:double-line}.

In the large-$\Nc$ limit, using the double-line notation, it is easy
to make sure that quark loops are always suppressed by $1/\Nc$ as
compared to gluon loops and thus diagrams with only gluon loops are
dominant (see the right of Fig.~\ref{fig:double-line}).  This is a
crucial feature in thinking of the thermodynamics of QCD in the
large-$\Nc$ approximation.  In other words, whenever we talk about the
quark and baryon properties in the large-$\Nc$ limit, we treat such
baryonic degrees of freedom as a probe into the gluonic system that is
not perturbed by the insertion of baryons (i.e.\ probe
approximation).

\begin{figure}
\begin{center}
 \includegraphics[width=0.65\textwidth]{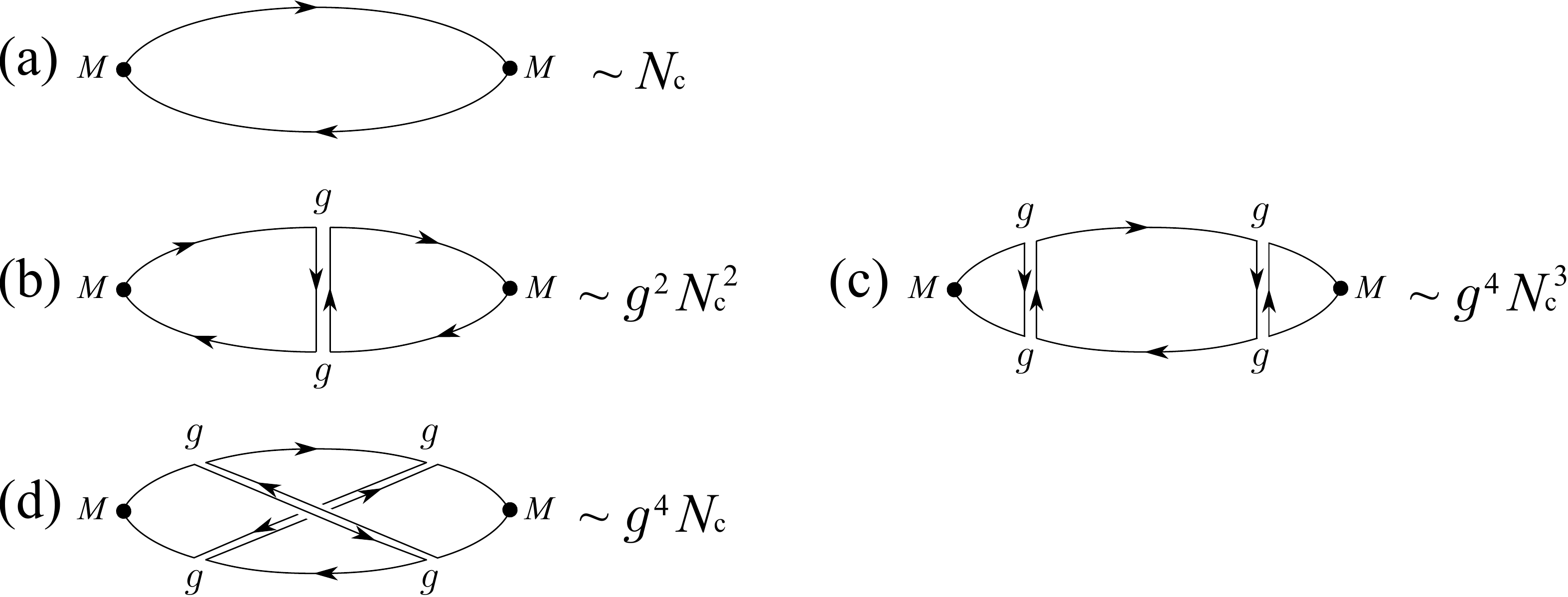}
\end{center}
 \caption{$\Nc$ counting for the meson two-point function.  (a)
   Tree-level diagram.  (b) First loop correction
   $g^2\Nc^2 \sim \Nc$.  (c) Next planar loop correction
   $g^4\Nc^3 \sim \Nc$.  (d) Non-planar loop correction
   $g^4\Nc \sim \Nc^{-1}$.}
 \label{fig:largeNcmeson}
\end{figure}

The next important observation in the large-$\Nc$ limit is that the
so-called non-planar diagrams are suppressed by at least $1/\Nc^2$ as
compared to the planar diagrams.  To see this, let us consider the
two-point function of the meson composite field
$M\sim \bar{\psi}\psi$ as drawn in Fig.~\ref{fig:largeNcmeson}.  It is
obvious that the tree-level diagram connecting two meson fields
consists of simple one-loop with two lines, making a contribution of
$\calO(\Nc)$ (see Fig.~\ref{fig:largeNcmeson}~(a)).  The next
contribution at the two-loop level is shown in
Fig.~\ref{fig:largeNcmeson}~(b), that has an extra factor by $g^2\Nc$
than the tree-level contribution.  Therefore, in this way, every time
the loop order is incremented, a factor $g^2\Nc$ at least is
multiplied.  Therefore, to construct the meson two-point function in a
non-perturbative and non-diverging way, the strong coupling constant
must scale as $g^2\sim 1/\Nc$ (or so-called 't~Hooft coupling
$\lambda\equiv g^2\Nc$ is kept fixed), and then the contribution from
Fig.~\ref{fig:largeNcmeson}~(b) can be of the same order as (a).  The
next loop order consists of the planar diagram (c) and the non-planar
diagram (d) with an overhead crossing.  Now that $g^2\sim 1/\Nc$, (c)
is at the same order as (b), while (d) is suppressed by $1/\Nc^2$.

The meson two-point function $\langle M(k) M(-k)\rangle$ is of order
of $\Nc$ as is clear from Fig.~\ref{fig:largeNcmeson}, that means that
the meson decay constant scales like $f_\pi \sim \sqrt{\Nc}$.  From
this, it is possible to identify the scaling of the inter-meson
coupling.  That is, let us consider the $n$-point meson correlation
function, that involves $(f_\pi)^n \sim \Nc^{n/2}$ and the $n$-point
interaction vertex $g_n$.  The whole planar diagrams should scale as
$\Nc$, and therefore, $g_n \sim \Nc^{1-n/2}$.  Hence, the inter-meson
vertices are all vanishing for $n\ge3$ in the limit of
$\Nc\to\infty$.  The baryon, on the other hand, is composed from $\Nc$
quarks and the baryon mass is thus of $\calO(\Nc)$ (there are
$\Nc(\Nc-1)$ pairs of inter-quark interactions, but the interaction
strength scales as $1/\Nc$ and then the total scaling is
$\calO(\Nc)$)~\cite{Witten:1979kh}.  Then, in QCD with $\Nc\to\infty$,
baryons are infinitely heavy and the theory is reduced to a system
with infinite towers of mesons and glueballs.

\subsubsection{Skyrme model}
\label{sec:skyrme}

One could construct the baryon as a bound state of $\Nc$ quarks.
Then, the baryon mass is naturally $\calO(\Nc)\sim\calO(1/g^2)$, while
the interaction of low-lying particles is
$\calO(1/\Nc)\sim\calO(g^2)$.  This inverse-correlation is reminiscent
of the properties of the soliton.  Indeed, it is possible to construct
the baryon as a soliton in terms of the low-lying fields.

\paragraph{Single skyrmion}
Here, we shall limit our discussions in the simplest example of the
$\mathrm{SU}(2)$ case.  The low-energy effective Lagrangian in the
non-linear representation again is
\begin{equation}
 \calL = \frac{f_\pi^2}{4}\tr\bigl(\partial_\mu U \partial^\mu U^\dagger\bigr)
  +\frac{1}{32e^2}
  \tr\bigl(\partial_\mu U U^\dagger,\partial_\nu U U^\dagger\bigr)^2 \;,
\end{equation}
if we neglect the infinite tower of mesons and pick up only the pion
field.  A parameter $e$ which is $\calO(\sqrt{\Nc})$ is dimensionless
and takes a real value.  The second term (sometimes called the Skyrme
term) is absent in the leading-order ChPT, and the presence of this
term is crucially important;  the soliton solution would collapse to
zero size without this fourth-order derivative term (Derrick's
theorem; this might be relaxed with inclusion of $\omega$ meson, see
Ref.~\cite{Adkins:1983nw}).

Introducing a current, $\Sigma_\mu = U\partial_\mu U^\dagger$, the
expectation of the Hamiltonian leads to the energy,
\begin{equation}
 E = \int d^3 x\, \tr\biggl[ \frac{f_\pi^2}{4}\Sigma_i \Sigma_i^\dagger
  +\frac{1}{16e^2}(\epsilon_{ijk}\Sigma_i\Sigma_j)
  (\epsilon_{abk}\Sigma_a\Sigma_b)^\dagger \biggr] \;,
\end{equation}
and the baryon current is expressed as $j_{\text{B}}^\mu = (1/24\pi^2)
\epsilon^{\mu\nu\alpha\beta}\tr\Sigma_\nu\Sigma_\alpha\Sigma_\beta$,
from which one can define the baryon number,
$B=\int d^3 x\,j_{\text{B}}^0(x)$.  With the following hedgehog Ansatz
using a spatial unit vector $\hat{\br}=\br/r$ for a chiral soliton,
\begin{equation}
 U(x) = \exp\bigl[ i F(r) \btau\cdot\hat{\br} \bigr] \;,
\end{equation}
the energy can be written as a functional of $F$, i.e.
\begin{equation}
 E[F] = 4\pi\int_0^\infty dr\, r^2 \biggl[\frac{f_\pi^2}{2}
  \Bigl( F'^2 + 2\frac{\sin^2 F}{r^2} \Bigr) + \frac{1}{2e^2}
  \frac{\sin^2 F}{r^2} \Bigl( \frac{\sin^2 F}{r^2}
  +2{F'}^2 \Bigr) \biggr] \;,
\label{eq:skyrme_energy}
\end{equation}
and the baryon number density is calculated into a form of
$j_{\text{B}}^0(x) = -F'\sin^2 F/(2\pi^2 r^2)$.  The baryon number is
then properly quantized as
\begin{equation}
 B = \frac{1}{2\pi} \bigl[ 2F(0) - \sin 2F(0) \bigr] \to 1 \;,
\end{equation}
if the boundary condition is chosen as $F(0)=\pi$, where we have used
$F(\infty)\to 0$.

\begin{figure}
\begin{center}
\begin{minipage}{0.3\textwidth}
 \includegraphics[width=\textwidth]{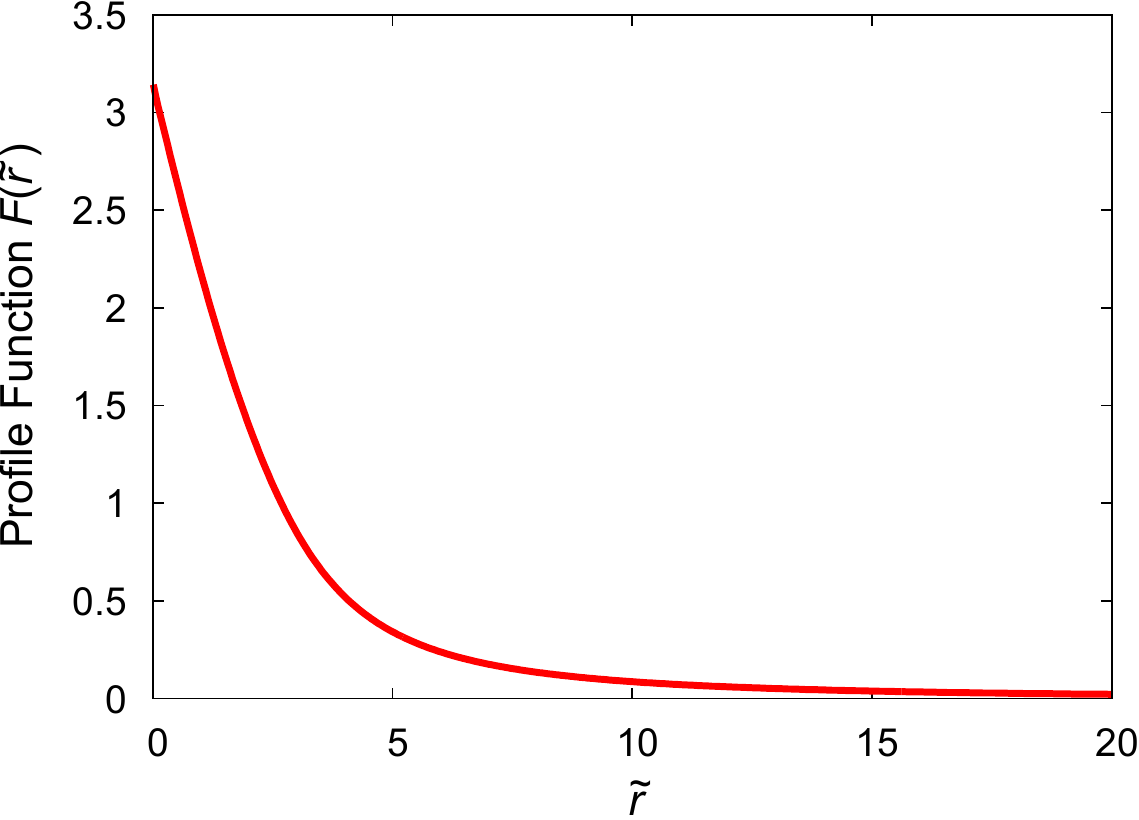}
\end{minipage} \hspace{5em}
\begin{minipage}{0.3\textwidth}
 \includegraphics[width=\textwidth]{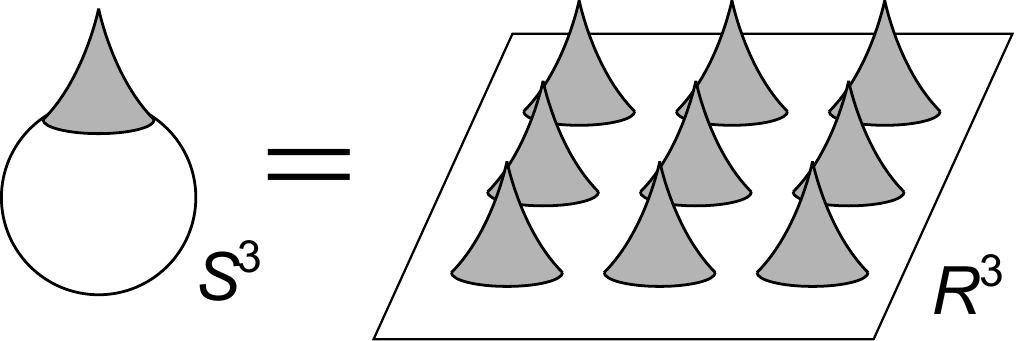}
\end{minipage}
\end{center}
 \caption{(Left) Profile function $F(\tilde{r})$ as a function of the
   dimensionless radius variable.  The slope at the origin is chosen 
   to satisfy the boundary condition; $F(\tilde{r}\to\infty)\to0$.
   (Right) Schematic picture of nuclear matter of skyrmions placed on
   $S^3$.}
 \label{fig:skyrme}
\end{figure}

The concrete form of the pion profile $F(r)$ is determined to minimize
the energy~\eqref{eq:skyrme_energy},
\begin{equation}
 \biggl( \frac{\tilde r^2}{4} + 2\sin^2 F \biggr) F''
 + \frac{\tilde r}{2} F' + F^{\prime 2} \sin 2F - \frac{\sin 2F}{4}
 - \frac{\sin^2 F \sin 2F}{\tilde r^2} = 0 \;.
\end{equation}
Here we used a dimensionless variable $\tilde r = r/R$ with $R^{-1} =
2e f_\pi$.  The numerical solution that satisfies the proper boundary
conditions is plotted in the left panel of Fig.~\ref{fig:skyrme}.  For
a more realistic description, one should include strangeness, quantize
the solution to give proper quantum numbers such as the angular
momentum and the iso-spin, which is beyond our scope of this review.

\paragraph{Nuclear matter as a Skyrme crystal}
This pion profile describes a single baryon, and it is necessary to
extend the formulation to be capable of including multi-baryons to
study the nuclear matter properties.  It is not easy to construct
skyrmions for $B\gg1$ directly, but a system at finite density can be
emulated with periodically placed skyrmions~\cite{Klebanov:1985qi}
since baryons are infinitely heavy and static in the large-$\Nc$
limit.  One can analyze this problem by putting a skyrmion on $S^3$,
which can be viewed as multi-skyrmions on $R^3$ as schematically
visualized in the right panel of Fig.~\ref{fig:skyrme}.  This idea has
been extended to the skyrmion in the holographic QCD
model~\cite{Nawa:2008uv}.

Interestingly enough, such a description of skyrmion matter might
suggest that nuclear matter in the large-$\Nc$ limit could have an
inhomogeneous structure in it.  Actually, the Skyrme model is based on
the non-linear representation of the chiral model as we saw above, and
it is impossible to reach chiral restoration, and yet, one can argue
chiral restoration as a spatial average of $\langle\tr U\rangle$.
This implies that the chiral condensate may have inhomogeneity there,
and this is indeed one of the characteristic features in quarkyonic
matter as we see in Sec.~\ref{sec:quarkyonic}.

\paragraph{Half-skyrmions and the Brown-Rho scaling}
In Ref.~\cite{Brown:1991kk} a universal scaling law in dense matter
was suggested;
\begin{equation}
 \Phi(\rho) = \frac{\mn^\ast}{\mn}
 = \frac{m_{\text{V}}^\ast}{m_{\text{V}}} = \frac{f_\pi^\ast}{f_\pi} \;,
\label{eq:BR}
\end{equation}
where the asterisks are for in-medium quantities and $\Phi$ is a
scaling function of density $\rho$.  This scaling called the Brown-Rho
(BR) scaling is achieved starting with the chiral Lagrangian with the
Skyrme term and modifying it such that the correct scaling property
expected from QCD is recovered.

The periodic skyrmion configuration is not necessarily optimal in
energy and more careful analysis on crystallography should be
necessary, which results in BCC and FCC crystal structures involving
half-skyrmions~\cite{Goldhaber:1987pb,Castillejo:1989hq}.  In fact, an
intermediate phase where a skyrmion turns into two half-skyrmions was
found at some critical density $\rho_{1/2}$ higher than the normal
nuclear density $\rho_0$~\cite{Park:2002ie,Lee:2003aq}, is also
observed in the holographic QCD models~\cite{Rho:2009ym}.  For an
intriguing implication to the QCD phase diagram, the state with
half-skyrmions indicates the presence of the pseudo-gap phase above the
chiral phase transition as is suggested in
Ref.~\cite{Zarembo:2001wr}.

It turned out that the meson and baryon masses behave differently in
increasing density and the scaling~\eqref{eq:BR} holds fairly well up
to the density $\rho_{1/2}$, but above $\rho_{1/2}$, it is
significantly deviated~\cite{Lee:2010sw}.  The baryon mass drops at a
slower rate than the meson mass with increasing density and stays
non-vanishing at chiral symmetry restoration, which matches the chiral
invariant mass $m_0$ discussed in Sec.~\ref{sec:mended}.  Its
consequences on the tensor forces and the EoS at high density are
discussed adopting a modified scaling motivated by the above
observation in Ref.~\cite{Lee:2010sw}.

\subsubsection{Sakai-Sugimoto model}
\label{sec:SSM}

Another non-perturbative approach developed in the large-$\Nc$ limit
is the holographic technique that is becoming more and more familiar
among QCD physicists nowadays.  There are countless number of works
using the holographic QCD models, and it is not realistic to make an
attempt to cover them all.  We will pay special attention to the phase
diagram disclosed by the Sakai-Sugimoto model, following the
computational steps of Ref.~\cite{Bergman:2007wp}.

\begin{figure}
\begin{center}
 \includegraphics[width=0.4\textwidth]{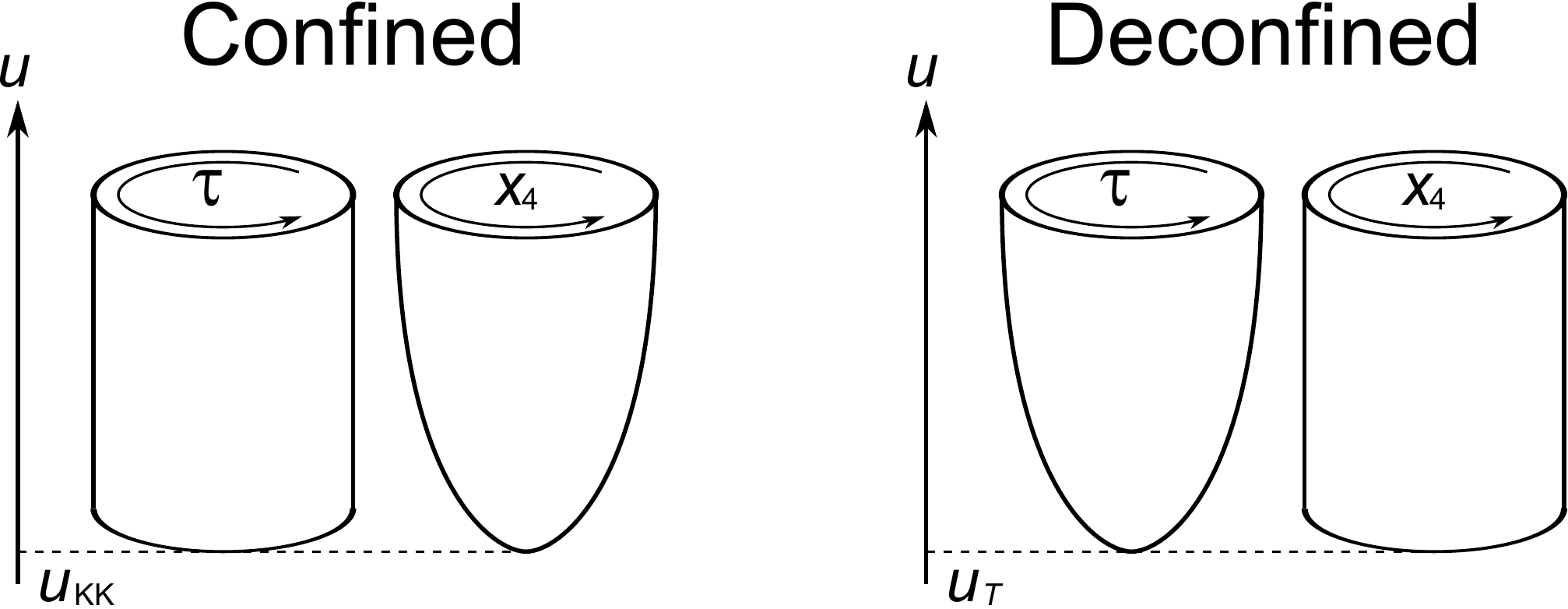}
\end{center}
 \caption{Geometries in the confined and the deconfined phases.  The
   roles of the (imaginary) time $\tau$ and the extra coordinate $x_4$
   are swapped from one to the other phase.}
 \label{fig:geometry}
\end{figure}

The basic idea of the holographic approach is based on a hypothetical
correspondence between a gravity theory in higher dimensions and a
field theory of our interest (see Ref.~\cite{Aharony:1999ti} for a
review).  The original conjecture was made for the $\mathcal{N}=4$
supersymmetric Yang-Mills theory, which at large-$\Nc$ and
large-'t~Hooft coupling may be equivalently described by the classical
solution (anti-de~Sitter metric) of the super-gravity theory.  For the
investigation of QCD, however, supersymmetry and conformal symmetry
are unwanted and should be gotten rid of.  To this end, one can
compactify one extra direction and impose the anti-periodic boundary
condition to fermionic super-particles, so that unphysical particles
become heavy and decouple from the low-lying dynamics~(see
Fig.~\ref{fig:geometry}).  This extra direction is denoted by the
coordinate $X_4$ here.

In this subsection, to simplify the notation, we use dimensionless
variables rescaled by the radius of the AdS space, $R$.  That is,
\begin{equation}
 \tilde{t} = \frac{t}{R}\;,\qquad
 \tilde{\bx} = \frac{\bx}{R}\;,\qquad
 \tilde{x}_4=\frac{X_4}{R}\;,\qquad
 \tilde{u} = \frac{u}{R}\;,
\end{equation}
where the first $(t,\bx)$ refer to the ordinary Minkowski
coordinates, $X_4$ is the compact direction, and $u$, together with
$(t,\bx)$, spans the 5-dimensional AdS space, and the Minkowski
space-time resides at the UV edge, $u=\infty$.  In what follows, we
omit writing the tilde for notational simplicity.

\paragraph{Deconfinement phase transition}
In the confined phase at low temperature, the bulk geometry
corresponding to the D4-brane background is expressed by the following
metric,
\begin{equation}
 \text{(Confining geometry)}~~~~~
 ds^2 = u^{3/2}\bigl[ - dt^2 + d\bx^2 + f(u) dx_4^2 \bigr]
  + \frac{du^2}{u^{3/2}f(u)} + u^{1/2} d\Omega_4^2 \;, \\
\label{eq:confining_metric}
\end{equation}
where $f(u) = 1-(\ukk/u)^3$ and $\ukk$ is fixed by the size of $x_4$
compactification.  This above geometry is singular at $u=\ukk$ and to
avoid the conical singularity at $u=\ukk$, in the same way as the
angle variable in the polar coordinates, the period of $x_4$ (denoted
by $\delta x_4$ here) is uniquely determined that is translated to a
mass scale called the Kaluza-Klein mass, i.e.
\begin{equation}
 \delta x_4 = \frac{4\pi}{3\ukk^{1/2}}
  \qquad\Rightarrow\qquad
 M_{\rm KK} = \frac{2\pi R^{-1}}{\delta x_4}
  = \frac{3}{2}\sqrt{\ukk}\, R^{-1} \;.
\label{eq:mkk}
\end{equation}
The physical meaning of $M_{\rm KK}$ is a cutoff scale above which
super-particles could get excited and the Sakai-Sugimoto model would
be no longer a QCD dual.  As we will see soon later, the deconfinement
transition temperature is smaller than $M_{\rm KK}$ by a factor and
the dual description of the phase diagram is indeed closed within the
validity region.

In the deconfined phase at finite $T$, on the other hand, the
so-called AdS-blackhole solution should be energetically preferred
with the metric,
\begin{equation}
 \text{(Deconfining geometry)}~~~~~
 ds^2 = u^{3/2}\bigl[ - f(u) dt^2 + d\bx^2 + dx_4^2 \bigr]
  + \frac{du^2}{u^{3/2}f(u)} + u^{1/2} d\Omega_4^2 \;,
\label{eq:deconfining_metric}
\end{equation}
where the singularity in $x_4$ is replaced in the temporal direction.
In Eq.~\eqref{eq:deconfining_metric} the definition of the function
$f(u)$ is changed to $f(u)=1-(u_T/u)^3$ with $u_T$ fixed by the period
as in Eq.~\eqref{eq:mkk}.  In Euclidean space-time with $\tau=it$ the
period of $\tau$, that is nothing but the temperature inverse $1/T$,
leads to
\begin{equation}
 \frac{1}{T} = \frac{4\pi}{3u_T^{1/2}} \;.
\end{equation}
Therefore, the location of the horizon at $u=u_T$ corresponds to the
physical temperature $T$.

One can then evaluate the free energy to identify which of the
confining and the deconfining metric is more favored and what the
critical temperature $\Td$ is.  We can easily make a theoretical guess
for $\Td$ from the corresponding geometries illustrated in
Fig.~\ref{fig:geometry}.  Obviously the energies (or actions) should
be identical when two periods coincide.  That is, the first-order
phase transition (Hawking-Page-type transition) should take
place~\cite{Herzog:2006ra} when
\begin{equation}
 \delta x_4 = \frac{1}{T} \qquad\Rightarrow\qquad
 \Td = \frac{3}{4\pi}\sqrt{\ukk}\,R^{-1} = \frac{M_{\rm KK}}{2\pi} \;.
\end{equation}
On the phenomenological level $M_{\rm KK}$ is fixed to be $949\MeV$ to
reproduce the vector meson spectrum, so that $\Td\simeq 151\MeV$ is
obtained, which is of the same order as the empirical value presented
in Sec.~\ref{sec:highlights}.  In the leading order of the large-$\Nc$
approximation this critical temperature for deconfinement is
insensitive to the presence of matter and, as discussed in
Sec.~\ref{sec:Nccounting}, this is simply because quarks are
suppressed by $1/\Nc$ as compared to gluons.

We here make a remark on the weak-coupling extrapolation with
non-trivial $\beta$ function.  This may lead to a possibility of
disconnection between deconfinement in the strong-coupling and the
weak-coupling regimes~\cite{Mandal:2011ws}.

\begin{figure}
\begin{center}
 \includegraphics[width=0.4\textwidth]{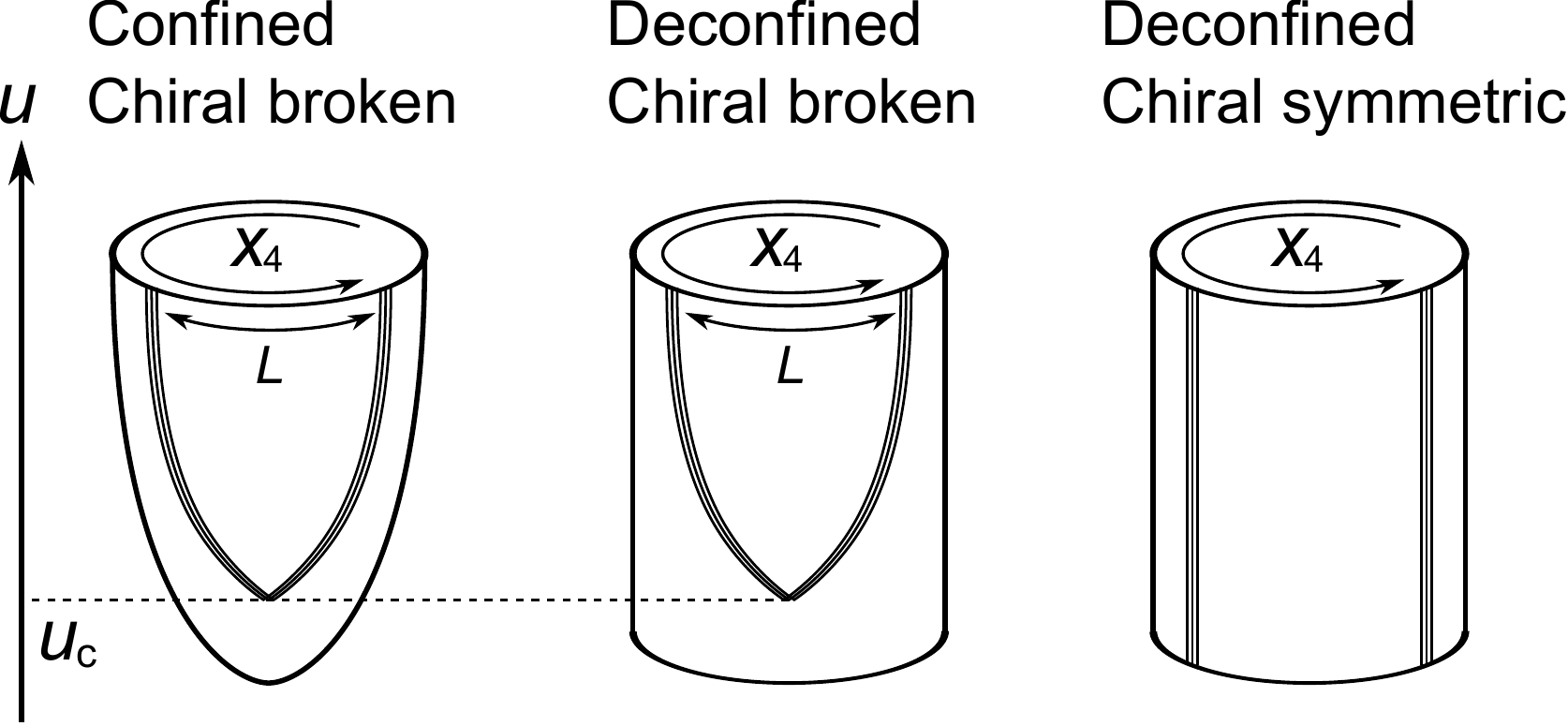}
\end{center}
 \caption{Flavor branes (D8 and $\overline{\rm D8}$ separated by the
   length parameter $L$) in the confined and the deconfined phases.
   In the confined phase chiral symmetry is inevitably broken, while
   in the deconfined phase both chiral broken and symmetric states are
   realized.}
 \label{fig:geometryD8}
\end{figure}

\paragraph{Flavor branes}
Let us now turn to chiral symmetry realized in the Sakai-Sugimoto
model~\cite{Sakai:2004cn} and the chiral phase transition.  The same
chiral symmetry as QCD, i.e.\
$\mathrm{U}_{\text{L}}(\Nf)\times\mathrm{U}_{\text{R}}(\Nf)$, can be
emulated by putting $\Nf$ D8-branes localized at $x_4=0$ and
$\Nf$ $\overline{\rm D8}$-branes at $x_4=L$.  This $L$ is a parameter
inherent in the Sakai-Sugimoto model and its physical interpretation
in QCD is still unclear.  For the meaning of the lower edge $u_c$, see
Fig.~\ref{fig:geometryD8}.  We will explain how to fix $u_c$ when we
incorporate the density source next.  In the large-$\Nc$ limit the
D4-brane is dominant and one can assume that the classical solutions
are intact regardless of flavor branes (that is, the probe
approximation), which corresponds to quark suppression by $1/\Nc$.

The induced metric on the D8-branes is (apart from the spherical
angular part, $\Omega_4$), thus,
\begin{align}
 & \text{(Confining geometry)}~~~~~~
 ds^2 = u^{3/2} \bigl( -dt^2 + d\bx^2 \bigr) + \biggl[
        u^{3/2}f(u){x_4'(u)}^2 + \frac{1}{u^{3/2}f(u)} \biggr]du^2 \;,
\\
 & \text{(Deconfining geometry)}~~~
 ds^2 = u^{3/2} \bigl( -f(u)dt^2 + d\bx^2 \bigr) + \biggl[
        u^{3/2}{x_4'(u)}^2 + \frac{1}{u^{3/2}f(u)} \biggr]du^2 \;.
\end{align}
Here $x_4(u)$ is solved with the equation of motion later.  To
introduce a finite density, we need a gauge field $A_0(u)$ that
becomes the quark chemical potential at $u=\infty$.  We can then
express the Dirac-Born-Infeld (DBI) action involving the gauge fields
on the $\Nf$ D8-brane, after integrating over the four-dimensional
angular part $\Omega_4$, as
\begin{equation}
 S_{\text{D8}}^{\text{DBI}} = -\frac{\calN}{V_4} \int dt\,d^3x\,du\,
  u^{1/4} \sqrt{-\det(g_{\alpha\beta}+F_{\alpha\beta})} \;.
\end{equation}
Here the dilaton potential is already included in the action and the
field strength tensor in the above is rescaled to eliminate the string
scale $l_s$ (see Refs.~\cite{Sakai:2004cn,Bergman:2007wp} for
details).  The normalization constant including the $\Omega_4$
integration is given as $\calN=V_4\cdot(\Nc\Nf/3)R^6/((2\pi)^5 l_s^6)$
with $V_4=\int dt\, d^3x$.

In Euclidean space-time at finite $T$, integrating further over the
Euclidean coordinates, we can express the action in a form of the
one-dimensional integration with respect to $u$.  Hereafter we drop
the irrelevant normalization $\calN$ from the action and then the
action simplifies, respectively, in the confined and the deconfined
phases as
\begin{align}
 &\text{(Confining geometry)}
 ~~~~~~~~~~~~ S_{\text{D8}}^{\text{DBI}} = \int du\,
  u^4\sqrt{f(u)x_4'(u)^2 + \frac{1}{u^3}\bigl( {f(u)}^{-1}
  - a_0'^2 \bigr)} \;,
\label{eq:Sc}\\
 &\text{(Deconfining geometry)}
 ~~~~~~~~~ S_{\text{D8}}^{\text{DBI}} = \int du\,
  u^4\sqrt{f(u)x_4'(u)^2 + \frac{1}{u^3}( 1-a_0'^2 )} \;,
\label{eq:Sd}
\end{align}
where $a_0(u)$ denotes the rescaled $A_0(u)$ that eventually
translates to the (dimensionless) chemical potential.  It is
convenient to use a density variable
$\rho(u)=-\delta S_{\text{D8}}^{\text{DBI}}/\delta a_0'(u)$, because
$\rho(u)$ turns out to be $u$-independent thanks to the equation of
motion.  Then, we can easily solve $a_0'(u)$ from the definition of
$\rho$ to find,
\begin{align}
 & \text{(Confining geometry)}~~~~~~~~~~~
 a_0'(u) = \rho\,\sqrt{\frac{u^3 f(u)^2 {x_4'(u)}^2 + 1}
  {f(u)(u^5 + \rho^2)}} \;,
\label{eq:a0c} \\
 & \text{(Deconfining geometry)}~~~~~~~~
 a_0'(u) = \rho\,\sqrt{\frac{u^3 f(u)\,{x_4'(u)}^2 + 1}
  {u^5 + \rho^2}} \;,
\label{eq:a0d}
\end{align}
from which the quark chemical potential $\muq=a_0(\infty)$ is obtained
in respective phases.

\paragraph{Introduction of the density source}
Here we shall explain how to fix the lower boundary $u_c$ displayed on
Fig.~\ref{fig:geometryD8}.  If the system is in the deconfined and
chiral symmetric phase at high $T$, D8 and $\overline{\rm D8}$ are
parallel and such a configuration can have a finite density without
source, and thus $u_c$ is not necessary and the $u$-integration starts
simply from $u_T$.

When chiral symmetry is spontaneously broken, on the other hand, there
should be a source for the density coming from the Chern-Simons
action.   To accommodate a finite density of \textit{baryonic} matter,
instantons or the D4-branes wrapped on the $S^4$ inside of the
D8-branes should be placed, as argued in Ref.~\cite{Witten:1998xy}, in
a similar way as the skyrmion in Sec.~\ref{sec:skyrme}.  In the
confined phase baryons are the unique source for the density, while
\textit{quark matter} is also possible in the deconfined phase.  Such
a density source originating from quarks can be formulated as strings
stretched from the D8-branes to the horizon.  It is known, however,
that such a density source leads to a larger action (i.e.\ larger free
energy) than the instanton source~\cite{Bergman:2007wp}, which means
that baryonic matter is \textit{always} favored than quark matter in
this model. This observation might be given a deeper insight from the
concept of quarkyonic matter that is the main subject of
Sec.~\ref{sec:quarkyonic}.

The point where the density source is placed along the $u$-direction
defines $u_c$ and its value is dynamically determined by the
zero-force condition.  In the presence of the source the flavor-branes
are pulled and $x_4(u)$ has a cusp at $u=u_c$, as illustrated in
Fig.~\ref{fig:geometryD8}, and the range of the $u$-integration should
be from $u_c$ to $+\infty$.  The zero-force condition describes a
situation that the tension of the D8-branes $f_{\text{D8}}$ should be
balanced by the force from the source D4-brane $f_{\text{D4}}$.  These
forces can be given by the derivative of corresponding actions with
respect to $u_c$.

\paragraph{Deconfined and chiral symmetric phase}

On the phase diagram we will find a first-order phase transition
associated with the restoration of chiral symmetry, and then the
density $\rho$ jumps discontinuously.  So, to distinguish the
densities in respective phases, let us denote the density in the
chiral symmetric phase as $\rhosym$, while we use $\rhobr$ in the
chiral broken phase.

When chiral symmetry is restored at sufficiently high temperature, D8
and $\overline{\rm D8}$ are straight and parallel (the far left of
Fig.~\ref{fig:geometryD8}).  In this case the solution of the equation
of motion is as simple as $x_4'(u)=0$.  Then, we do not have to
introduce a source term but the horizon at $u=u_T$ can take care of
the role as a density source.  The physical chemical potential is
obtained from $\muq=a_0(\infty)$, that is,
\begin{equation}
 \muq = \rhosym \int_{u_T}^\infty du\,
   \frac{1}{\sqrt{u^5 + \rhosym^2}} \;,
\end{equation}
from Eq.~\eqref{eq:a0d}.  The thermodynamic potential is then
calculated from the action~\eqref{eq:Sd} as
\begin{equation}
 \Omega_{\text{sym}} = \int_{u_T}^\infty du\, \frac{u^5}
  {\sqrt{u^5 + \rhosym^2}} \;.
\label{eq:Omega_s_SS}
\end{equation}

\paragraph{Deconfined and chiral broken phase}

The treatment of the chiral broken phase is much more complicated than
the chiral symmetric phase because the configuration of D8-branes
should be numerically solved from the Euler-Lagrange equation from
the action~\eqref{eq:Sd} as
\begin{equation}
 x_4'(u) = \frac{\sqrt{f(u_0)(u_0^8 + \rhobr^2 u_0^3)}}
  {u^{3/2}\sqrt{f(u)\bigl[f(u)(u^8+\rhobr^2 u^3)
  -f(u_0)(u_0^8+\rhobr^2 u_0^3) \bigr]}} \;,
\label{eq:x4_decon}
\end{equation}
where $u_0$ is defined by a point where $x_4'(u_0)$ diverges.  Then,
in this phase, the chemical potential is expressed from
Eq.~\eqref{eq:a0d} as
\begin{equation}
 \muq = \rhobr \int_{u_c}^\infty du\, \sqrt{\frac{f(u) u^3}
  {f(u)(u^8+\rhobr^2 u^3)-f(u_0)(u_0^8+\rhobr^2 u_0^3)}}
  + \frac{u_c}{3}\sqrt{f(u_c)} \;.
\label{eq:chem_br}
\end{equation}
We note that the second term in the right-hand side is added by the
D4-brane contribution in the presence of density, and this determines
the threshold $\muq$ for finite density.  Such an onset of the density
corresponds to $\mn-B/A$ in nuclear matter;  we have already seen
$\rho=0$ as long as $\muB<\mn-B/A$ (see the right panel of
Fig.~\ref{fig:walecka-t}).  In other words, if $\muq$ is smaller than
this threshold value, the only possible solution is $\rhobr=0$ and the
ground state is matter-free empty.  We should concretely figure out
the lower edge of the integration, $u_c$, to proceed in the
calculation.  The zero-force condition, $f_{\text{D8}}=f_{\text{D4}}$,
yields~\cite{Bergman:2007wp},
\begin{equation}
 \sqrt{\frac{f(u_c)(u_c^8 +\rho^2 u_c^3)
  - f(u_0)(u_0^8+\rho^2 u_0^3)} {f(u_c) u_c^3}}
 = \frac{\partial}{\partial u_c}\,\frac{\rho}{3}\,
  u_c \sqrt{f(u_c)} \;.
\label{eq:balance}
\end{equation}
The thermodynamic potential in the chiral broken phase is, thus,
obtained from the action, which amounts to
\begin{equation}
 \Omega_{\text{broken}} = \int_{u_c}^\infty du\,
  \frac{\sqrt{f(u)\, u^3}\, u^5}
  {\sqrt{f(u)(u^8+\rhobr^2 u^3)
   -f(u_0)(u_0^8+\rhobr^2 u_0^3)}} \;.
\label{eq:Omega_b_SS}
\end{equation}

The above-mentioned setup suffices for the determination of the chiral
phase transition point by the comparison between
Eqs.~\eqref{eq:Omega_s_SS} and \eqref{eq:Omega_b_SS}.  The concrete
procedure for this is as follows:  First, for a given density
$\rhobr$, the zero-force condition~\eqref{eq:balance} gives
$u_c(u_0,T)$, and this renders the condition
$L=2\int_{u_c}^\infty x_4'(u)du$ expressed as a function of $u_0$.
This enables us to solve $u_0(\rhobr,L,T)$, so that
Eq.~\eqref{eq:chem_br} directly relates $\muq$ and $\rhobr$.  Then we
get $\Omega_{\text{broken}}(L,T,\muq)$ finally.  In the same (and much
simpler) way, we can calculate $\Omega_{\text{sym}}(L,T,\muq)$, and we
can locate the first-order Hawking-Page transition of chiral
restoration by looking for $T$ and $\muq$ that makes
$\Omega_{\text{broken}}=\Omega_{\text{sym}}$.  In this way,
particularly at vanishing $\muq$, the critical temperature turns out
to be $T_\chi = 0.154/L$, which was first found in
Ref.~\cite{Aharony:2006da}.

\paragraph{Density onset and the phase diagram}
The phase diagram is not yet complete.  We should consider another
line corresponding to the liquid-gas phase transition as presented
explicitly in the conventional description in the right panel of
Fig.~\ref{fig:walecka-t}.

In the confined phase chiral symmetry is always broken.  The equation
of motion for $x_4'(u)$ is slightly different from
Eq.~\eqref{eq:x4_decon} and is given by
\begin{equation}
 x_4'(u) = \sqrt{\frac{f(u_0)(u_0^8 + \rhobr^2 u_0^3)}{u^3 f(u)^2
  [ f(u)(u^8+\rhobr^2 u^3) - f(u_0)(u_0^8+\rhobr^2 u_0^3)]}} \;,
\label{eq:x4c}
\end{equation}
from which the zero-force condition results in a similar form to
Eq.~\eqref{eq:balance} in which the right-hand side is replaced with
$\partial(\rho u_c/3)\partial u_c$.  The onset at which a finite
density is turned on can be found analytically with $\rhobr=0$ plugged
into the above equations.  Then, we simply have $u_c=u_0$, and the
corresponding chemical potential is $\muq=u_0/3$.  At the onset we can
approximate \eqref{eq:x4c} with $f(u)\simeq 1$ to derive $u_0$
analytically, and the critical $\muq$ is then obtained as
$(4\pi/3)(\Gamma[9/16]/\Gamma[1/16])^2\simeq 0.175$ in the unit of
$L$.  In the deconfined phase, on the other hand, the density onset
lies at $\muq=(u_0/3)\sqrt{f(u_0)}$, which is $T$-dependent.  Then,
the phase boundary is not a straight vertical line but it has a minor
dependence on $T$.

\begin{figure}
\begin{center}
 \includegraphics[width=0.5\textwidth]{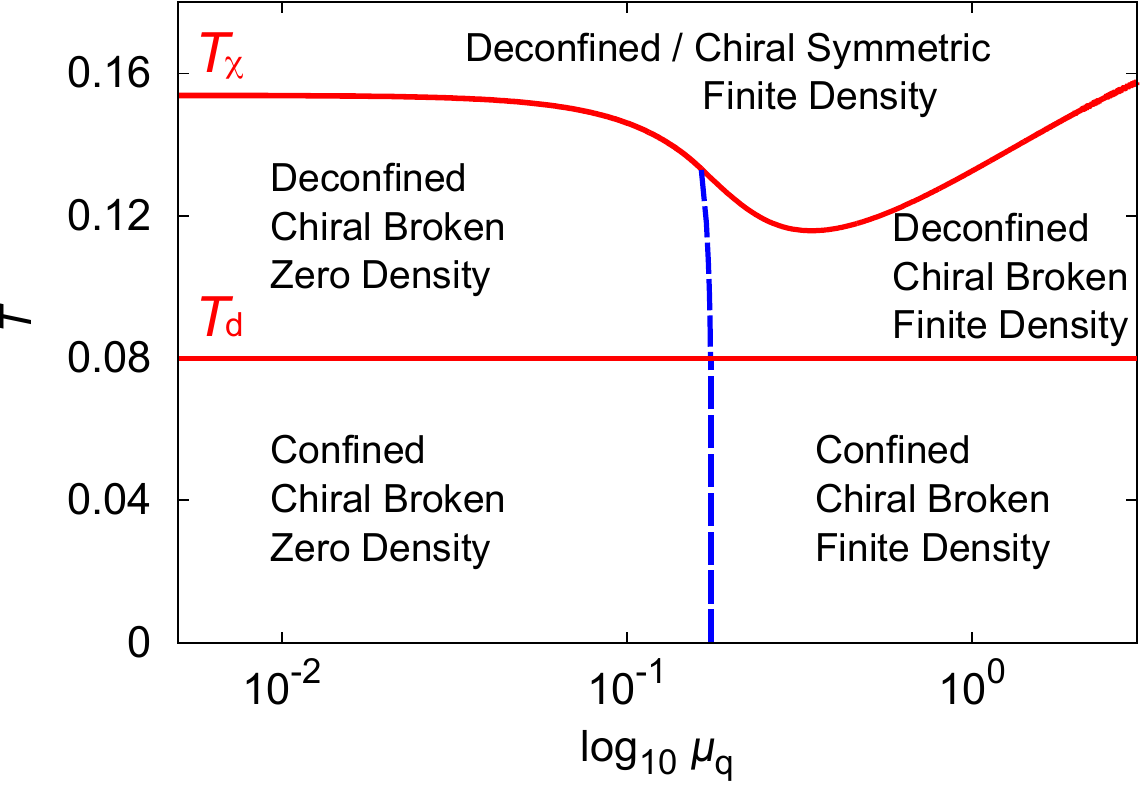}
\end{center}
 \caption{Phase diagram at finite temperature and density in the
   Sakai-Sugimoto model.  The deconfinement temperature is chosen at
   $\Td=0.08$ and the scale is set by $L=1$ according to the choice of
   Ref.~\cite{Bergman:2007wp}.  The vertical dashed line represents
   the onset of baryon density.}
 \label{fig:phaseSS}
\end{figure}

Some numerical calculations are needed to clarify the whole phase
structure, and the resulting phase diagram is depicted in
Fig.~\ref{fig:phaseSS}, which first appeared in
Ref.~\cite{Bergman:2007wp}.  This is a quite suggestive phase diagram,
but we shall defer a more detailed comparison between
Fig.~\ref{fig:phaseSS} and the expected phase structure of large-$\Nc$
QCD till Sec.~\ref{sec:quarkyonic}.

We here raise some caveats;  the above treatment is based on the probe
approximation that corresponds to the quench limit of QCD, and the
$1/\Nc$ corrections should be considered to take account of
back-reactions from matter.  For more details including other
holographic models than the Sakai-Sugimoto model, see a recent
review~\cite{Kim:2012ey}.  This is not yet the end of the story about
the phase diagram in the Sakai-Sugimoto model.  Recently there has
been an important recognition;  an instability of gauge fluctuations
at non-zero momenta was discovered at finite
density~\cite{Chuang:2010ku,Ooguri:2010xs}, and so, the phase diagram
in Fig.~\ref{fig:phaseSS} must be supplemented with additional phase
boundaries associated with spatial modulations (see Fig.~1 of
Ref.~\cite{Chuang:2010ku} for instance).

It is quite interesting to notice that this sort of spatially
inhomogeneous structure was once discussed actively in the context of
nuclear physics, and nowadays, is becoming one of the central issues
in quark matter research.  Here, for a while, it should be worth
revisiting the possibility of inhomogeneous condensation in nuclear
matter.

\subsection{Inhomogeneity: pion condensation}
\label{sec:picon}

In a medium at finite density there are collective excitations made
from particle and hole ($p$-$h$) in the same channel as pions.  The
interaction between $N$ and $\pi$ is, however, repulsive in the
$s$-wave channel and the pion self-energy is positive then.  It was
recognized later in Refs.~\cite{Sawyer:1973fv,Migdal:1973zm} that the
$p$-wave interaction is attractive, which causes a resonant state
$\Delta$ in the channel of $L=1$, $J=3/2$, $T=3/2$, and it would
render the pion energy decrease in matter.  This attractive
interaction is attributed to underlying chiral symmetry.

The in-medium pion propagator can be written with the self-energy as
\begin{equation}
 D_\pi^{-1}(\omega,\bp) = \omega^2 - \bp^2 - m_\pi^2
  - \Pi(\omega,\bp) \;,
\end{equation}
where the self-energy $\Pi(\omega,\bp)$ should involve the $p$-$h$ and
$\Delta$-$h$ contributions;
$\Pi=-p^2[U_N(\omega,\bp)+U_\Delta(\omega,\bp)]$.  In the symmetric
nuclear matter with $Z=N$ that is of our main interest, the collective
mode is pushed down at finite $\bp$ and there appears a condensate of
$\pi^0$, while in neutron matter which is relevant to the
astrophysical application the collective $\pi^+$ (Migdal's $\pi_s^+$)
and $\pi^-$ are spontaneously generated~\cite{Migdal:1973zm}.

When the $\pi^0$ condensation occurs in symmetric nuclear matter at
the momentum $\bp_c$, the expectation value should behave as
\begin{equation}
 \langle\pi^0(\bx,\bp_c)\rangle \;\sim\; \cos(\bp_c\cdot\bx) \;.
\label{eq:pi0}
\end{equation}
The physical implication of such condensate to nuclear matter has been
discussed (see Ref.~\cite{Tamagaki:1993pu} and other contributions in
this volume).

One can find the critical density for the $\pi^0$ condensation, for
instance, adopting the linear sigma model including $p$, $n$, and
$\Delta$'s.  Once the density exceeds the threshold, however, the
system is unstable because only the attractive force is taken into
account and the short-range (shorter than the pion Compton length)
repulsive interaction between baryons should be considered.  In the
Fermi liquid theory such short-range $NN$ interactions (proportional
to $\delta(\bx_1\!-\!\bx_2)$) are incorporated through the Landau
parameters in the following form;
\begin{equation}
 F_{NN} = f_{NN} + f'_{NN}\btau_1\cdot\btau_2
  + g_{NN}\bsigma_1\cdot\bsigma_2
  + \biggl(\frac{f_{\pi NN}^2}{m_\pi^2}\biggr) g'_{NN}
    (\bsigma_1\cdot \bsigma_2)(\btau_1\cdot\btau_2) \;.
\end{equation}
Here $\btau_1$ and $\bsigma_1$ (and $\btau_2$ and $\bsigma_2$) refer
to the isospin and the spin of nucleon 1 (and nucleon 2,
respectively) in the $NN$ system.  In fact this last term is required
to cancel Dirac's delta function appearing in the one-pion exchange
potential (OPEP) (if $g_{NN}'=1/3$ is chosen, the attractive delta
function in the OPEP is exactly canceled).  Such a cancellation is a
universal phenomenon; the Lorentz-Lorenz effect in dielectric media is
the most famous example.  In the same manner, including the mixing
with isobars, one should further consider the Landau-Migdal parameters
as
\begin{equation}
 F_{N\Delta} = \biggl(\frac{f_{\pi NN}f_{\pi N\Delta}}{m_\pi^2}\biggr)
  g'_{N\Delta} (\bsigma_1\cdot \bsigma_2^\Delta)
  (\btau_1\cdot\btau_2^\Delta) \;,\quad
 F_{\Delta\Delta} = \biggl(\frac{f_{\pi N\Delta}^2}{m_\pi^2}\biggr)
  g'_{\Delta\Delta} (\bsigma_1^\Delta\cdot \bsigma_2^\Delta)
  (\btau_1^\Delta\cdot \btau_2^\Delta) \;,
\end{equation}
all of which contribute to $U_N(\omega,\bp)$ and
$U_\Delta(\omega,\bp)$.  Because $g_{NN}'$, $g_{N\Delta}'$, and
$g_{\Delta\Delta}'$ represent the effect of repulsive interactions,
they would disfavor the pion condensation.  Therefore, the precise
determination of these parameters is fatally important to judge
whether the pion condensation can occur or not in reality.

Today it is widely believed that the possibility of the pion
condensation in nuclear matter was somehow ruled out.  This was
concluded from the experimental data of the giant Gamow-Teller (GT)
states which are collective modes with respect to the spin-isospin
fluctuations, so that their strength can constrain $g_{NN}'$,
$g_{N\Delta}'$, and $g_{\Delta\Delta}'$ (see Ref.~\cite{Oset:1981ih}
for a review).  Since the decomposition is difficult, it used to be
often the case to assume the universality,
$g_{NN}'=g_{N\Delta}'=g_{\Delta\Delta}'=g'$, and then the experimental
data suggests $g'=0.6\sim0.9$, which is much larger than $1/3$ and the
pion condensation cannot occur in a reasonable range of baryon
density.  Later on, in theory and experiment, different values of
$g'$ have been investigated~\cite{Shiino:1986kp}, and as
comprehensively reviewed in Ref.~\cite{Tatsumi:2003fa}.  Although the
pion condensation is unlikely at normal nuclear density $\rho_0$, the
fact is that \textit{there is still a good deal of chance for the
  $p$-wave pion condensation to develop in nuclear matter at
  $\rho\gtrsim 2\rho_0$}.  According to Ref.~\cite{Tatsumi:2003fa},
the critical density $\rho_c$ is calculated with $g_{NN}'=0.59$ and
$g_{N\Delta}'=0.18+0.05g_{\Delta\Delta}'$ fixed from the experimental
data, which results in $\rho_c=(1.2\sim 2)\rho_0$ for the undetermined
parameter $g_{\Delta\Delta}'=0\sim 1$.  This makes a sharp contrast to
$\rho_c$ under the universality Ansatz, which rapidly grows up to
unrealistic density.  Definitely, the possibility of the pion
condensation should deserves more investigations.  As we will see
later in Sec.~\ref{sec:inhomo}, an interesting form of the
inhomogeneous condensate analogous to the condensate~\eqref{eq:pi0} is
a likely candidate for the ground state of quark matter.

\section{Quark/Quarkyonic Matter}
\label{sec:quark}

In this section we address the properties and the phase transition in
deconfined quark matter assuming that the ground state of high density
matter is composed from quasi-quarks.  First we will make a quick
review over a phenomenological description of deconfinement, and then
continue our discussions utilizing chiral effective models.  Finally
we will closely discuss the possibility of inhomogeneous chiral
condensate and the general mechanism that would derive inhomogeneity
at high density.

\subsection{Statistical bootstrap and the Hagedorn limiting temperature}
\label{sec:bootstrap}

In quantum field theories we are still far from satisfactory
understanding of confinement and deconfinement.  In phenomenology,
however, it is quite useful (and probably a right picture) to think of
the Hagedorn spectrum.  Let us denote the density of states
$\rho(m^2)$ so that $\rho(m^2)dm$ counts the number of hadronic states
in the mass range $m\sim m+dm$.  The idea of the statistical bootstrap
model is that hadronic states are composed from a convolution of
smaller hadrons.  This picture is embodied as the following
statistical bootstrap equation for the hadron mass
spectrum~\cite{Hagedorn:1965st},
\begin{equation}
 H \rho(p^2) = H \theta(p_0)\delta(p^2-m_{\rm in}^2)
  +\sum_{N=2}^\infty \frac{1}{N!} \int\delta^{(4)}
  (p-\sum_i^N p_i) \prod_{i=2}^N H\rho(p_i^2) d^4 p_i \;,
\label{eq:boot}
\end{equation}
where $H$ represents a proper hadronic volume and $m_{\rm in}$ is an
input parameter corresponding to the ``elementary'' mass in the
bootstrap picture.  Interestingly, it is possible to fix the
asymptotic form of $\rho(m^2)$ semi-analytically from the above
equation.  Introducing the Laplace transforms of Eq.~\eqref{eq:boot}
as
$G(\beta)=\int e^{-\beta p} H \rho(p^2) d^4 p$ and
$\varphi(\beta)=\int e^{-\beta p} H \theta(p_0)
\delta(p^2\!-\!m_{\rm in}^2) d^4 p$, where $\beta$ corresponds to the
temperature in the Boltzmann factor, the bootstrap
equation~\eqref{eq:boot} is rewritten into a simple form;
\begin{equation}
 G(\beta) = \varphi(\beta) + e^{G(\beta)} - G(\beta) - 1\;.
\label{eq:boot2}
\end{equation}
Although it is not easy to solve $G$ as a function of $\varphi$,
Eq.~\eqref{eq:boot2} already solves $\varphi$ in terms of $G$;
$\varphi=2G+1-e^G$, which has a maximum and we shall denote this
extremal point by $\varphi_0$, and this $\varphi_0$ can be converted
to corresponding $\beta_0$.  Then, $G(\beta)$ can be expanded around
$\beta_0$, i.e.\ 
$G(\beta)\simeq G(\beta_0) + \text{(const.)}(\beta-\beta_0)^{1/2}$.
Such a type of square-root singularity can be reproduced by the
density of states,
\begin{equation}
 \rho(m^2) = C\, m^{-3} e^{m/\Th} \;,
\label{eq:Hagedorn}
\end{equation}
where $\Th$ is a temperature parameter (related to $\beta_0$) and
called the Hagedorn temperature.  This exponential form of the density
of hadronic states is an important observation in understanding the
physics of deconfinement at the phenomenological level.

The physical interpretation of the Hagedorn temperature is quite
instructive.  The partition function, $\int\rho(m)e^{-m/T}dm$, would
diverge when $T>\Th$ and thus the physical temperature cannot exceed
$\Th$ as long as the hadronic description is valid.  In this sense
$\Th$ used to be considered as the limiting temperature.  In
Ref.~\cite{Cabibbo:1975ig} it was argued that $\Th$ should be the
critical temperature where the hadronic description breaks down and
the system goes through a phase transition into deconfined quarks.
Actually the very first prototype of the QCD phase diagram was
conjectured in Ref.~\cite{Cabibbo:1975ig} based on this interpretation
of $\Th$.  In a modern language stemming from the Hagedorn picture,
the deconfinement phenomenon is correctly captured by the thermal
statistical model or the hadron resonance gas model (see also
Sec.~\ref{sec:def_quarkyonic}).

\subsection{Chiral effective models}

Once quark deconfinement occurs, the most suited description of the
state of matter should be of interacting quarks rather than mesons and
baryons.  It is still challenging to tackle the QCD phase transitions
from the first principle unless the the quark density is so small that
the lattice-QCD is at work.  Then, for the pragmatic studies on the
QCD phase diagram, one can adopt an effective model that shares common
physical features with QCD, and chiral symmetry is the guiding
principle for the model building.

\subsubsection{Nambu--Jona-Lasinio model}
\label{sec:NJL}

At sufficiently high density it is conceivable that the physical
degrees of freedom should be quarks rather than baryons, and then it
makes sense to consider a picture of quasi-quarks as a result of
integrating out heavy gluonic degrees of freedom.  A recent attempt
along this line is found in Ref.~\cite{Kondo:2010ts}.

In this context the Nambu--Jona-Lasinio (NJL) model is one of the most
preferable choices of the low-lying QCD models (for reviews, see
Refs.~\cite{Hatsuda:1994pi,Vogl:1991qt}).  In principle, the
integration over gluons would lead to an infinite number of
interactions in terms of quark fields, which can be expanded according
to the power of $\bar{\psi}$ and $\psi$ as
\begin{equation}
 \calL = \bar{\psi}(i\gamma_\mu\partial^\mu - m_0)\psi
  + \calL_{\text{int}}^{(4)}(\bar{\psi},\psi)
  + \calL_{\text{int}}^{(6)}(\bar{\psi},\psi) + \cdots\;,
\end{equation}
where $\calL_{\rm int}^{(4)}$ represents the interaction terms at the
quartic order in terms of quark fields, $\calL_{\rm int}^{(6)}$ at the
sixth order, etc.  In the vicinity of the chiral phase transition only
$\calL_{\rm int}^{(4)}$ is predominantly important in the mean-field
analysis (if it is of second order and the mean-field mass is
vanishingly small), and we will keep this leading-order interaction
only in this article (see Ref.~\cite{Kashiwa:2006rc} for an extension
with higher-order interactions).  If the above $\calL$ is designed to
mimic the QCD dynamics, $\calL$ should be invariant under the
symmetries of the QCD Lagrangian as exposited in Sec.~\ref{sec:QCD}.
Motivated by the presence of the chiral condensate in the scalar
channel, we should retain the interaction involving the scalar channel
in $\calL_{\rm int}^{(4)}$, and naturally, $\calL_{\rm int}^{(4)}$
should contain the interaction in the pseudo-scalar channel as the
chiral partner, which leads to
\begin{equation}
 \calL_{\text{int}}^{(4)} = \frac{g_s}{2}\bigl[(\bar{\psi}\lambda_a\psi)^2
  +(\bar{\psi}i\gamma_5\lambda_a\psi)^2\bigr]
  -g_v(\bar{\psi}\gamma_\mu\psi)^2 \;.
\label{eq:NJL4}
\end{equation}
Here the scalar part $(\bar{\psi}\lambda_a\psi)^2$ and the
pseudo-scalar part $(\bar{\psi}i\gamma_5\lambda_a\psi)^2$ are both
necessary to be consistent with chiral symmetry, whilst the second
term $(\bar{\psi}\gamma_\mu\psi)^2$ is chiral symmetric as it is.  In
the NJL model study in the vacuum, this vector interaction has only a
minor effect except for the vector mesons, but at finite density we
must not neglect it.  As a matter of fact, this vector interaction
plays an equivalent role as the term
$(m_\omega^2/2)\omega_\mu\omega^\mu$ in the Walecka
model~\eqref{eq:Lag_w}.

In QCD at the quantum level $\UA$ symmetry is explicitly broken due to
the axial anomaly, which should give rise to the instanton-induced
interaction, as already derived in Eq.~\eqref{eq:inst_induce}.
Because the determinant in flavor space should be taken, the
instanton-induced interaction is the four-quark interaction for
$\Nf=2$ and the six-quark interaction for $\Nf=3$, and written then as
\begin{equation}
 \calL_{\text{int}}^{(6)} = g_d\bigl[\det\bar{\psi}(1-\gamma_5)\psi
  +\det\bar{\psi}(1+\gamma_5)\psi\bigr] \;.
\label{eq:NJL6}
\end{equation}
The model parameters, $g_s$, $g_v$, $g_d$, and the UV cutoff $\Lambda$
are fixed so as to reproduce the hadron properties in the vacuum.  In
general these parameter should vary with changing $T$ and $\muq$, but
there is no reliable strategy for the determination of such medium
dependence.  Thus, in the model studies, one usually assumes that one
can keep using the same parameters as fixed at $T=\muq=0$, which
brings in the most unmanageable uncertainty in theory.

\paragraph{Mean-field approximation in the vacuum}

The thermodynamic potential is composed from three individual
contributions in the simplest mean-field approximation of the NJL
model.  That is,
\begin{equation}
 \Omega = \Omega_{\text{cond}} + \Omega_{\text{zero}}
  + \Omega_{\text{quark}} \;,
\end{equation}
where the three contributions come from the condensation energy, the
zero-point oscillation energy, and the quasi-quark partition
function.  Let us take a look at respective contributions in order
below.

The condensation energy comes from the vacuum expectation value of the
quark composite operator.  The four-point interaction vertex of
$u$-quarks, for example, can be approximated as
\begin{equation}
 (\bar{u}u)^2 =
  (\bar{u}u-\langle\bar{u}u\rangle+\langle\bar{u}u\rangle)^2
 \simeq \langle\bar{u}u\rangle + 2\langle\bar{u}u\rangle
  (\bar{u}u-\langle\bar{u}u\rangle)
 = -\langle\bar{u}u\rangle^2 + 2\langle\bar{u}u\rangle\bar{u}u\;.
\label{eq:uu_mf}
\end{equation}
Here we have ignored second-order infinitesimals,
$(\bar{u}u-\langle\bar{u}u\rangle)^2$.  Then, in the final expression
above, $-\langle\bar{u}u\rangle^2$ is the condensation energy and the
second term amounts to the dynamical mass term for $u$-quarks, which
should be taken into account in the quasi-quark partition function as
the constituent quark mass.  It should be mentioned that there is no
mixing between different flavors like $\bar{u}u\bar{d}d$ from
$\calL_{\rm int}^{(4)}$ but the flavor mixing interaction appears
solely from the anomaly terms;  for $\Nc=3$ for instance, the
determinant interaction in Eq.~\eqref{eq:NJL6} causes the flavor
mixing.  Including the terms $\calL_{\rm int}^{(6)}$ for $\Nf=3$, the
condensation energy amounts in total to
\begin{equation}
 \Omega_{\text{cond}} = g_s(\langle\bar{u}u\rangle^2
  +\langle\bar{d}d\rangle^2 + \langle\bar{s}s\rangle^2)
  +4g_d \langle\bar{u}u\rangle\langle\bar{d}d\rangle
  \langle\bar{s}s\rangle \;.
\end{equation}

In the NJL model the zero-point oscillation energy is essential to
bring about the spontaneous breaking of chiral symmetry.  Because the
NJL model is non-renormalizable, the momentum integration needs a UV
cutoff $\Lambda$.  In a na\"{i}ve momentum cutoff scheme the diverging
zero-point oscillation energy is expressed as
\begin{equation}
 \Omega_{\text{zero}} = -2\cdot 2 \Nc\sum_i \int^\Lambda
  \frac{d^3 p}{(2\pi)^3}\, \frac{\varepsilon_i(p)}{2} \;,
\end{equation}
where the overall factor $2\cdot 2$ is from the spin and the
particle and anti-particle degeneracy, and $\Nc=3$ is the number of
colors.  For a more sophisticated treatment with non-local
interactions that tame the UV divergence, see
Refs.~\cite{Schmidt:1994di,Hell:2008cc,Hell:2009by,Radzhabov:2010dd}.
We also note that the non-local formulation is already close to the
Dyson-Schwinger approach~\cite{Horvatic:2010md}.  The quasi-particle
energy $\varepsilon_i(p)$ depends on the flavor which is indicated by
the index $i$.  In the mean-field approximation, together with the
second term of Eq.~\eqref{eq:uu_mf}, the quasi-quarks should have the
constituent quark masses;
\begin{equation}
 M_i = m_i - 2g_s\langle\bar{\psi}_i\psi_i\rangle
  -g_d\,\epsilon_{ijk}\langle\bar{\psi}_j\psi_j\rangle
  \langle\bar{\psi}_k\psi_k\rangle \;.
\end{equation}
It is possible to carry out the momentum integration analytically.
In particular in the chiral limit (where $\mmu=\mmd=\mms=0$), an
expanded form suffices for the investigation in the vicinity of chiral
restoration;
\begin{equation}
 \Omega_{\text{zero}} = -\frac{\Nc\Nf\Lambda^4}{8\pi^2} \biggl[
  \sqrt{1+\xi^2}(2+\xi^2) + \frac{\xi^4}{2}\ln\Bigl|
  \frac{\sqrt{1+\xi^2}-1}{\sqrt{1+\xi^2}+1} \Bigr| \biggr]
 \simeq \frac{\Nc\Nf\Lambda^4}{4\pi^2} (1+\xi^2) + \calO(\xi^4) \;,
\label{eq:Omega_zero}
\end{equation}
where we have defined a dimensionless variable $\xi=\Mq/\Lambda$ with
$\Mq=M_{\text{u}}=M_{\text{d}}=M_{\text{s}}$ in the chiral limit and
$\calO(\xi^4)$ denotes the neglected higher-order terms.  The
logarithmic term has a special effect on the analysis of the order of
the phase transition.  Together with the condensation energy, the
total energy is expanded up to the quadratic order as
\begin{equation}
 \Omega_{\text{cond}} + \Omega_{\text{zero}} \simeq
  -\frac{\Nc\Nf\Lambda^4}{4\pi^2}\biggl[ 1+\Bigl(1-\frac{\pi^2}
  {\Nc\Nf g_s \Lambda^2}\Bigr) \xi^2 \biggr] + \calO(\xi^4)\;.
\end{equation}
We note here that the instanton-induced interaction of $\calO(\xi^3)$
is negligible in the present approximation.  Then, if the potential
curvature is negative, $\xi$ or the constituent quark mass
spontaneously has a finite expectation value, and the symmetry
breaking condition reads~\cite{Nambu:1961tp},
\begin{equation}
 g_s \Lambda^2 > g_s^\ast\Lambda^2 \equiv \frac{\pi^2}{\Nc\Nf} \;.
\label{eq:cri_g}
\end{equation}
The existence of the critical value of the coupling strength is
presumably an artifact of non-confining model.

\begin{figure}
 \begin{center}
 \includegraphics[width=0.24\textwidth]{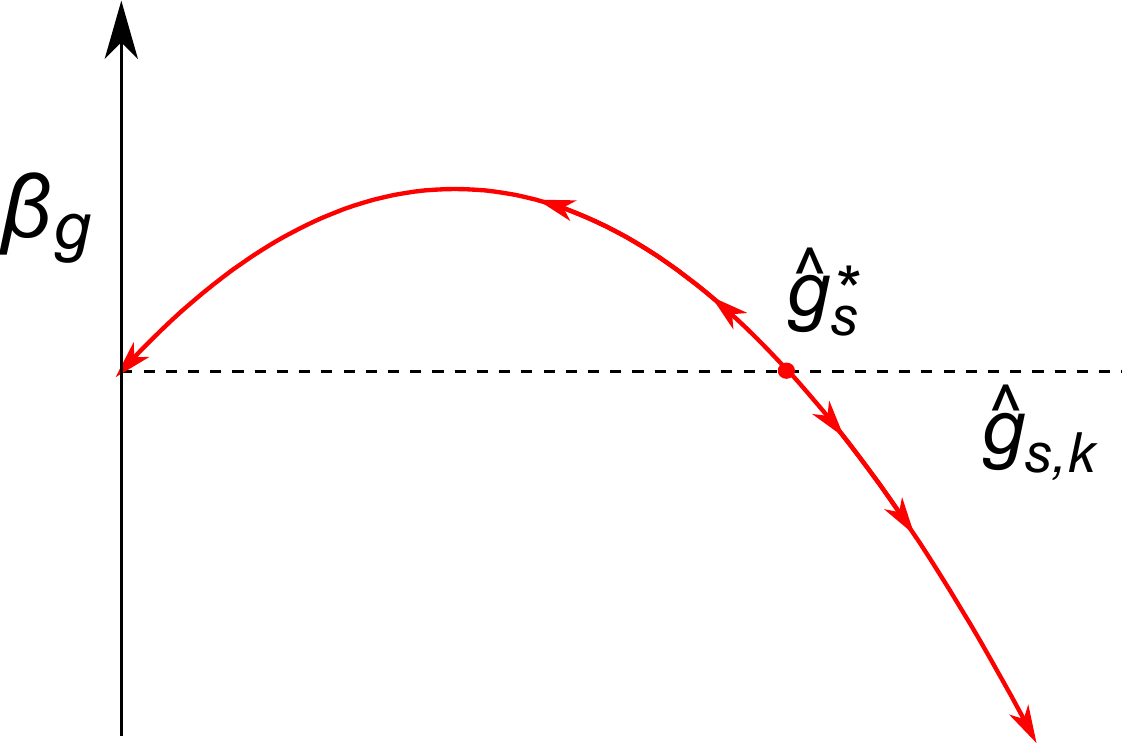} \hspace{3em}
 \includegraphics[width=0.65\textwidth]{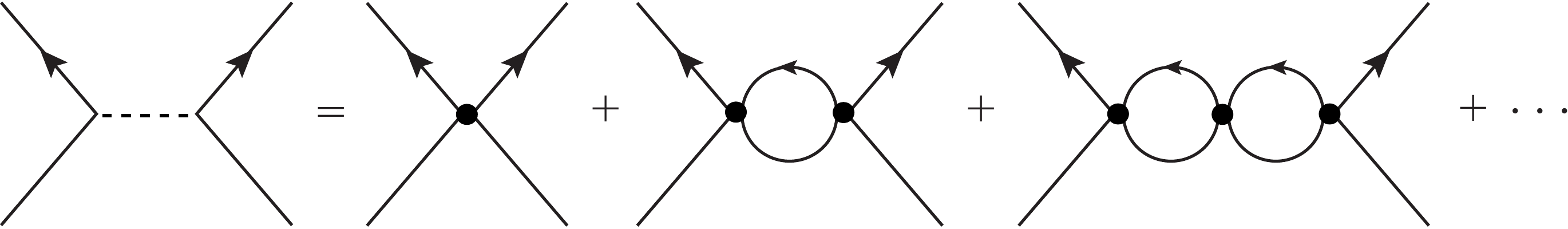}
 \end{center}
 \caption{(Left) Schematic picture of the RG flow of the coupling
   constant for the four-fermion interaction $g_s$.  The $\beta$
   function crosses zero at $\hat{g}_s^\ast$.  (Right) Bubble
   resummation for the four-fermion coupling $g_s$.}
 \label{fig:flow_NJL}
\end{figure}

The physical interpretation of this condition~\eqref{eq:cri_g} from
the RG flow is quite informative (for a review, see
Ref.~\cite{Braun:2011pp}).  For this purpose let us introduce the
dimensionless coupling $\hat{g}_{s,k}\equiv g_{s,k}k^2$.  The coupling
constant in the NJL Lagrangian should be regarded as the bare one at
the UV scale, i.e.\ $g_s = g_{s,\Lambda}$ at $k=\Lambda$.  The $\beta$
function of $\hat{g}_{s,k}$ (denoted by $\beta_g$ here) behaves as
shown schematically in the left panel of Fig.~\ref{fig:flow_NJL}.  If
the dimensionless NJL coupling $\hat{g}_s$ is smaller than the
fixed-point $\hat{g}_s^\ast$ that makes $\beta_g$ vanish, $\beta_g$ is
positive and $\hat{g}_{s,k}$ goes smaller with decreasing $k$.  If
$\hat{g}_s$ is larger than $\hat{g}_s^\ast$, on the other hand,
$\hat{g}_{s,k}$ gets larger and eventually diverges at $k\to0$.  This
diverging behavior of $\hat{g}_{s,0}$ signals that chiral symmetry
should be spontaneously broken.  As a matter of fact, the RG flow is
generated through the inclusion of fluctuations and $\hat{g}_{s,0}$
should contain all higher-loop diagrams in the four-fermion channel.
In the one-loop order of $\beta_g$, after integrating with respect to
$k$, $\hat{g}_{s,0}$ resums all bubble-type diagrams as sketched in
the right panel of Fig.~\ref{fig:flow_NJL}.  Thus, there appear
$\sigma$ and $\pi$ mesons in the intermediate states (dotted line in
the right panel of Fig.~\ref{fig:flow_NJL}) in $\hat{g}_{s,0}$, and
$\hat{g}_{s,0}\to\infty$ at $k\to0$ exactly corresponds to the
(zero-momentum) propagation of massless mesons at the critical point.
In this way, $g_s^\ast\Lambda^2$ in Eq.~\eqref{eq:cri_g} can be given
a clear interpretation.

With gauge field fluctuations, this $\beta_g$ is pushed down and if
the gauge coupling is large enough, $\beta_g$ would be entirely
negative, and chiral symmetry is spontaneously broken regardless of
the starting point of the RG flow~\cite{Braun:2011pp}.  At finite
temperate, on the other hand, $\beta_g$ is pushed up, and
$\hat{g}_s^\ast$ moves toward a larger value.  Then, when
$\hat{g}_s^\ast$ becomes greater than a given $\hat{g}_s$ at a certain
temperature, the second-order phase transition of chiral restoration
occurs, as we see below.

\paragraph{Phase diagram in the mean-field approximation}

\begin{figure}
 \begin{center}
 \includegraphics[width=0.4\textwidth]{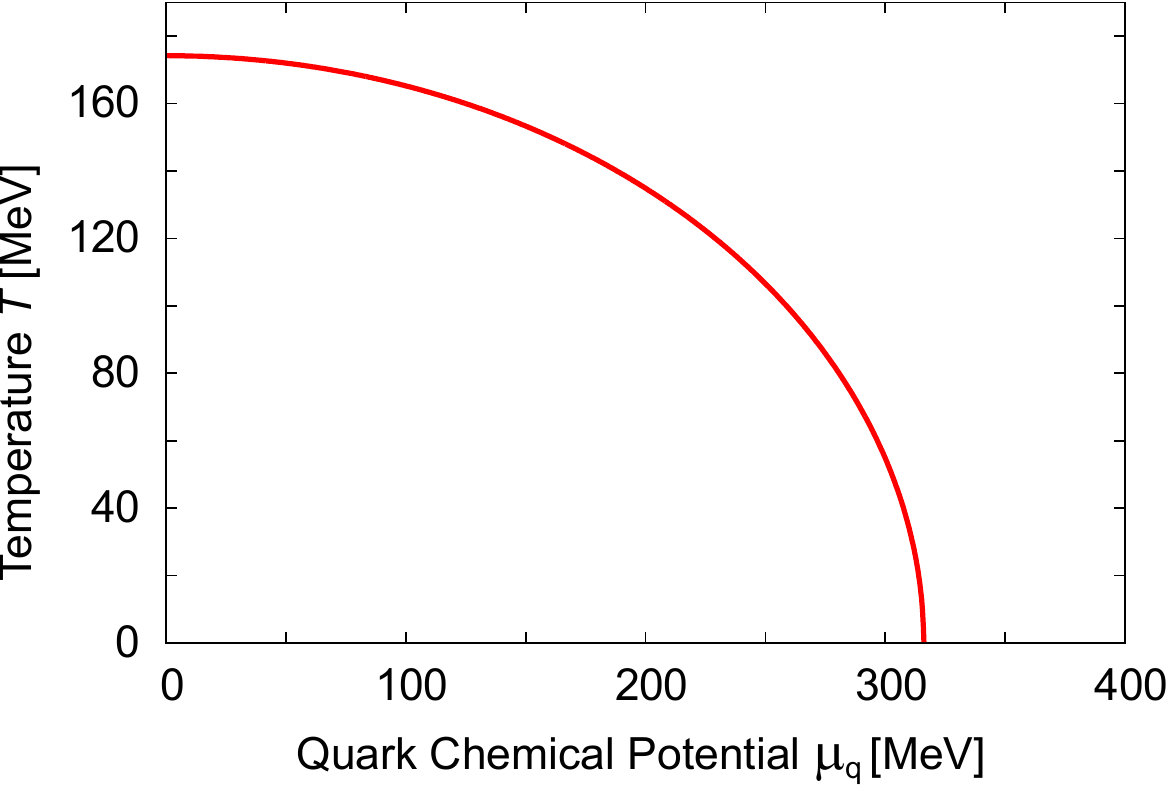}
 \end{center}
 \caption{Typical shape of the phase diagram in the NJL model.  The
   curve shows the analytical result~\eqref{eq:NJLphase} with
   $\Lambda=631\MeV$ and $g_s=0.24\fm^2$ used for the $\Nf=2$ case.}
 \label{fig:NJL}
\end{figure}

At finite temperature and density the partition function of the
quasi-particle excitations should be added to the thermodynamics
potential, i.e.
\begin{equation}
 \Omega_{\text{quark}} = -2\Nc T\sum_i \int\frac{d^3 p}{(2\pi)^3}
  \biggl\{ \ln\bigl[ 1 + e^{-(\varepsilon_i(p)-\muq)/T} \bigr]
  + \ln\bigl[ 1 + e^{-(\varepsilon_i(p)+\muq)/T} \bigr] \biggr\}\;,
\label{eq:Omega_q}
\end{equation}
which induces a phase transition.  In the chiral limit, again, we can
find the critical temperature in the analytical way.  That is, near
the critical point, we can assume $T\gg \Mq$ without loss of
generality.  Then, the high-$T$ expansion leads to,
\begin{equation}
 \beta\Omega \simeq = -\frac{\Nc\Nf\Lambda^4}{4\pi^2}\biggl\{
  1 + \Bigl[ 1-\frac{\pi^2}{\Nc\Nf g_s \Lambda^2}-\frac{\pi^2}{3}
  \Bigl(\frac{T}{\Lambda}\Bigr)^2 - \Bigl(\frac{\muq}{\Lambda}
  \Bigr)^2 \Bigr] \xi^2 \biggr\}
  + \calO(\xi^4) \;.
\end{equation}
The potential curvature has a modification from the finite-$T$
effects, and from the condition for the curvature to vanish at the
critical point, the critical temperature is deduced as
\begin{equation}
 \Tc = \sqrt{\frac{3(\Lambda^2-\muq^2)}{\pi^2} - \frac{3}{\Nc\Nf g_s}}\;.
\label{eq:NJLphase}
\end{equation}
This analytical result leads to a typical shape of the phase diagram
in the $\muq$-$T$ plane as depicted in Fig.~\ref{fig:NJL}, where the
Hatsuda-Kunihiro choice of the model parameters is used;
$\Lambda=631\MeV$ and $g_s=0.214\fm^2$ for $\Nf=2$.

So far, our discussions are limited to the special case of the
second-order phase transition, and a possibility of the first-order
phase transition cannot be ruled out from the above analysis.  In
fact, the phase diagram like the one presented in Fig.~\ref{fig:NJL}
should be more carefully examined and could be modified by a
first-order phase boundary in the high-density region.  This
possibility and the underlying mechanism will be the subject of
Sec.~\ref{sec:liquid_gas}.

\paragraph{Coupling with the Polyakov loop}
One of the major obstacles in the application of the NJL model to the
finite-$T$ study is that this model may allow for unphysical quark
excitations even below $\Tc$.  Such unphysical excitations should be
diminished by the effect of confinement in reality.  In other words,
the thermal weight $e^{-(\varepsilon_i(p)\mp \muq)/T}$ for quark
excitations in Eq.~\eqref{eq:Omega_q} should be suppressed whenever
the Polyakov loop $\Phi$ takes a small value.  This can be easily
implemented by the alteration of Eq.~\eqref{eq:Omega_q}
into~\cite{Fukushima:2003fw,Ratti:2005jh}
\begin{align}
 \Omega_{\text{quark}} &\to -2T\sum_i \int\frac{d^3p}{(2\pi)^3}
  \Bigl\{ \tr\ln\bigl[ 1 + L\, e^{-(\varepsilon_i(p)-\muq)/T} \bigr]
  + \tr\ln\bigl[ 1 + L^\dagger e^{-(\varepsilon_i(p)+\muq)/T} \bigr]
  \Bigr\} \;. \notag\\
 &= -2T\sum_i \int\frac{d^3 p}{(2\pi)^3} \Bigl\{ \ln\bigl[
  1 + 3\ell\,e^{-(\varepsilon_i(p)-\muq)/T}
  + 3\ell^\ast e^{-2(\varepsilon_i(p)-\muq)/T}
  +e^{-3(\varepsilon_i(p)-\muq)/T} \bigr] \notag\\
 &\qquad\qquad\qquad\qquad\; + \ln\bigl[
  1 + 3\ell^\ast e^{-(\varepsilon_i(p)+\muq)/T}
  + 3\ell\, e^{-2(\varepsilon_i(p)+\muq)/T}
  +e^{-3(\varepsilon_i(p)+\muq)/T} \bigr] \Bigr\} \;.
\label{eq:l_coupling}
\end{align}
In fact, in the field-theoretical derivation, one can arrive at the
above coupling of Eq.~\eqref{eq:l_coupling} assuming that the Dirac
operator has a constant $A_4$ background~\cite{Ratti:2005jh}.  The
integration along the temporal direction results in the Wilson line
and thus the Polyakov loop as defined in Eq.~\eqref{eq:Polyakov}.

This analytical structure itself is quite useful to understand some
features of hot and dense QCD, especially the sign
problem~\cite{Fukushima:2011jc}.  One can also find a reason why the
strange quark (or heavy quark generally) number susceptibility can
serve as a deconfinement order parameter.  If the quark mass of the
heavy flavor is sufficiently large, we can expand
Eq.~\eqref{eq:l_coupling} and keep only the lowest-order term with
respect to small Boltzmann weight.  Then, differentiating the
partition function with respect to the heavy-flavor chemical potential
$\mu_{\text{s}}$ (which is eventually taken to be zero), we can derive
the heavy-quark number $n_{\text{s}}\propto \langle\ell-\ell^\ast \rangle
\,e^{-M_{\text{s}}/T}$ and the susceptibility $\chi_{\text{s}}\propto
\langle\ell+\ell^\ast \rangle\,e^{-M_{\text{s}}/T}$.  Since
$\chi_{\text{s}}$ is proportional to the Polyakov loop,
$\chi_{\text{s}}$ is qualified as an order parameter, which explains
the right panel of Fig.~\ref{fig:suscept}.

To make concrete calculations, one needs to specify the effective
potential for $\Phi$.  Such a potential, $\Omega_{\text{glue}}[\Phi]$,
is calculable in the perturbation theory that leads to what is called
the Weiss potential~\cite{Weiss:1980rj,Weiss:1981ev}.  Non-perturbative
evaluation is necessary, however, to describe a phase transition.  The
following Ansatz is well motivated from the strong-coupling expansion
on the lattice;
\begin{equation}
 \Omega_{\text{glue}}[\Phi] = -\frac{a(T)}{2}\bar{\Phi}\Phi
  + b(T)\ln\Bigl[ 1-6\bar{\Phi}\Phi+4(\bar{\Phi}^3+\Phi^3)
  -3(\bar{\Phi}\Phi)^2 \Bigr] \;,
\label{eq:Omega_g}
\end{equation}
where the logarithmic term has an origin in the Haar measure or the
Jacobian associated with the variable change from $L$ to $A_4$.  (The
potential~\eqref{eq:Omega_g} is a function of the traced Polyakov loop
that can be expressed using $A_4$, and thus the Jacobian from the
group integration of $L$ to $A_4$ appears.)  This Haar measure
potential has cubic terms reflecting the color SU(3) nature.  The
$T$-dependent coefficients are chosen so that this
$\Omega_{\text{glue}}[\Phi]$ can reproduce the thermodynamics of the
pure Yang-Mills theory.  One might wonder that $\Phi$ or $A_4$ cannot
saturate the full thermodynamics and physical gluons should be
rather two $A_T$'s (transverse components).  The point is that $A_T$'s
also couple to the Polyakov loop in the color adjoint representation
and their thermal excitations can be controlled fully as a function of
$\Phi$~\cite{Fukushima:2010zz,Sasaki:2012bi}, which is also manifest
in the non-perturbative construction of $\Omega_{\text{glue}}[\Phi]$
by means of the inverted Weiss
potential~\cite{Braun:2007bx,Fukushima:2012qa}.

Suppose that $\Omega_{\text{glue}}[\Phi]$ is known, the Polyakov loop
expectation value and the pressure are read from
\begin{equation}
 \frac{\partial\Omega_{\text{glue}}[\bar{\Phi},\Phi]}{\partial\Phi}
  \biggr|_{\Phi=\Phi_0,\bar{\Phi}=\bar{\Phi}_0}
 \!\!\!\!\! =
 \frac{\partial\Omega_{\text{glue}}[\bar{\Phi},\Phi]}{\partial\bar{\Phi}}
  \biggr|_{\Phi=\Phi_0,\bar{\Phi}=\bar{\Phi}_0}
 \!\!\!\!\! = 0\;, \qquad
 P = -\Omega_{\text{glue}}[\Phi_0,\bar{\Phi}_0]/V \;.
\end{equation}
We note that $\bar{\Phi}=\langle\ell^\ast\rangle$ is independent of
$\Phi$ at finite density.  For $\muq>0$ generally $\bar{\Phi}>\Phi$ is
energetically favored.  Once the expectation values $\bar{\Phi}_0$ and
$\Phi_0$ are known, one can evaluate all other thermodynamic
quantities from $\Omega_{\text{glue}}$.  For the parametrization of
$a(T)$ and $b(T)$ in Eq.~\eqref{eq:Omega_g} the following choice was
inspired from the strong coupling expansion~\cite{Fukushima:2008wg},
\begin{equation}
 a(T) = 12\Nc^2\beta T\Lambda^3 e^{-\alpha \Lambda/T} \;,\qquad
 b(T) = -\beta T\Lambda^3 \;,
\end{equation}
where $\alpha=1.052$ and $\beta=0.03$ are fixed to set the crossover
temperature around $T=150 \sim 200\MeV$.  Another frequently-used
parametrization~\cite{Roessner:2006xn} is motivated to fit the
thermodynamics of the pure Yang-Mills theory as
\begin{equation}
 \frac{a(T)}{T^4} = 3.51 - 2.47 \biggl(\frac{T_0}{T}\biggr)
  + 15.2\biggl(\frac{T_0}{T}\biggr)^2 \;,\qquad
 \frac{b(T)}{T^4} = -1.75\biggl(\frac{T_0}{T}\biggr)^3
\end{equation}
with $T_0$ being a scale parameter that characterizes the critical
(crossover) temperature.  Blue curves in Fig.~\ref{fig:purelattice}
show a quantitative comparison between the lattice data and the
potential fit from this $\Omega_{\text{glue}}$.  The agreement is
pretty good in the entire temperature range.  One might have thought
that the agreement is simply a result of the fit, but it is a highly
non-trivial question whether the entire $T$ range can be covered with
only three fitting parameters.

It is a straightforward extension of the NJL model with this modified
$\Omega_{\text{quark}}$ of Eq.~\eqref{eq:l_coupling} and the Polyakov
loop potential $\Omega_{\text{glue}}$ of Eq.~\eqref{eq:Omega_g}.  Such
a model setup is called the Polyakov loop extended Nambu--Jona-Lasinio
(PNJL) model (there are hundreds of application works; see
e.g.\ Refs.~\cite{Blaschke:2007np}.  Here let us make a technical remark;  we still need to
introduce a further approximation in the PNJL model for the treatment
of Eq.~\eqref{eq:l_coupling}.  In many calculations the traced
Polyakov loops are replaced as $\ell\to\Phi$ and
$\ell^\ast\to\bar{\Phi}$, but this procedure assumes
$\langle\ell^n\rangle\to\Phi^n$ and $\langle\ell^{\ast
  n}\rangle\to\bar{\Phi}^n$, which can be acceptable unless the
Polyakov loop fluctuations become significant as compared to the
expectation values.  It is possible to go beyond this approximation to
make use of a matrix model of the Polyakov
loop~\cite{Dumitru:2003hp,Megias:2004hj,Abuki:2009dt}.

\begin{figure}
 \begin{center}
 \includegraphics[width=0.33\textwidth]{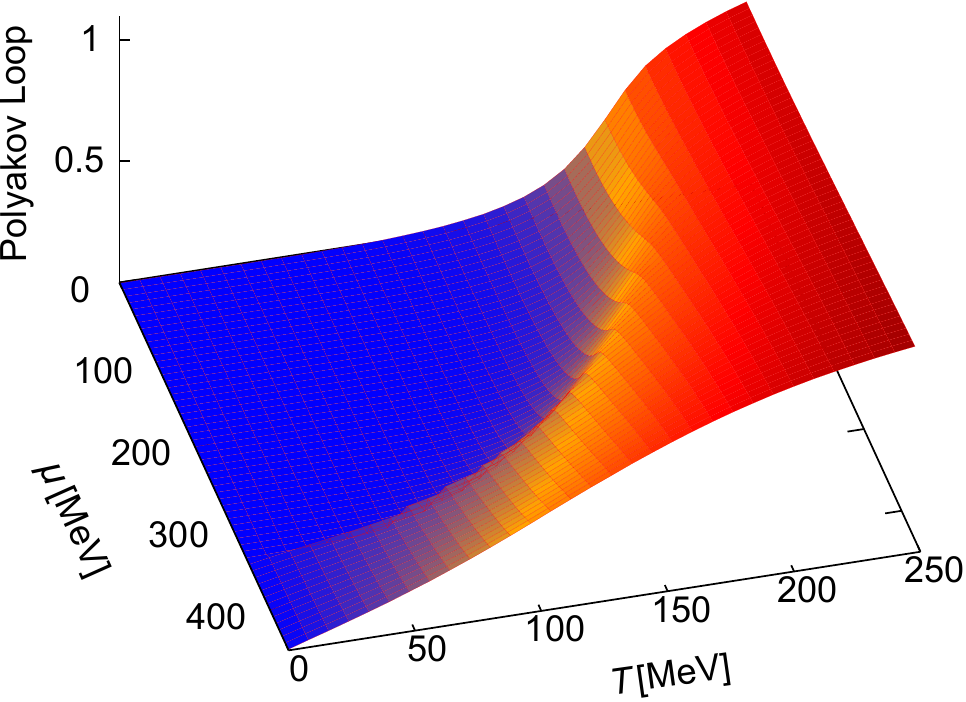} \hspace{4em}
 \includegraphics[width=0.33\textwidth]{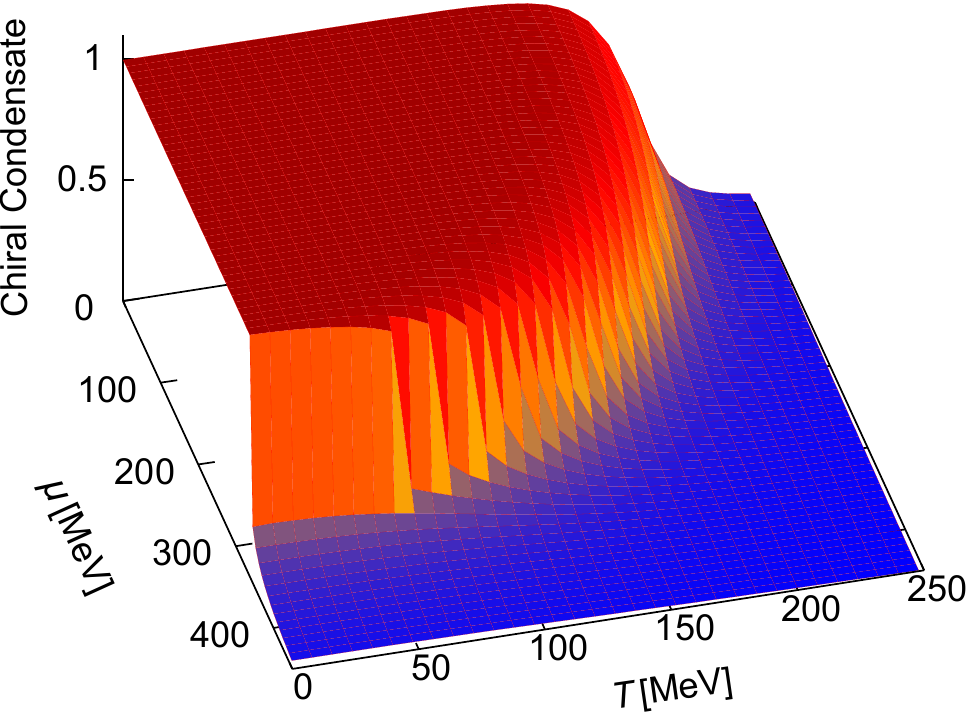}
 \end{center}
 \caption{Order parameters calculated in the (2+1)-flavor PNJL model
   in the mean-field approximation.  Figures adapted from
   Ref.~\cite{Fukushima:2008wg}.}
 \label{fig:order_pnjl}
\end{figure}

Figure~\ref{fig:order_pnjl} shows typical results from the mean-field
approximation of the (2+1)-flavor PNJL model.  The order parameters,
i.e.\ the Polyakov loop and the normalized chiral condensate are
plotted as functions of $\muq$ and $T$.  The NJL model parameters are
fixed according to the Hatsuda-Kunihiro choice;  $\Lambda=631\MeV$,
$\mmu=\mmd=5.5\MeV$, $\mms=135.7\MeV$, $g_s\Lambda^2=3.67$,
$g_d\Lambda^5=-9.29$.  In the market there are several different (and
consistent) choices available;  for example $\Lambda=602.3\MeV$,
$\mms=140.7\MeV$, $g_s\Lambda^2=3.67$,
$g_d\Lambda^5=-12.36$~\cite{Fu:2007xc} leading to qualitatively
consistent results.  We here make a remark that the PNJL model has
been applied to color-superconductivity as
well~\cite{Roessner:2006xn,GomezDumm:2008sk}, but one elaborates an
improved treatment of the Polyakov loop~\cite{Abuki:2009dt}, otherwise
the color density behaves unphysically.

We note that the trend of two simultaneous crossovers is
semi-quantitatively reproduced in the whole $\muq$-$T$ region.  There
is seen a discontinuous jump in the chiral condensate in the
high-density region, which signals a first-order phase transition.
Actually it is quite non-trivial what is going on in the high-density
region.  One can see that the Polyakov loop stays small at low
temperatures, though a jump is still visible which presumably arises
from the mixing with the chiral condensate.  One may be tempted to
make a conclusion from the smallness of $\Phi$ that confinement must
persist even in the high-density region.  This smallness is
intuitively understood from the interpretation of $\Phi$ as a
single-quark free energy $f_q$, i.e., $\Phi\sim e^{-f_q/T}$.  Quark
confinement means $f_q\to\infty$ and so $\Phi\to 0$.  However, even
with a finite $f_q$, we can have $\Phi\to 0$ in the limit of $T\to0$.
Thus, $\Phi$ is no longer a good measure for deconfinement at low
temperature.  The physical picture of confinement and deconfinement
of quarks still has subtleties, which has inspired theoretical
deliberations on quarkyonic matter.

Finally, we make a remark on the sign problem within the framework of
the PNJL model.  The quasi-quark contribution of
Eq.~\eqref{eq:l_coupling} corresponds to the logarithm of the Dirac
determinant in the presence of constant $A_0$ from which $L$ and
$L^\dagger$ emerge.  In the color SU(3) case the traced Polyakov
loops, $\ell$ and $\ell^\ast$, generally take a complex number.  When
the density is zero with $\muq=0$, the Dirac determinant is positive
definite.  At $\muq\neq0$, however, $\Omega_{\text{quark}}$ can be
complex, or its real-part can be negative.  In the mean-field
approximation in which $\tr L$ and $\tr L^\dagger$ are replaced with
$\Phi$ and $\bar{\Phi}$, the energy becomes real and the sign problem
seemingly disappears.  However, we can show that the thermodynamic
potential $\Omega[\Phi,\bar{\Phi}]$ is unstable in the direction of
growing $\delta\Phi=\bar{\Phi}-\Phi$.  The solution of the gap
equation turns out to sit at the saddle point of the
potential~\cite{Fukushima:2006uv}.  Such weird behavior of the
thermodynamic potential is a manifestation of the difficulty of the
sign problem within the mean-field approximation.

\subsubsection{Quark-meson model}
\label{sec:quark-meson}

In the (P)NJL model it is technically complicated to include the
effects of meson fluctuations.  In such microscopic approaches the
meson propagator is dynamically generated through quark loops, and
thus, the momentum dependence is highly non-trivial.  Moreover, the
validity of the model is severely restricted by the Mott threshold for
unphysical meson decay into a quark and an anti-quark.

Instead of the NJL model, depending on the problem, it would be
sometimes more suitable to adopt another chiral model including
meson degrees of freedom as point particles, namely, the linear sigma
model.  The linear sigma model usually has fermionic degrees of
freedom as color-singlet baryons, but to investigate the phase
transitions at high $T$ and/or $\muq$, quarks may be able to enter the
dynamics, and then, a model with quarks and mesons could make more
sense.  Such a model setup with quarks and mesons is called the
quark-meson (QM) model.

\paragraph{Problems in the mean-field approximation}

In the two-flavor case, the QM model consists of quarks and light
mesons, $\sigma$ and $\pi$, with the following Lagrangian density,
\begin{equation}
 \calL = \bar{\psi}\bigl[i\gamma_\mu \partial^\mu - g(\sigma
  + i\gamma_5 \btau\cdot\bpi) \bigr]\psi
  + \frac{1}{2}(\partial_\mu \sigma)^2
  + \frac{1}{2}(\partial_\mu \vec{\pi})^2
  - \frac{\lambda}{4}(\sigma^2+\vec{\pi}^2 -v^2)^2
  + c\,\sigma \;.
\label{eq:QM}
\end{equation}
With the potential in this Lagrangian density, chiral symmetry is
spontaneously broken and the last term with $c=f_\pi m_\pi^2$ (which
is constrained by the PCAC~\eqref{eq:PCAC}) makes the potential
slightly tilted so that $\sigma$ acquires an expectation value and
three $\bpi$'s become the nearly massless NG bosons.  In the chiral
limit ($c=0$) one can perform the mean-field approximation simply in
an analytical way, which is, however, troublesome as compared to the
NJL model.

At $T=\muq=0$ the chiral symmetry breaking is almost trivial.  The
potential is so designed that $\sigma$ can have an expectation value
at $v$, which turns out to be $f_\pi$.  The constituent quark mass is
then $\Mq=g\sigma$.  Unlike the NJL model the zero-point oscillation
energy is discarded because the potential terms are supposed to take
care of it.  Then, at finite $T$, chiral restoration can be caused by
meson excitations even without quarks, and this fact that one can treat
meson loop effects is the advantage of the QM model over the NJL
model.  However, it is known that the simple mean-field approximation
(or the Hartree approximation) would lead to an unphysical first-order
phase transition due to the approximation
artifact~\cite{Baym:1977qb}.

The presence of quarks would make the situation even worth.  It is
known by now that the conventional treatment of discarding
Eq.~\eqref{eq:Omega_zero} is inappropriate for the finite-$T$
investigation especially in the chiral limit~\cite{Skokov:2010sf}.  To
understand the problem, let us take a close look at
Eq.~\eqref{eq:Omega_zero}.  We have seen that this expanded form in
terms of $\xi=\Mq/\Lambda$ correctly describes a second-order phase
transition in the NJL model.  Strictly speaking, one should check if
the quartic term $\sim\xi^4$ has a positive coefficient, and in the
case of Eq.~\eqref{eq:Omega_zero}, indeed, the coefficient of the
$\xi^4$ term has a logarithmic singularity
$\sim -\ln|\sqrt{1+\xi^2}-1|\to +\infty$, and it is always positive
for sufficiently small $\xi$.  We note that this logarithmic
singularity appears from the IR sector of massless quarks.
Interestingly enough, at $T\neq0$, this logarithmic singularity is
exactly canceled by terms from Eq.~\eqref{eq:Omega_q} because quarks
are fermions without Matsubara zero-mode at finite $T$.  Therefore,
the IR singularity should disappear whenever a finite $T$ is turned
on.  This implies that Eq.~\eqref{eq:Omega_q} alone without
Eq.~\eqref{eq:Omega_zero} would leave an uncanceled singularity,
$+\ln|\sqrt{1+\xi^2}-1|\to -\infty$, for the coefficient of the
$\xi^4$ term.  So, a first-order phase transition occurs when the
$\xi^2$-term approaches zero.  This is obviously an artificial phase
transition.

We can overcome this problem in the following way.  One easy
resolution is to introduce physical quark masses, and then the chiral
phase transition is a smooth crossover.  Such a treatment might be
useful for qualitative studies at $\muq=0$, but the tendency to favor
a first-order phase transition is overestimated, which is fatal for
the QCD critical point search.  For a more serious remedy, the latter
problem is not so difficult to remove;  because the logarithmic
singularity is analytically known, one can cancel it by hand, or one
can simply add the zero-point oscillation energy and readjust the
model parameters.  More cumbersome is the problem of artificial
first-order transition;  one should improve the approximation to go
beyond the Hartree resummation.

\paragraph{Renormalization group improvement}
It is a tedious calculation to handle the Fock and the higher-loop
diagrams such as the sunset type~\cite{Nishikawa:2003js}.
Fortunately, nowadays, a sophisticated formulation for
non-perturbative resummation has been developed well, which completely
resolves the problem of artificial first-order phase transition.

Let us elaborate how to perform the calculation.  The starting point
is the functional RG equation, which is also known as the Wetterich
equation~\cite{Wetterich:1992yh}, that is concisely written as
\begin{equation}
 \frac{\partial \Gamma_k}{\partial k} = \frac{1}{2}\Tr
  \bigl( R_k \, G_k \bigr) \;,\qquad
 G_{k\,ij}^{-1}(q) = \frac{\delta^2\Gamma_k}
  {\delta\phi_i(-q)\delta\phi_j(q)} + R_k(q) \delta_{ij} \;,
\label{eq:Wetterich}
\end{equation}
for generic scalar field theories.  Here $i$, $j$ indicate the meson
species such as $\sigma$ and $\pi$.  The trace denoted by $\Tr$
involves the momentum integration as well as the sum over $i$, $j$.
The scale $k$ refers to the RG scale and the effective action
$\Gamma_k$ is coarse-grained up to this IR scale $k$.  The important
step to reach Eq.~\eqref{eq:Wetterich} is to introduce a
scale-dependent mass that satisfies the following boundary condition,
\begin{equation}
 R_k(q) \to \left\{ \begin{array}{lp{1em}l}
  0 && (q^2 \gg k^2) \\
  k^2 && (q^2 \ll k^2)
  \end{array} \right.
 \qquad \Longrightarrow \qquad
 \Gamma_k \to \left\{ \begin{array}{lp{1em}l}
  \Gamma && (k\to 0) \\
  S && (k=\Lambda)
  \end{array} \right.
\label{eq:Rkreq}
\end{equation}
Then, the theory has $k$-dependence through $R_k(q)$, and one can
define the generating functional $Z_k[J]$, and $W_k[J]$ for connected
diagrams in the standard method in the quantum field theory.  The
$k$-dependent effective action, $\Gamma_k[\phi]$, is a Legendre
transform of $W_k[J]$.  We note that the regulator term should be
properly subtract by $-\frac{1}{2}\int d^4q\,R_k(q)\phi(-q)\phi(q)$ in
the definition of $\Gamma_k[\phi]$ (see Ref.~\cite{Berges:1998nf} for
a pedagogical derivation).  Since $R_k(q)$ vanishes in the limit of
$k\to0$, the coarse-grained action amounts to the standard effective
action $\Gamma$ with full quantum effects, while
$\Gamma_{k\to\Lambda}\to S$ where $S$ is the classical action and the
flow start at the UV scale $\Lambda$.  Actually
Eq.~\eqref{eq:Wetterich} is the exact formula to \textit{define}
quantum field theories containing equivalent information as the
functional quantization method.  Therefore, of course, it is an
impossible task to integrate Eq.~\eqref{eq:Wetterich} unless the
theory is solvable.

A frequently-used assumption to go into the practical calculations is
that the non-locality in the effective action is neglected so that
$\Gamma_k[\phi]$ can be decomposed into the tree-level kinetic term
and the local potential.  Corresponding to the Lagrangian density of
the QM model in Eq.~\eqref{eq:QM} the Ansatz in this local potential
approximation (LPA) is taken as
\begin{equation}
 \Gamma_k[\sigma,\pi] = \int d^4x \biggl\{ \bar{\psi}\bigl[ i\gamma_\mu
  \partial^\mu - g(\sigma + i\gamma_5\btau\cdot\bpi) \bigr]
  \psi + \frac{1}{2}(\partial_\mu \sigma)^2 + \frac{1}{2}
  (\partial_\mu\bpi)^2 + \Omega_k[\sigma,\pi] \biggr\} \;,
\end{equation}
and then one can convert the RG equation for $\Gamma_k$ into that for
$\Omega_k$.  The RG equation turns out to be surprisingly simple
under the choice of the so-called optimized
regulator~\cite{Litim:2000ci},
\begin{equation}
 R_k(q) = (k^2 - \bq^2) \theta(k^2 - \bq^2) \;,
\end{equation}
which is a useful choice for the finite-$T$ calculation, while the
covariant form would be more appropriate for the $T=0$ case.  One can
immediately confirm that this simple choice satisfies the
requirement~\eqref{eq:Rkreq}.  Thanks to the Heaviside theta function,
one can carry the $q$-integration out analytically to find,
\begin{equation}
 \frac{\partial \Omega_k}{\partial k} = \frac{k^4}{12\pi^2}
  \biggl[ \frac{3}{E_\pi}\coth\Bigl(\frac{E_\pi}{2T}\Bigr)
  +\frac{1}{E_\sigma}\coth\Bigl(\frac{E_\sigma}{2T}\Bigr)
  -\frac{24}{E_q}\bigl\{ 1-\nF(E_{\text{q}}\!-\!\muq)
  -\nF(E_{\text{q}}\!+\!\muq) \bigr\} \biggr] \;,
\label{eq:QMrg}
\end{equation}
where the quark part has a coefficient given by twice of the fermionic
degrees of freedom, $2\Nc\Nf=12$.  Here $\nF(E)$ is the Dirac-Fermi
distribution function at $T$, and the energies in the denominator are
$E_i=\sqrt{k^2+M_i^2}$ with
\begin{equation}
 \Mq^2 = g^2\sigma^2 \;,\qquad
 M_\sigma^2 = \Omega_k'' \;,\qquad
 M_\pi^2 = \frac{\Omega_k'}{\sigma} \;.
\end{equation}
The $\sigma$ meson mass is naturally the curvature of the effective
potential, and the pion mass is also the curvature with respect to the
$\pi$ field, which can be expressed by the $\sigma$-derivative in an
assumption that $\Omega_k$ is a function of $\sigma^2+\bpi^2$ (except
for the explicit symmetry breaking term $c\sigma$).

There are two representative strategies to integrate the RG
equation~\eqref{eq:QMrg} with respect to $k$.  One is to expand
$\Omega_k(\sigma)$ around the minimum at each $k$.  The whole shape
of $\Omega_k(\sigma)$ is not necessary, but the meaningful observable
is the potential value right at the physical point.  Another method is
to discretize $\Omega_k(\sigma)$ on the $\sigma$-grid and solve
hundreds of differential equations numerically.  In this latter
formulation one does not have to find a minimum of the potential at
each $k$.

We shall make some comments on the RG equation~\eqref{eq:Wetterich}.
In the traditional application of the RG analysis, the critical
phenomena associated with the second-order phase transition have been
of central interest and the critical exponents have been calculated
around the IR fixed point.  The applicability of the
formula~\eqref{eq:Wetterich} is, however, not restricted only around
the second-order phase transition.  As we mentioned,
Eq.~\eqref{eq:Wetterich} can be an alternative of the quantization
scheme, and the full effective potential can be retrieved whether or
not the system is far from the critical point.  For the purpose to
examine the first-order phase transition, the grid method is powerful
enough to carve the double-well potential shape.  In practical
calculation, one should stop the $k$-integration at some small
$k$~\cite{Berges:1996ib};  otherwise the potential becomes flat due to
the convexity.

It is a straightforward extension of the QM model to introduce the
coupling with the Polyakov loop in the same way as in
Eq.~\eqref{eq:l_coupling}, and this extended model is called the PQM
model.  In the PQM model the resulting phase diagram has turned out to
be quite similar to what we have already discussed using the PNJL
model.  In particular chiral restoration in the high density region
has a general tendency to favor the first-order phase transition.  If
the existence of the first-order boundary is a robust feature of the
phase diagram, it ends up with a terminal that is to be identified as
the second-order phase transition point.  Then, around the critical
region where fluctuations are enhanced, the (P)QM model should by far
surpass the PNJL model.  For the phase diagram and the critical
properties around the QCD critical point in the PQM model, see
Refs.~\cite{Schaefer:2007pw,Schaefer:2009ui,Schaefer:2011ex,Herbst:2010rf}
and references therein.  The question is, then, whether the
first-order phase transition really exists or not.

\subsection{Liquid-gas phase transition of quark matter}
\label{sec:liquid_gas}

For the first-order phase transition at high density there is no
universality argument.  This is a dynamical problem depending on the
details of the microscopic theory.  Nevertheless, most of chiral
models (without repulsive vector interactions) predict the first-order
phase transition in the high-density region, and there must be some
intuitive explanation for this general
tendency~\cite{Fukushima:2008is,Fukushima:2012mz}.

Let us start our heuristic argument with a simple setup; cold and
dense quark matter in a quasi-particle description.  This means that
we assume a Fermi liquid of quark matter, which should be valid for
bulk thermodynamic quantities as long as $T$ is small enough and the
Landau damping is a minor effect.  Then, the thermodynamic potential
coming from a hot and dense medium is expressed as a function of the
quasi-particle mass $\Mq$ as
\begin{equation}
 \Omega_{\text{matter}}(\Mq) = -\int_0^{\muq} d\mu' \rhoq(\mu')
  -4\Nc\Nf\, T\int\frac{d^3 p}{(2\pi)^3} \ln\bigl(1+e^{-\omega_p/T}\bigr) \;,
\label{eq:Omega_matter}
\end{equation}
where $\rhoq(\muq)$ is the quark number density;
$\rhoq(\muq)=2\Nc\Nf\int\frac{d^3 p}{(2\pi)^3}
[n_{\text{F}}(\omega_p-\muq) - n_{\text{F}}(\omega_p+\muq)]$ and the
dispersion relation is $\omega_p=\sqrt{p^2+\Mq^2}$ in the
quasi-particle picture.  This form~\eqref{eq:Omega_matter} is common
in any chiral models such as the (P)NJL and the (P)QM models.  It is
then the vacuum part that is severely contaminated by model
uncertainties.

In the spirit of the Ginzburg-Landau expansion, we can postulate the
vacuum part as a polynomial form,
\begin{equation}
 \Omega_0(\Mq) = a(M_0^2-\Mq^2)^2 -b \Mq - c \Mq^3 \;.
\label{eq:GL}
\end{equation}
In the parametrization~\eqref{eq:GL} $a$ is the curvature of the
potential at the origin, $b$ represents the effect of the explicit
symmetry breaking due to current quark mass, and $c$ takes care of the
instanton-induced interaction for $\Nf=3$.  Here, for simplicity, we
take $c=0$;  we are interested in the reason why a first-order phase
transition is dynamically favored at high density, and if $c\neq0$ the
tendency toward the first-order transition would be just
strengthened by the tree-level contribution.  The potential minimum is
located around $\Mq\simeq M_0$, so that $M_0$ is a parameter to be
identified as the constituent quark mass in the vacuum.

Although this setup is extremely simple or even primitive, we
emphasize that the above $\Omega_0+\Omega_{\text{matter}}$ can grasp
the essential properties of all chiral quark models, except for the
uncanceled logarithmic singularity in Eq.~\eqref{eq:Omega_matter},
which we will ignore here because it is dangerous only near the chiral
limit.  Moreover, in the regime at high $T$, meson fluctuations may
give rise to $T$-dependent coefficients.  Therefore, the analysis in
this subsection is valid only for $\muq\gg T$.

\begin{figure}
\begin{center}
 \includegraphics[width=0.3\textwidth]{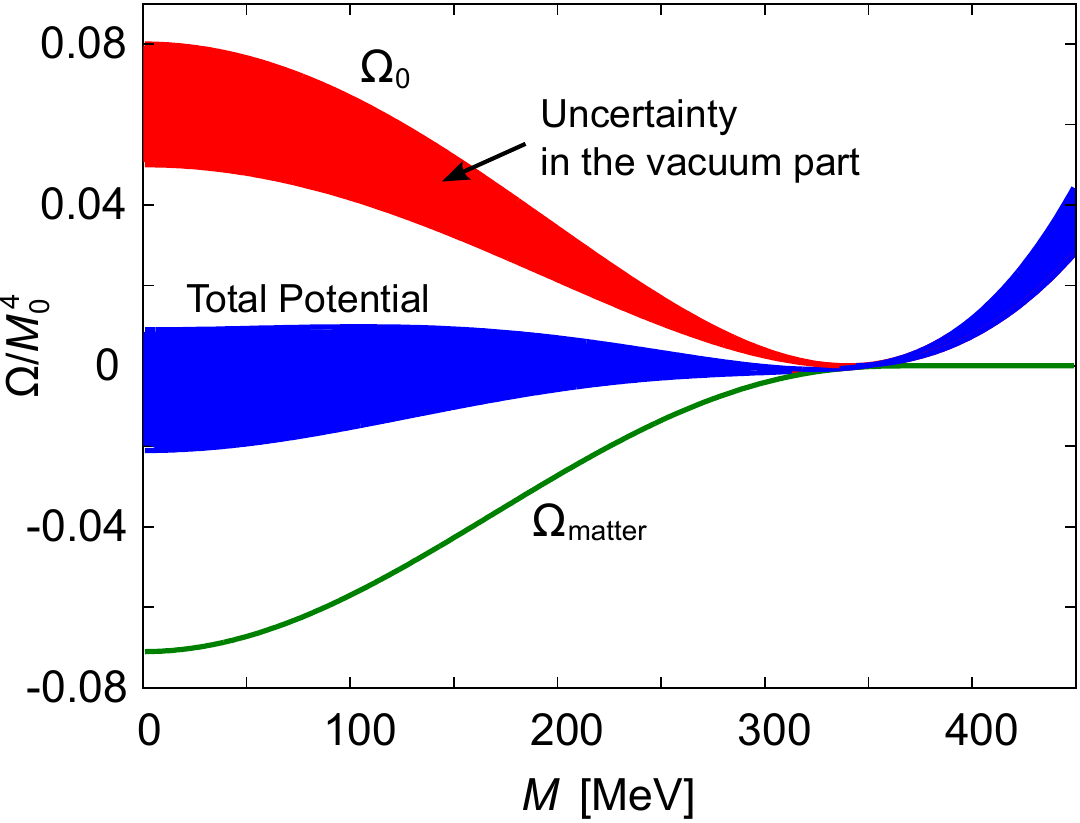} \hspace{5em}
 \includegraphics[width=0.33\textwidth]{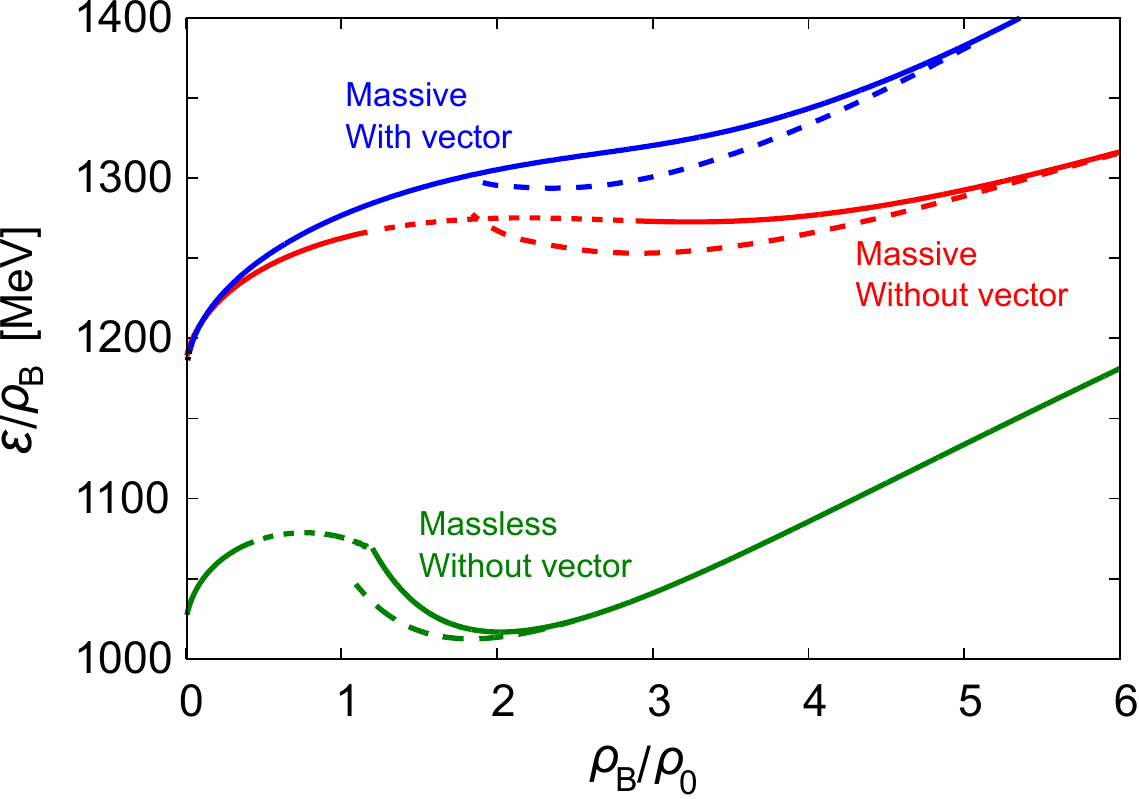}
\end{center}
 \caption{(Left) Potential shapes from Eqs.~\eqref{eq:Omega_matter} and
   \eqref{eq:GL} at $T=0$ and $\muq=370\;\text{MeV}$.
   $\Omega_{\text{matter}}$ is fairly model independent, while
   uncertainty is unavoidable in $\Omega_0$.  The parameters in
   $\Omega_0$ are chosen as $M_0=340\;\text{MeV}$, $a=0.05\sim0.08$,
   $b=c=0$.  (Right) The saturation curve of the energy per baryon as
   a function of the baryon density.  From the bottom to the top, the
   solid curves represent the results in the chiral limit without the
   vector interaction, with massive quarks without the vector
   interaction, and with massive quarks with the vector interaction.
   The dotted parts are unstable and the dashed curves represent the
   branches corresponding to the chiral spiral.  Figures adapted from
   Ref.~\cite{Fukushima:2012mz}.}
 \label{fig:potential}
\end{figure}

The left panel of Fig.~\ref{fig:potential} shows the typical behavior
of the effective potential.  The matter part $\Omega_{\text{matter}}$
always has a minimum at $\Mq=0$ because the baryon density is the
largest when quasi-particles are massless.  Then, together with
$\Omega_0$ that has a minimum at $\Mq=M_0$, it is conceivable to
expect two minima in the total $\Omega$, and the absolute minimum
jumps from one to the other at the first-order phase transition.  This
is the mechanism why the density effect tends to favor the first-order
phase transition.  At the same time, this argument can explain why the
location and even the existence of the first-order boundary on the
phase diagram are such model-dependent.  The largest uncertainty comes
from $a$;  in Fig.~\ref{fig:potential} uncertainty associated with
$a=0.05\sim0.08$ (empirical values in the NJL and the QM models) is
shown.

In the same way as the left panel of Fig.~\ref{fig:walecka-t} for
nuclear matter, it is useful to make a plot for the energy per baryon,
$\varepsilon/\rho_{\rm B}$ at $T=0$, which is presented in the right
panel of Fig.~\ref{fig:potential}.  As we already discussed in
Sec.~\ref{sec:Walecka}, the minimum of this curve tells us the
saturation density and the binding energy.  Because we are now working
for quark matter, the saturation density can differ from $\rho_0$, and
moreover, there may not be a minimum at all.

The existence of the first-order boundary of chiral restoration
corresponds to the appearance of the minimum in the saturation curve.
Thus, if quark droplets are possible~\cite{Buballa:1996tm}, as is the
case in Fig.~\ref{fig:potential} for massless $b=0$, the first-order
phase transition and the QCD critical point can be concluded.  From
these analyses it is evident that the first-order phase transition in
quark matter has exactly the same origin as the liquid-gas phase
transition of nuclear matter.  Even when droplets are only
meta-stable, as is the case in Fig.~\ref{fig:potential} for
$b=0.08M_0^3$ (massive, without vector), the first-order phase
transition occurs.

We have not mentioned on the impact of the vector interaction as
introduced in the Lagrangian density~\ref{eq:NJL4}.  The role of this
type of the interaction is to add,
\begin{equation}
 \Omega_{\text{vec}}(\Mq) = g_v \rhoq^2 \;,
\label{eq:vector}
\end{equation}
to the thermodynamic potential at the mean-field level.  It should be
noted that the mean-field calculation needs not only the tree-level
$-g_v\rhoq^2$ term but also the shifted chemical potential
$\muq^\ast$ as seen in Sec.~\ref{sec:w_calc}.  If expanded by the
shift $g_v\rhoq$, this latter effect amounts to $2g_v\rhoq^2$, and
Eq.~\eqref{eq:vector} results together with the tree-level term.

It is obvious from the saturation curve that the repulsive ($g_v>0$)
vector interaction always disfavors the first-order phase transition.
The energy $\varepsilon$ is pushed up by $\propto \rho_{\rm B}^2$ from
the vector interaction, and eventually the local minimum is completely
washed out with increasing $g_v$, as exemplified in
Fig.~\ref{fig:potential} for $b=0.08M_0^3$ and $g_v=0.12/M_0^2$
(massive, with vector).  These parameters are arbitrarily chosen here
for demonstration.  Once the absence of the local minimum becomes the
situation with sufficiently large $g_v$, there no longer appears a
first-order phase boundary, and the QCD phase diagram has only smooth
crossover entirely.

Above mentioned is the simplest and the most intuitive explanation
about the impact of the vector interaction on the order of the chiral
phase transition (at $T=0$) that was first addressed in a model
study~\cite{Kitazawa:2002bc} and discussed
repeatedly~\cite{Fukushima:2008wg,Sasaki:2006ws,Bratovic:2012qs,Contrera:2012wj}.
See also Ref.~\cite{Lourenco:2012yv} for an attempt to estimate the
strength of the vector interaction using the PNJL model.

\subsection{Inhomogeneous chiral condensates}
\label{sec:inhomo}

The existence of the QCD critical point is easily affected by unknown
factors, while the inhomogeneous chiral condensate is more robust and,
to the best of our knowledge, no counter-example against the
inhomogeneous state has been found.  Since the discovery of the
relation between the QCD critical point and the Lifshitz
point~\cite{Nickel:2009ke}, more and more attention is being attracted
to inhomogeneous quark matter, though there were earlier theoretical
studies~\cite{Shuster:1999tn,Park:1999bz,Rapp:2000zd} and also the
crystalline color-superconductors were intensely investigated in the
higher density region (see Ref.~\cite{Anglani:2013gfu} and references
therein).

The simplest way to introduce inhomogeneity in the chiral condensate
is to adopt the Ansatz of the chiral spiral, and actually, the chiral
spiral is very useful to deepen the understanding of the driving force
for spatial modulation.  Under the 1-dimensional chiral-spiral
Ansatz~\cite{Schon:2000qy}, the scalar and the pseudo-scalar
condensates form a spiral (along the $z$-direction in our choice) as
\begin{equation}
 \langle\bar{\psi}\psi\rangle = \chi\cos(2qz) \;,\qquad
 \langle\bar{\psi}\gamma_5\tau_3\psi\rangle=\chi\sin(2qz)
\end{equation}
with a wave-number $q$.  This is nothing but the quark-matter analogue
of the $p$-wave pion condensation~\eqref{eq:pi0} in nuclear matter.
Also, this Ansatz is sometimes referred to as the dual chiral-density
wave~\cite{Nakano:2004cd}.

The technical advantage of the chiral spiral is that this
inhomogeneity is readily described by the redefinition of the basis;
$\psi=e^{i\gamma_5 \tau_3 q z}\psi'$ with a homogeneous condensate
$\chi=\langle\bar{\psi}'\psi'\rangle$.  Such a rotation in the chiral
limit can remove the $\muq$-term in the Dirac operator, and then, the
quasi-particle dispersion relation in the $\psi'$-basis is expressed
as
\begin{equation}
 E_p = \sqrt{p_\perp^2+(\sqrt{p_z^2+\Mq^2}\pm q)^2}
\label{eq:disp_q}
\end{equation}
with dynamical mass $\Mq$ on the $\psi'$-basis.  This type of simple
inhomogeneity pattern has been considered repeatedly in various
contexts such as large-$\Nc$
QCD~\cite{Deryagin:1992rw,Shuster:1999tn}, the Overhauser
instability~\cite{Park:1999bz,Nakano:2004cd}, the quarkyonic chiral
spiral~\cite{Kojo:2009ha}, and so on.  The dispersion
relation~\eqref{eq:disp_q} should be plugged into
$\Omega_{\text{matter}}$ in Eq.~\eqref{eq:Omega_matter} to evaluate
the thermodynamic potential.

The most important observation is that a large part of the mass effect
can be absorbed if $q\sim \Mq$.  Then, the density $\rhoq$ is not
suppressed even with large $\Mq$.  In fact, the (1+1)-dimensional
system is this extreme example;  $\rhoq$ has an origin in the quantum
anomaly and is completely insensitive to $\Mq$ but dependent only on
the chemical potential.

\begin{figure}
\begin{center}
 \includegraphics[width=0.4\textwidth]{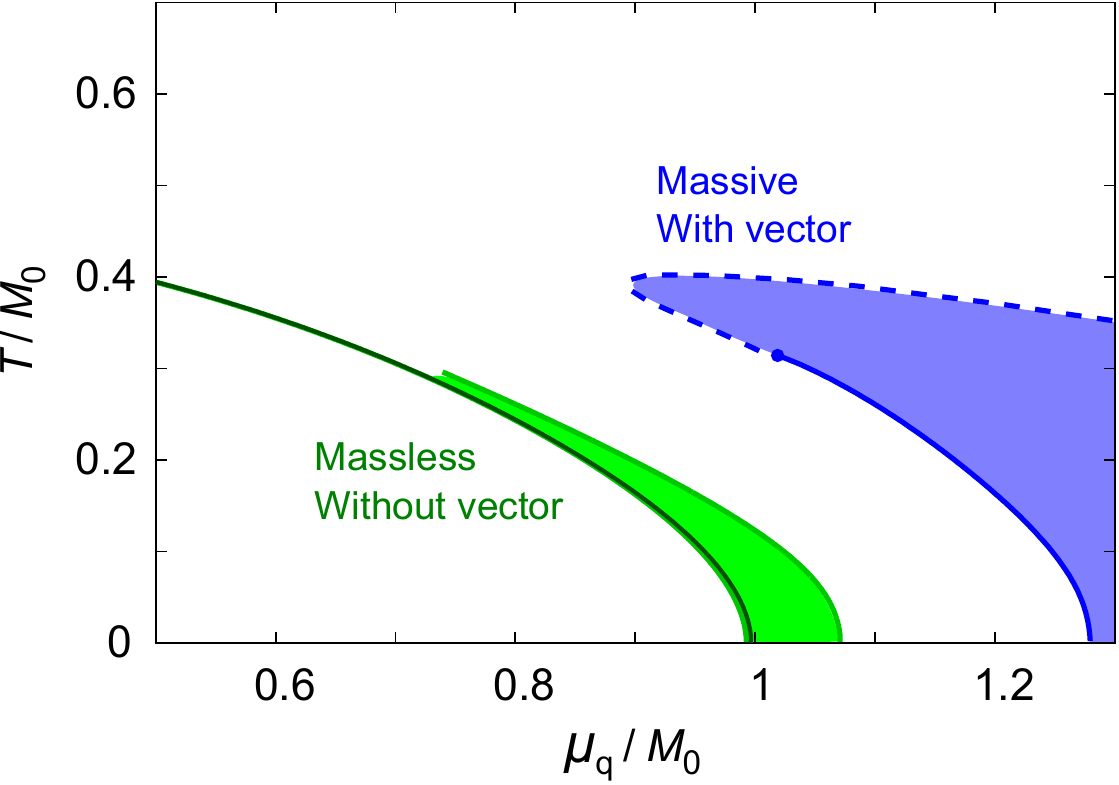}
\end{center}
 \caption{Typical phase diagrams with the chiral spiral for two cases.
   The solid curves represent a first-order phase transition.  The
   shaded region in the left shows the chiral spiral for the massless
   ($b=0$) case without the vector interaction ($g_v=0$), and that in
   the right for the massive ($b=0.08M_0^3$) case with the vector
   interaction ($g_v=0.12/M_0^2$) with $M_0$ being a typical scale in
   the potential~\eqref{eq:GL}.  Figure adapted from
   Ref.~\cite{Fukushima:2012mz}.}
 \label{fig:diagram}
\end{figure}

This simple observation from Eq.~\eqref{eq:disp_q} can explain how the
chiral spiral with $q\sim\Mq$ can lower the total thermodynamic
potential.  If one makes a plot for $\Omega_{\text{matter}}(\Mq,q)$,
one can find that this mechanism works efficiently and
$\Omega_{\text{matter}}(\Mq,q)$ goes arbitrarily smaller for
$q\sim\Mq\to\infty$.  This is a very simple and robust mechanism for
the energy gain from finite $\q$.  This large energy gain should
overcome the energy loss from the kinetic term in $\Omega_0(\Mq,q)$,
which can be expanded as
\begin{equation}
 \Omega_0(\Mq,q) = \Omega_0(\Mq,q=0)
 + (\alpha \Mq^2 + \beta b) q^2 \;.
\end{equation}
The first term with $\alpha>0$ represents the kinetic term.  One could
estimate $\alpha$ using a chiral model, but one should be careful not
to pick unphysical terms up from gauge-variant regularization.  The
term $\propto \beta$ is also necessary to take account of the effect
of the current quark mass that would disfavor the chiral spiral.
Quantitative details depend on $\alpha$ and $\beta$, but this setup
with $\Omega_0(\Mq,q)+\Omega_{\text{matter}}(\Mq,q)$ suffices to
capture the model-independent essence regardless of the choice of
$\alpha$ and $\beta$.

Figure~\ref{fig:diagram} shows typical phase diagrams resulting from
this simple description with a choice, $\alpha=0.25$ and
$\beta=0.25/M_0$.  The chiral spiral is favored in the shaded region
surrounded by the phase transitions of first order or second order.

We would stress that the vector interaction chosen here
($g_v=0.12/M_0^2$) is large enough to make the QCD critical point
disappear from the phase diagram.  Yet, the chiral spiral persists for
any $g_v$ as confirmed in Fig.~\ref{fig:diagram} even for the massive
case with strong vector interaction.  Therefore, the possibility of
inhomogeneous states in quark matter is a more robust prediction from
the model than the QCD critical point.  The corresponding saturation
curves are overlaid on the right panel of Fig.~\ref{fig:potential}
with dashed lines.  As noticed in this figure, it is possible that a
local minimum is revived, so that a first-order phase boundary can
come back with inclusion of the chiral spiral structure.

For more realistic ground states, it is desired to optimize the
spatial structure beyond the Ansatz of the chiral spiral.  This has
been done by minimizing the Ginzburg-Landau potential and the
solitonic solution is derived, which is known to have a lower energy
than the chiral spiral in concrete model
calculations~\cite{Nickel:2009wj,Carignano:2010ac}, which is also
systematically investigated using the Ginzburg-Landau
theory~\cite{Abuki:2011pf}.  The saturation curves associated with
such solitonic solutions are discussed in Ref.~\cite{Buballa:2012vm}.
Even though the chiral spiral is an oversimplified form as compared to
the soliton, it is still useful to exemplify that the inhomogeneous
state takes over the homogeneous one.  More theoretical studies are
needed to clarify the stability of inhomogeneous states with other
interactions (analogous to the effect of $g'$ in the $p$-wave pion
condensation) or with meson fluctuations in the RG method.

\subsection{Quarkyonic matter}
\label{sec:quarkyonic}

\begin{figure}
 \begin{center}
  \includegraphics[width=0.35\textwidth]{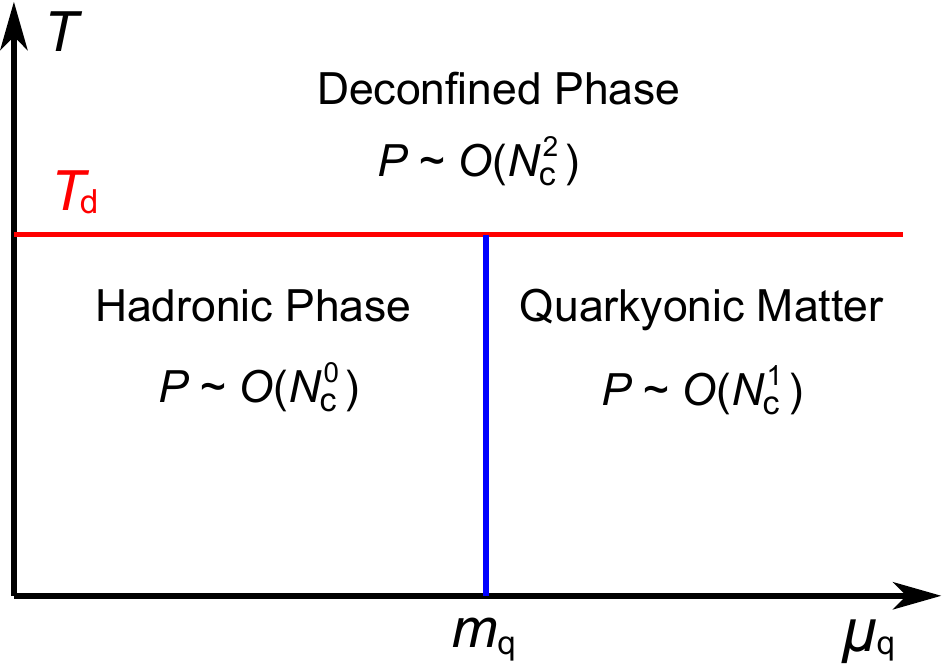} \hspace{6em}
  \includegraphics[width=0.14\textwidth]{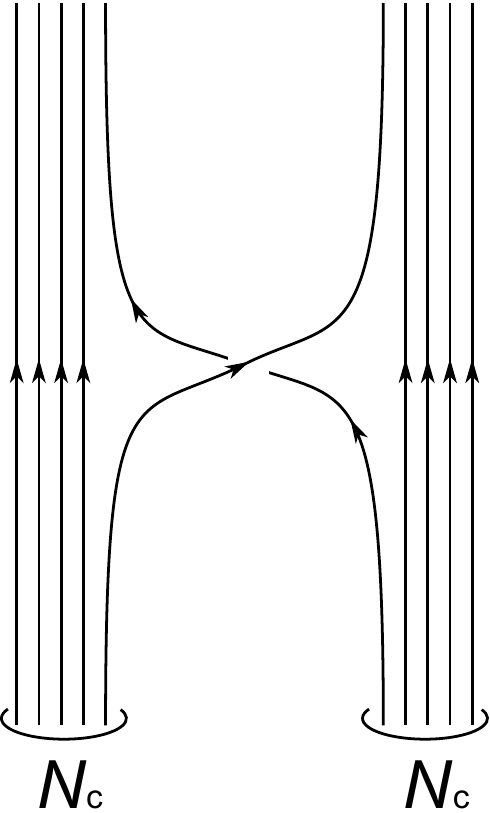}
 \end{center}
 \caption{(Left) Schematic phase diagram of large-$\Nc$ QCD.\ \ The
   pressure is $\calO(\Nc^0)$ in the hadronic phase, $\calO(\Nc^2)$ in
   the deconfined phase, and $\calO(\Nc^1)$ in quarkyonic matter.
   (Right) Baryon-baryon interaction of $\calO(\Nc)$ in the
   large-$\Nc$ limit.  The exchanged quarks should have the same
   color.  If they have the different color, one gluon exchange is
   necessary.}
 \label{fig:quarkyonic}
\end{figure}

In the large-$\Nc$ limit the deconfinement phase transition is
well-defined and the phase structure associated with deconfinement
is quite simple.  As we have seen in Sec.~\ref{sec:largeNc} quark
loops are suppressed by $1/\Nc$ as compared to gluons, and yet, matter
with finite baryon density may appear if the chemical potential is large
enough.  What does nuclear matter look like in the large-$\Nc$ limit?
We already have sufficient ingredients to tackle this question -- we
have discussed a Skyrme crystal in Sec.~\ref{sec:skyrme} and
discovered a large-$\Nc$ phase diagram in Fig.~\ref{fig:phaseSS}.  We
have also clarified the order parameter behavior in
Fig.~\ref{fig:order_pnjl} obtained in the PNJL model in which the
large-$\Nc$ limit is implicitly assumed~\cite{McLerran:2008ua}.

\subsubsection{Phase diagram and the pressure}

Apart from the fate of chiral symmetry, the phase diagram of
large-$\Nc$ QCD is as simple as in the left panel of
Fig.~\ref{fig:quarkyonic}, which is equivalent to
Fig.~\ref{fig:phaseSS} without a curve representing $T_\chi$.  The
horizontal line at $T=\Td$ is the deconfinement phase boundary below
which only glueballs exist, the pressure of which should be of
$\calO(\Nc^0)$.  Because there are $(\Nc^2-1)$ gluons in the
deconfined phase, the pressure sharply jumps from $\calO(\Nc^0)$ in
the glueball (or hadronic) phase to $\calO(\Nc^2)$ in the deconfined
phase.  An onset for a finite quark density is located at $\muq=\mq$
and, if finite-density matter behaves as a free quark gas, its
pressure is $\sim \Nc\muq^4$.  Hence, the deconfinement phase boundary
cannot be affected by quarks unless $\Nc\muq^4$ becomes comparable to
$\Nc^2$, namely, $\muq\sim \calO(\Nc^{1/4})$.  Eventually, for
$\muq\sim \calO(\Nc^{1/2})$, quarks are no longer suppressed by
$1/\Nc$ which is compensated by $\muq^2$ from the quark 
loop, and the gluon interactions are screened by dynamical quarks.

The above argument suggests that the pressure in the right-bottom part
is of $\calO(\Nc^1)$, which is indeed the case if the state of matter
is a (nearly) free quark gas and $\muq\sim\calO(\Nc^0)$.  Such an
argument is, however, too na\"{i}ve;  as is clear from
Fig.~\ref{fig:phaseSS} in the Sakai-Sugimoto model, the ground state in
the low-$T$ and high-$\muq$ region is identified as nuclear matter
rather than quark matter.  This is also consistent with model studies
as in Fig.~\ref{fig:order_pnjl} which implies quark confinement with
small $\Phi$ in this region.  In fact, gluons are confined below
$\Td$, and it would be quite reasonable to assume that quarks are also
confined there.

In Ref.~\cite{McLerran:2007qj} it was pointed out that the pressure of
large-$\Nc$ \textit{nuclear} matter is of $\calO(\Nc^1)$ and it
resembles a pressure of quark matter.  In nuclear matter at large
$\Nc$ nucleons are infinitely heavy and static, so that their kinetic
energy is suppressed by $1/\MN\sim\calO(\Nc^{-1})$ and the dominant
pressure contribution comes from the $NN$ interaction.  Such a
situation is correctly incorporated as a Skyrme crystal, which should
be an approximate description of quarkyonic matter.  Baryons at large
$\Nc$ interact strongly and the $NN$ interaction energy should be of
$\calO(\Nc)$, which is understood immediately from a diagram in the
right panel of Fig.~\ref{fig:quarkyonic};  the interaction is of order
of the combinatorial factor to pick exchanged quarks up among $\Nc$
quarks inside of baryon.  One can make sure that not only the two-body
but also the multi-baryon interactions generally scale as
$\calO(\Nc)$.  In summary, in the large-$\Nc$ limit, nuclear matter is
a system of strongly interacting baryons, and it looks like knowing
quark degrees of freedom in it.  In Ref.~\cite{McLerran:2007qj} a name
was given to such a baryonic and quark-like state of matter,
i.e.\ \textit{quarkyonic matter}.

If quarkyonic matter is a new state of matter, there must be some
order parameter to characterize it.  This is a natural question but is
a source of confusion about the physical interpretation of quarkyonic
matter.  An intuitive picture of quarkyonic matter is the following;
quarks deep inside the Fermi sphere are weakly interacting, because it
is hard to excite these quarks above the Fermi sea due to Pauli
blocking.  On the other hand, the quarks near the Fermi surface with a
shell-width $\sim\LQCD$ are not affected so much from the Pauli
blocking and can interact strongly.  Thus, the bulk thermodynamics
such as the pressure, entropy and so on are dominated by the quarks
inside of the Fermi sphere, while the physical excitations on top of
the Fermi surface are dominated by color-singlet baryons.

\subsubsection{Characterization of quarkyonic matter}
\label{sec:def_quarkyonic}

A frequently-asked question is;  what is the definition of quarkyonic
matter?  In other words, what is the order parameter for quarkyonic
matter?

\paragraph{Formal definition}
In accord with the original argument in Ref.~\cite{McLerran:2007qj},
the definition of quarkyonic matter could be \textit{a state of
  matter that satisfies the McLerran-Pisarski conjecture}.  Here we
introduced a term -- the McLerran-Pisarski conjecture -- which claims
that a system of dense baryons be dual to a system of quarks.  Thus,
in quarkyonic matter, both descriptions of baryonic matter and quark
matter can work to capture correct thermodynamic properties.

In Sec.~\ref{sec:SSM} we mentioned on two density sources; the D4
instantons corresponding to baryons and the string sources
corresponding to quarks in the Sakai-Sugimoto model.  Although the
baryonic state has a lower energy in the analysis in
Ref.~\cite{Bergman:2007wp}, such a treatment of the density sources
may have some subtlety (for example the validity of the DBI action
with a cusp singularity, see Ref.~\cite{Harada:2011aa}).  If the
baryonic and quark sources turn out to be indistinguishable after all,
it would be the clearest holographic example of quarkyonic matter.

\paragraph{Phenomenological signature}
This above formal definition is not radical but quite understandable
in view of the success of the thermal statistical model or the hadron
resonance gas (HRG) model to capture the QCD
thermodynamics~\cite{Tawfik:2004sw}.  (For details about the model
setup, see Refs.~\cite{Becattini:2005xt,Andronic:2008gu} and also
Ref.~\cite{letessier2002hadrons} for a textbook.)  In the HRG model it
is assumed that hot and dense matter consists of non-interacting
mesons, baryons, and all resonating states.  This is a modern
renovation of Hagedorn's old idea (see Sec.~\ref{sec:bootstrap}).
Suppose that all observed states obey the Hagedorn
spectrum~\eqref{eq:Hagedorn}, the partition function blows up at
$T=\Th$, and the growing behavior of the pressure near $\Tc$ fits well
with the one known from the lattice-QCD simulation as in
Fig.~\ref{fig:fulllattice}.  This observation is quite consistent with
the behavior of QCD at large $\Nc$.  As we discussed in
Sec.~\ref{sec:largeNc}, the system is reduced to infinite towers of
non-interacting mesons in the large-$\Nc$ limit of QCD.

It is also known from the chiral effective models that the
thermodynamic properties near $\Tc$ are well understood in terms of
quasi-quarks and gluons~\cite{Fukushima:2012qa}.  Therefore, thanks to
a smooth crossover from the hadronic phase to the quark-gluon plasma,
one can view the state of matter in this transitional region as both
hadronic and quark-like.  In a broad sense, the crossover region of
QCD at zero density satisfies the McLerran-Pisarski conjecture, and
the state of matter is quark-mesonic instead of
quark-baryonic=quarkyonic.

In a narrow sense, quarkyonic matter should be distinguished from
quark-mesonic matter.  Baryons interact strongly, and this situation
is so different from the description by the HRG model.  In the thermal
statistical model the interaction effects are indirectly taken into
account via the Van~der~Waals force with a parameter of the excluded
volume~\cite{Andronic:2012ut}.  However, in physical processes, the
interactions should encompass the quark degrees of freedom as depicted
in the diagram of Fig.~\ref{fig:quarkyonic}, and the excluded volume
factor is not adequate to go into the regime of quarkyonic matter.
Thus, in the beam-energy scan of the relativistic heavy-ion collision,
if any deviation between the experimental data and the HRG model fit
becomes more and more appreciable, it would hint quarkyonic matter.

\paragraph{Chiral property}
Nuclear physicists would still wonder how quarkyonic matter can be
different from ordinary nuclear matter in the real world with
$\Nc=3$.  There is no clear-cut distinction.  In nuclear matter $\kF$
is as small as the pion mass and the system is dilute (in a sense that
one can perform the expansion with respect to $\kF$), while quarkyonic
matter is dense by definition.

One possible characterization of quarkyonic matter is how it breaks
chiral symmetry.  In quarkyonic matter with infinitely large $\Nc$
inhomogeneity is expected~\cite{Shuster:1999tn,Kojo:2009ha}.  This
spatial modulation is attributed to the pseudo 1-dimensional nature on
top of the Fermi surface at high density.  Interestingly, as briefly
noted in Sec.~\ref{sec:SSM}, the Sakai-Sugimoto model exhibits
instabilities with respect to spatial
modulations~\cite{Chuang:2010ku,Ooguri:2010xs}.  However, this sort of
inhomogeneity is nothing peculiar for nuclear physics;  as addressed
in Sec.~\ref{sec:picon}, the $p$-wave pion condensation is essentially
the same as the (quarkyonic) chiral spiral.  Then, we have to conclude
that quarkyonic matter is not really distinct from nuclear matter, or
the $p$-wave pion condensation may be smoothly connected to the
quarkyonic chiral spiral and vice versa.

\subsubsection{Implications to the phase diagram}

\begin{figure}
 \begin{center}
  \includegraphics[width=0.28\textwidth]{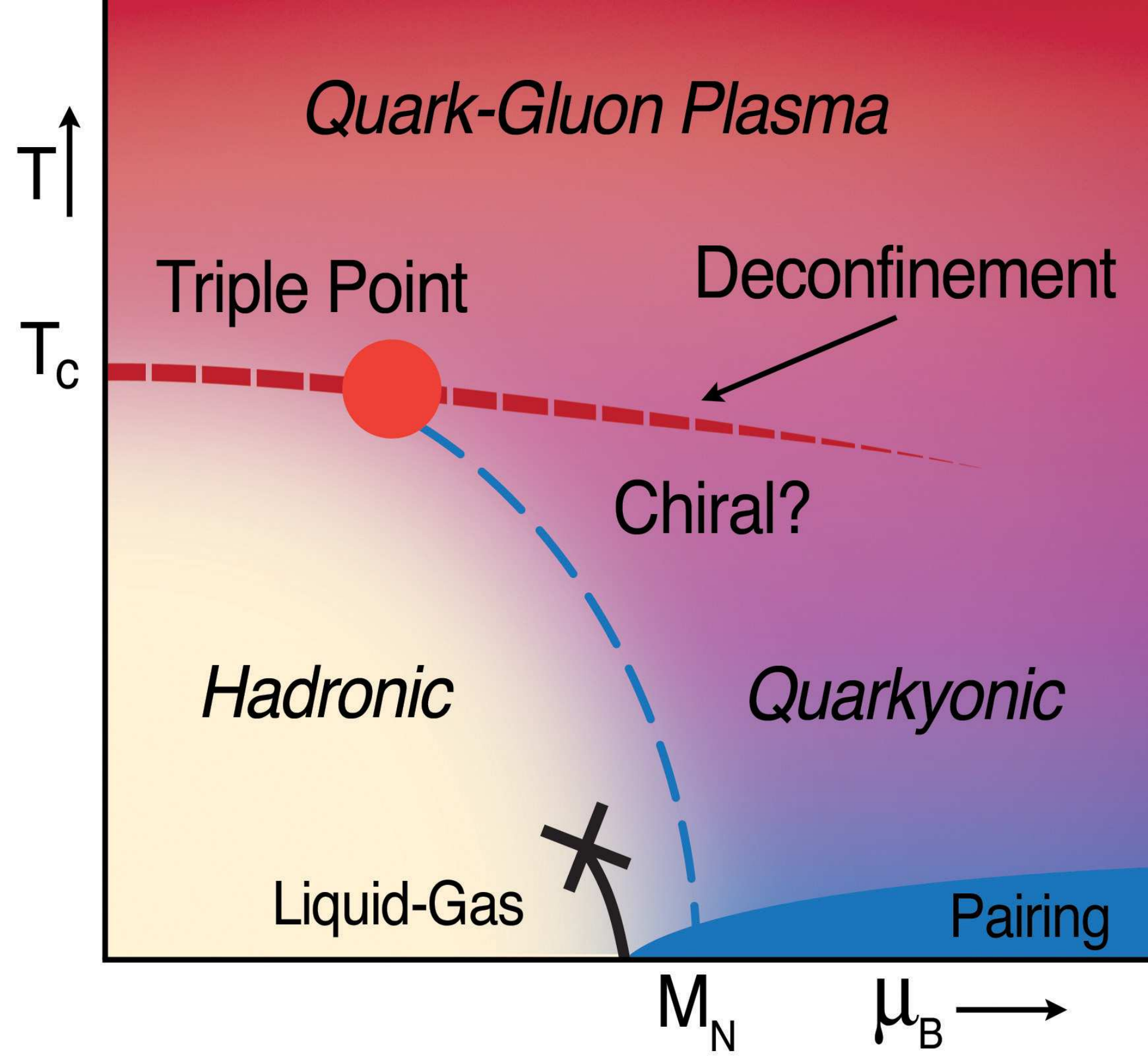} \hspace{6em}
  \includegraphics[width=0.35\textwidth]{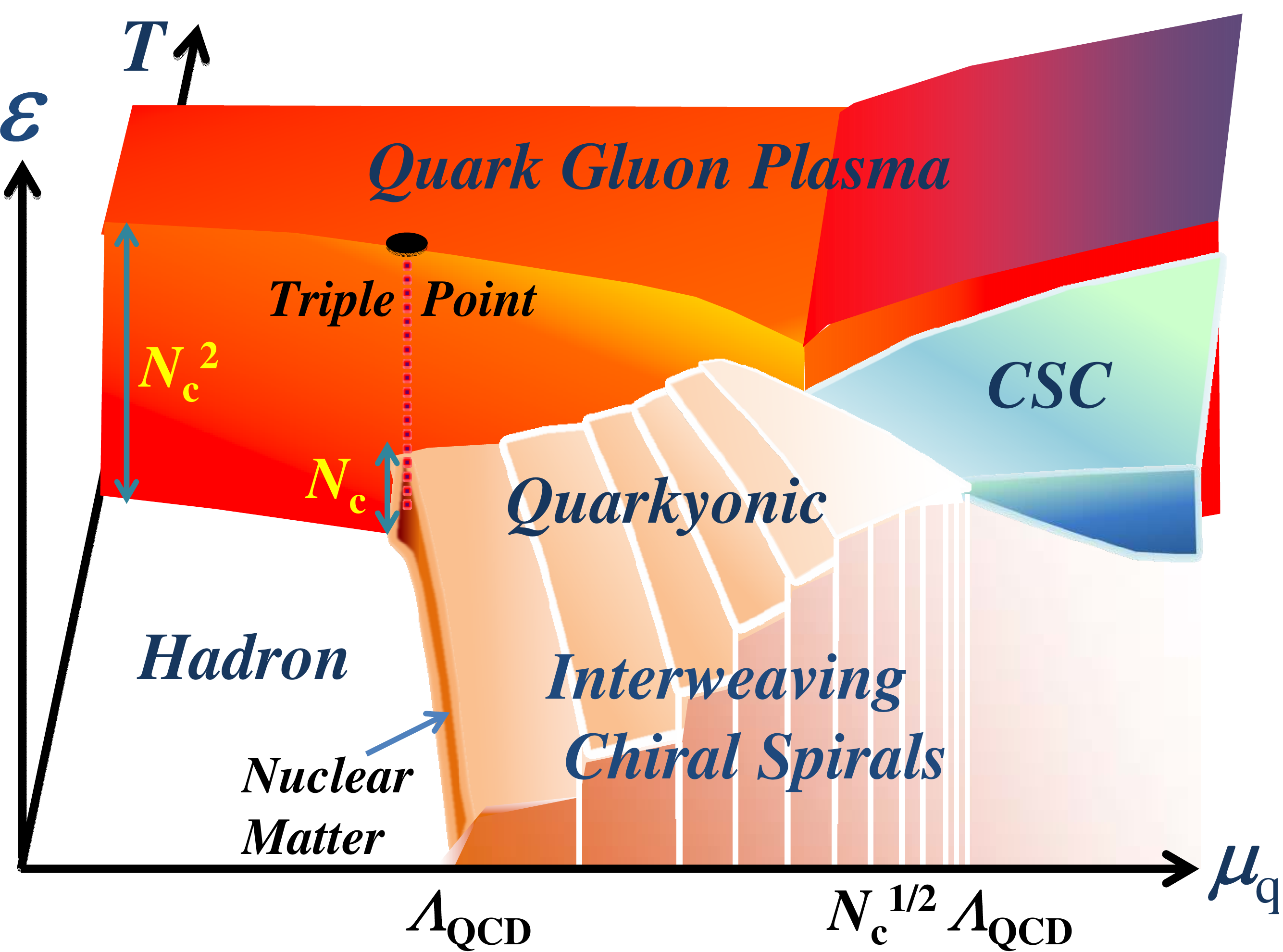}
 \end{center}
 \caption{(Left) Phase diagram with a region that looks like a triple
   point where the hadronic phase, the deconfined phase (quark-gluon
   plasma), and quarkyonic matter meet.  Figure adapted from
   Ref.~\cite{Andronic:2009gj}.  (Right)  Phase diagram with
   interweaving chiral spirals.  Figure adapted from
   Ref.~\cite{Kojo:2011cn}.}
 \label{fig:weaving}
\end{figure}

With quarkyonic matter in addition to the hadronic phase and the
quark-gluon plasma, the phase diagram should take a structure of the
left panel of Fig.~\ref{fig:weaving}, and then there may appear a
region that looks like an approximate triple-point as speculated in
Ref.~\cite{Andronic:2009gj} (see also the lattice-QCD results in
Ref.~\cite{Fodor:2007vv}).

Later on, in Ref.~\cite{Kojo:2011cn}, this phase diagram was brushed
up into the form of the right panel of Fig.~\ref{fig:weaving}.  Here
the essential alteration is to add the phase structures associated
with the quarkyonic chiral spirals.  The chiral spiral is favored due
to the pseudo 1-dimensionality, and the Fermi surface should be
covered with such pseudo 1-dimensional patches.  Because the
transverse size of each patch is expected to be $\sim\LQCD$, the
number of patches should increase with increasing $\muq$ (and
increasing area of the Fermi surface).  This ``interweaving'' picture
naturally leads us to the speculation of successive phase transitions
associated with the number of patches.  In other words, the most
optimal crystalline structure should depend on the density and the
structural changes would divide the inhomogeneous region (as indicated
by the shaded area in Fig.~\ref{fig:diagram}) into finer substates
named the interweaving chiral spirals~\cite{Kojo:2011cn}.

The most serious concern on reality of inhomogeneous quarkyonic matter
is the validity of the extrapolation from $\Nc=\infty$ to $\Nc=3$,
which is discussed for example in Ref.~\cite{Torrieri:2010gz}.  We
finally point out that there exist quite a few theoretical arguments
and pictures that are essentially indistinguishable from the
McLerran-Pisarski conjecture and quarkyonic matter at $\Nc=3$.  A
related duality (at zero density) has already been discussed in
Ref.~\cite{Wetterich:1999vd}.  At high density the quark-hadron
continuity in the context of
color-superconductivity~\cite{Schafer:1998ef,Alford:1999pa} is not
really different from the McLerran-Pisarski conjecture (for the
confinement properties in color-superconductor, see
Ref.~\cite{Rischke:2000cn}).  At the phenomenological level the
percolation model may provide us with an interpretation of quarkyonic
matter at $\Nc=3$~\cite{Lottini:2011zp,Lottini:2012as}.

\section{Experimental Prospects}
\label{sec:experiment}

There are many theoretical speculations and conjectures on QCD at
high density, and some of them should be filtered out by experimental
data.  Here we concisely discuss experimental implications to explore
the properties of dense QCD matter.

\subsection{Liquid-gas phase transition and the critical point}
\label{sec:liquid-gas}

\subsubsection{Experimental signals in nuclear matter}

For general discussions on the liquid-gas phase transition, see
Ref.~\cite{Tatsumi:2011tt} and references therein.  Here we enumerate
experimental signals for the critical point in nuclear matter.  The
extensive study of nuclear multi-fragmentation has been stimulated by
the suggestion that this process might be a critical
phenomenon~\cite{Bondorf:1985mv,Bondorf:1985mp,Bondorf:1995ua,Ma:1999qp}.
In particular the produced fragment distributions follow a power law
of the form $P(A) \sim A^{-\tau}$ with the atomic number $A$, which
suggests that a class of ``universality'' exists behind
it~\cite{Finn:1982tc}.  Available multi-fragmentation data in
nucleus-nucleus and hadron-nucleus reactions provide a distinct change
of its character.  Given these signals, the observation of the
liquid-gas phase transition in a finite nuclear system is strongly
supported~\cite{Panagiotou:1984rb,Viola:2003hy,Viola:2006fh,Chomaz:2003dz,Ma:2004ey}.

The caloric curve, i.e.\ the heat versus temperature, is one of the
important observables to distinguish the critical phenomena.
Temperature of the system has to be reconstructed from observable
quantities.  In chemical and thermal equilibrium the temperature is
extracted from the yields of isotope ratios~\cite{Albergo:1985zz}.
The observed isotope temperature as a function of the excitation
energy per nucleon $E/A$ shows a plateau over the range of relatively
low $E/A$~\cite{Pochodzalla:1995xy,Natowitz:2001cq}.  The observed
caloric curve agrees qualitatively with predictions of the
multi-fragmentation model~\cite{Bondorf:1995ua} which is an
event generator as a generalized liquid-drop model for hot nuclei,
and the properties are consistent with a liquid-gas phase transition.

Another experimental signal of the phase transition is the appearance
of a negative specific heat capacity:  The entropy has typically a
convex structure when a system experiences a first-order phase
transition and this results in the specific heat $C$ being negative.
This effect has been extracted from analysis of the energy
fluctuations and the presence of a negative $C$ branch was indeed
found~\cite{D'Agostino:1999kp}.

The location of the critical point $\Tc$ for the nuclear liquid-gas
phase transition comes out from the observables.  At this point the
isotherm in the phase diagram has an inflection point.  The major
source of experimental information for $\Tc$ is the fragment yield.
However, the procedures to extract $\Tc$ are highly scheme-dependent
and the obtained values lie in a wide range $\Tc \sim 5$-$20\MeV$,
indicating a severe model-dependence to be resolved (see
e.g.\ Ref.~\cite{Karnaukhov:2003vp}).  The critical exponents have
also been extracted from the moments of the fragmentation charge
distributions~\cite{Hufner:1985hz,Gilkes:1994gc,Elliott:1994zz}.  The
analysis concludes that the phase transition is of second order and
finds the critical exponents consistent with the $\mathrm{Z}_2$
universality class.

\subsubsection{Toward the QCD critical point}
\label{sec:QCDcp}

First-order phase transitions for cold and dense quark matter have
been predicted in several approaches using chiral
models~\cite{Asakawa:1989bq,Berges:1998rc,Halasz:1998qr},
Dyson-Schwinger
equations~\cite{Barducci:1989wi,Barducci:1989eu,Taniguchi:1995vf},
and lattice-QCD in the strong coupling 
limit~\cite{Ilgenfritz:1984ff,Damgaard:1985bn,Karsch:1988zx,Fukushima:2003vi,Nishida:2003fb}.
Given the observation of a crossover at zero chemical potential from
lattice  QCD computations, this might suggest an additional critical
point other than that of the nuclear liquid-gas in the QCD phase
diagram.

The order of the QCD phase transition at low temperature and high
density is not established yet and thus the existence of a QCD
critical point by itself remains an issue under debate. If it exists,
the critical point must exhibit the critical exponents of the
3-dimensional Ising model that belongs to the $\mathrm{Z}_2$
universality class~\cite{Wilczek:1992sf}.  A potential signal for the
QCD phase transition is modifications in the magnitude of fluctuations
or the corresponding susceptibilities.  In particular, fluctuations
related to conserved quantities (baryon number $B$, quark number
$q=3B$, electric charge $Q$, strangeness $S$, energy $E$, etc) play an
important role since they are directly accessible in
experiments~\cite{Stephanov:1998dy}.

A phase transition can be probed with response of a medium to
temperature and chemical potential. The $n$-th order cumulant of a
conserved quantity $X$, such as $B$, $q$, $Q$, $S$, etc, is computed
from the partition function $Z$ via
\begin{equation}
 \chi_n^X = \frac{1}{VT^3} \frac{\partial^n\ln Z}{\partial(\mu_X/T)^n}\;.
\end{equation}
By taking the following cumulant ratios a volume factor can be
eliminated;
\begin{equation}
 R_{n,m}^X = \frac{\chi_n^X}{\chi_m^X}\;.
\end{equation}
The ratios $R_{4,2}^q$ and $R_{4,2}^Q$ are shown to be sensitive to
quark deconfinement at $\muq=0$ and they exhibit qualitatively
different behavior from those of the HRG model~\cite{Ejiri:2005wq}.
At finite $\muq$ higher order cumulants of baryon number are
particularly important since they diverge stronger on the chiral phase
boundary in the chiral limit as well as at the critical
point~\cite{Stephanov:2008qz}.  Higher moments and their signs have
also been proposed as sensitive probes to the phase
transitions~\cite{Asakawa:2009aj}.

Recently it has been shown that the fourth-order cumulant and the ratio
$R_{4,2}$ (kurtosis) become universally negative when the critical
point is approached from the crossover side~\cite{Stephanov:2011pb}.
Here the variance $\sigma_X$, the skewness $S_X$, and the kurtosis
$\kappa_X$ are defined as
\begin{equation}
 \sigma_X^2 = VT^3\chi_2^X \;, \qquad
 S_X\sigma_X = \frac{\chi_3^X}{\chi_2^X} \;, \qquad
 \kappa_X\sigma_X^2 = \frac{\chi_4^X}{\chi_2^X} \;.
\end{equation}
A study using scaling functions shows that the fourth-order cumulant
$\chi_4^B$ changes its sign around the crossover region from positive
at lower temperature to negative at higher temperature, as a generic
feature~\cite{Friman:2011pf}.  The temperature at which $\chi_4^B$
becomes negative is not universal since it depends on the regular part
of the partition function.  This suggests that the negative kurtosis
is not a unique indicator of the critical point.

The STAR collaboration reported the ratio
$R_{4,2}^B$~\cite{Aggarwal:2010wy}, which is consistent with the HRG
model result~\cite{Karsch:2010ck,Gavai:2010zn}.  More recent data of
the moment products, $\kappa_X\sigma_X^2$ and $S_X\sigma_X$, of the
net proton number and electric charge
distributions~\cite{Luo:2012sd,McDonald:2012ts,Mitchell:2012mx} differ
from the HRG model result.  This might be due to such simplified
modeling of the QCD medium, although the thermal statistical model or
the HRG model works well with various particle number ratios.  So far,
no experimental indication of the critical point is found.

If any signal of a first-order phase transition were observed, this
would become an indication of the presence of the critical point as in
the case of nuclear matter.  In this respect, it is also challenging
in heavy-ion collisions to observe the proposed consequences of the
spinodal decomposition associated with a would-be first-order phase
transition at high density, such as an enhancement of baryon and
strangeness
fluctuations~\cite{Bower:2002cr,Koch:2005pk,Randrup:2003mu} and a
negative quark number susceptibility~\cite{Sasaki:2007db}.

\subsection{Dilepton measurements and chiral symmetry restoration}
\label{sec:dilepton}

Dileptons are considered to be promising probes to study for changes
of hadron properties in matter since they pass through the fireball
created in heavy-ion collisions without hadronic interactions.  The
short-lived vector mesons like the $\rho$ mesons are expected to decay
into dileptons inside the hot and dense matter.  An enhancement in the
dilepton rates below the $\rho/\omega$ resonance -- indicating the
medium modification of the vector mesons -- has been observed in many
experiments.  Although a major number of experiments detected a strong
width broadening and no ``mass shift'', no perfect agreement among
different experiments is reached
yet~\cite{Hayano:2008vn,Leupold:2009kz,Rapp:2009yu}.  A shift of the
$\rho/\omega$ peak in the in-medium spectral function is often
regarded as a ``mass shift'' of the vector meson.  However, this is
clearly an oversimplified interpretation since complicated reaction
processes, numbers of resonances in a medium, and their interactions
are involved and those complications are integrated into the final
spectra.  Therefore, picking up a maximum of such broad distributions
does not readily give us the in-medium mass of the vector meson.

Below we will briefly argue some theoretical issues related to
the dilepton measurements.

\subsubsection{Is vector dominance fulfilled in a medium?}

A lepton pair is emitted from the hot and dense matter through a
decaying virtual photon.  The differential production rate in the
medium for fixed temperature $T$ or baryon density $\rho_{\rm B}$ is
calculated with the imaginary part of the photon self-energy
$\text{Im}\Pi$ via
\begin{equation}
 \frac{dN}{d^4 q}(q_0,\vec{q};T,\rho_{\rm B})
 = \frac{\alpha^2}{\pi^3 M^2} \, \frac{1}{e^{q_0/T}-1}\,
   \text{Im}\Pi (q_0,\vec{q};T,\rho_B) \;,
\end{equation}
where $\alpha = e^2/4\pi$ is the electromagnetic coupling constant,
$M$ is the invariant mass of the produced dilepton, and
$q_\mu=(q_0,\vec{q})$ denotes the momentum of the virtual photon.  We
will focus on an energy region around the $\rho$-meson mass scale in
the following argument.  In this energy range the photon self-energy
is expected to be dominated by the two-pion process and its imaginary
part is related to the pion electromagnetic form factor
$\mathcal{F}(s;T)$ through
\begin{equation}
 \text{Im}\Pi(s;T,\rho_{\rm B}) = \frac{1}{6\pi\sqrt{s}}
  \biggl( \frac{s - 4m_\pi^2}{4} \biggr)^{3/2}
  \bigl| \mathcal{F}(s;T,\rho_{\rm B}) \bigr|^2
\end{equation}
with the pion mass $m_\pi$ and the form factor given by
\begin{equation}
 \mathcal{F}(s;T,\rho_{\rm B})
 = g_{\gamma\pi\pi}(T,\rho_{\rm B})
 + \frac{g_\rho(T,\rho_{\rm B}) \cdot g_{\rho\pi\pi}(T,\rho_{\rm B})}
   {m_\rho^2(T,\rho_{\rm B}) - s - \theta(s - 4m_\pi^2) \, i
    m_\rho(T,\rho_{\rm B}) \Gamma_\rho(s;T,\rho_{\rm B})} \;.
\end{equation}

In matter-free space, a single photon mainly couples to two pions via
the vector meson exchange, known as vector meson dominance (VD).  This
is assumed to hold in a hot and dense medium in many model
calculations for the in-medium vector spectrum.  One should, however,
keep in mind that this is not a priori justifiable.  A reliable
theoretical framework to handle the VD is a hidden local symmetric
(HLS) approach as discussed in Sec.~\ref{sec:hls}; the vector meson
properties are essentially controlled by one parameter $a$ and a
particular choice, $a=2$, leads to vanishing photon-pion coupling,
\begin{equation}
 g_{\gamma\pi\pi}(T=\rho_{\rm B}=0) = 0 \;.
\end{equation}
A consistent achievement of chiral symmetry restoration at a given set
of $T$ and $\rho_{\rm B}$ requires that the parameter $a$ should
evolve toward unity with $T$ and
$\rho_{\rm B}$~\cite{Harada:2001it,Harada:2001qz,Harada:2003wa}.  The
immediate consequence is a {\it strong violation} of the VD in matter:
The coupling $g_{\gamma\pi\pi}$ is an increasing function of thermal
parameters whereas the vector-meson--photon coupling is a decreasing
function of them toward chiral symmetry restoration as
\begin{equation}
 g_{\gamma\pi\pi} \to \frac{1}{2}\;, \qquad
 g_\rho \to 0\;.
\end{equation}
This indicates more dilepton rates from two-pion annihilation
\textit{not involving the in-medium vector mesons} when the system
approaches the chiral phase transition.  Thus the dilepton yields
become suppressed there~\cite{Harada:2006hu}.

\subsubsection{Has the BR scaling been excluded by the NA60 dimuon data?}

The NA60 collaboration reported a strong in-medium broadening of the
$\rho$ meson and no mass shift in dimuon
measurements~\cite{Arnaldi:2006jq,Damjanovic:2006bd}.  The data were
compared with the theoretical predictions based on phenomenological
Lagrangians (width broadening due to hadronic many-body effects versus
dropping $\rho$ mass), and the broadening with no mass shift is
apparently favored.  The measurement has excluded the scenario with a
na\"{i}ve dropping mass as in the form,
\begin{equation}
 m_\rho(n) \sim m_\rho(n=0)\biggl(1 - \alpha\frac{n}{n_0}\biggr)\;,
\label{eq:shift}
\end{equation}
where $n$ represents either temperature or density, $\alpha$ a
constant, and $n_0$ an appropriate scale for normalization.

If partial restoration of chiral symmetry modifies the vector meson
property, it would appear not only in the mass but also all the
interactions involving the vector meson.  In principle those
modifications should be incorporated systematically within a given
effective theory for relevant hadrons.  Therefore, a na\"{i}ve
replacement of the vector meson mass with Eq.~\eqref{eq:shift} in the
quantities such as the spectral function is incomplete.  Also, the VD
is shown to be violated at a certain $n$ once the vector meson, if its
mass drops, feels \textit{partial} chiral symmetry
restoration~\cite{Harada:2001it,Harada:2001qz,Harada:2003wa}.  On top
of the in-medium broadening and the suppression of the
vector-meson--photon coupling due to the VD violation, it was
argued~\cite{Brown:2005ka,Brown:2005kb,Brown:2009az} that the signal
of the Brown-Rho (BR) scaling is totally hidden in matter-free
contributions essentially coming from the vector mesons which
\textit{do not feel} partial restoration and thus peaked around the
vacuum mass, $m_\rho = 770\MeV$.  Thus, a comparison of the data with
the ``dropping mass scenario'' needs to be made with great caution.

Aiming to characterize a broad spectrum without clear resonance
structure, an ``average mass'' can be introduced
as~\cite{Kwon:2008vq,Kwon:2010fw}
\begin{equation}
 \bar{m}_{V,A}^2  \equiv
  \frac{\int ds\, s\, \rho_{V,A}}{\int ds\, \rho_{V,A}}\;.
\end{equation}
At order $T^2$ the vector meson pole mass does not receive any thermal
corrections~\cite{Dey:1990ba,Eletsky:1994rp} and the numerical
analysis for $\bar{m}_V$ using QCD sum rules was shown to be
consistent with this statement.  The vector meson ``mass'' $\bar{m}_V$
stays almost constant up to $T \sim m_\pi$, whereas the axial-vector
meson ``mass'' $\bar{m}_A$ decreases with rising temperature.
Therefore, in this range of the temperature a dropping $\rho$ mass is
not supported~\cite{Kwon:2010fw}.

On the other hand, in-medium $\rho$ mesons in cold nuclear matter
exhibit a different feature:  The calculated $\bar{m}_V$ making use of
the in-medium spectral function follows~\cite{Kwon:2008vq}
\begin{equation}
 \frac{\bar{m}_V^\ast}{\bar{m}_V^{\rm vac}}
 \sim \frac{f_\pi^\ast}{f_\pi}
 \sim 1 - 0.15\,\frac{\rho_{\rm B}}{\rho_0}\;.
\end{equation}
This suggests that the tendency of the BR scaling is indeed visible
with the in-medium spectral function which shows a quite broad
distribution.  This observation is,
\textit{contrary to the first impression}, a good example of the
``hidden'' BR scaling in the broad spectrum in matter.

\subsubsection{Chiral mixing in vector and axial-vector spectral functions}

There is no doubt that dilepton measurements have observed in-medium
modifications in the vector-meson channel.  Yet, it remains unclear
how such broadened vector spectra would eventually be linked to chiral
symmetry restoration.  Although the axial-vector spectral function is
hardly measurable, axial-vector mesons come in to the vector spectrum
via a mixing to pions in hot/nuclear
matter~\cite{Dey:1990ba,Krippa:1997ss}.  The mixing between the vector
and axial-vector spectral functions, $\rho_V$ and $\rho_A$, is derived
in a model-independent way at low $T$ and $\rho_{\rm B}$, arranged in
a low-temperature/density theorem,
\begin{equation}
 \begin{split}
 \rho_V(s;n) &= (1-\epsilon) \rho_V(s;0) + \epsilon \rho_A(s;0)\;,\\
 \rho_A(s;n) &= (1-\epsilon) \rho_A(s;0) + \epsilon \rho_V(s;0)\;,
 \end{split}
\label{eq:mixing}
\end{equation}
where chiral mixing is characterized by the parameter 
$\epsilon = T^2/6f_\pi^2$ or $4\rho_{\rm B}\sigmaN/3f_\pi^2 m_\pi^2$
with the pion-nucleon sigma term $\sigmaN$.  The spectral functions
with the mixing Eq.~\eqref{eq:mixing} at finite temperature has been
explored in the QCD sum
rules~\cite{Marco:2001dh,Kwon:2010fw,Hohler:2012xd,Holt:2012wr} where
one finds an indication of the expected tendency of the spectra
$\rho_V$ and $\rho_A$ becoming degenerate.

Properties of the vector and axial-vector mesons constrained by
current algebra are summarized in the sum rules for the spectral
functions, known as Weinberg sum rules~\cite{Weinberg:1967kj},
\begin{equation}
 \int_0^\infty \frac{ds}{s} \bigl[ \rho_V(s) - \rho_A(s) \bigr]
  = f_\pi^2\;,
 \qquad
 \int_0^\infty ds \bigl[ \rho_V(s) - \rho_A(s) \bigr] = 0\;.
\end{equation}
Those sum rules extended to a finite temperature have been
explored~\cite{Kapusta:1993hq} where the following relation was
deduced assuming the pole-saturated forms of $\rho_V$ and $\rho_A$;
\begin{equation}
 \frac{f_\pi^2(T)}{f_\pi^2} = 2Z_\rho(T) \biggl(
  \frac{m_\rho^2}{m_\rho^2(T)} - \frac{m_\rho^2}{m_{a_1}^2(T)} \biggr)
\end{equation}
with the pole residue $Z_\rho$.  Clearly, when
$m_\rho(T) = m_{a_1}(T)$, chiral symmetry is restored with the
vanishing order parameter $f_\pi(T) \to 0$.  The sum rules by
themselves do not distinguish how the masses behave with increasing
temperature.  Possible scenarios are
(i) both decrease,
(ii) both increase,
(iii) $m_\rho(T)$ increases whereas $m_{a_1}(T)$ decreases.
Using an $N_f=2$ linear sigma model, $m_\rho(T)$ is shown to go up by
chiral restoration temperature 
\textit{if the VD holds in hot matter}~\cite{Pisarski:1995xu}.  (See
also Refs.~\cite{Urban:2001ru,Parganlija:2010fz} for the linear sigma
model with global chiral symmetry, which can be extended to non-zero
temperature.)  Non-linear sigma model calculations also yield the same
tendency at low temperature where the VD is well
satisfied~\cite{Harada:2003wa}.  If the VD is violated, there is no
unique prediction unless one imposes a certain set of conditions for
the model parameters being consistent with restoration of chiral
symmetry.

The spectral functions $\rho_V$ and $\rho_A$ become degenerate when
the mixing parameter reaches $\epsilon = 1/2$.  However, such a
na\"{i}ve extrapolation of Eq.~\eqref{eq:mixing} to higher $n$
requires the vector and axial-vector meson masses staying their vacuum
values and thus they are not necessarily degenerate in contradiction
to the desired property of chiral partners.  An approach which does
not rely on Eq.~\eqref{eq:mixing} is to use chiral effective theories
of the vector and axial-vector mesons explicitly.  The mixing and its
evolution in matter are generated from processes involving the
$\pi\,\rho\,a_1$ interaction. Due to the mixing the axial-vector meson
appears as a bump in $\rho_V$ around $\sqrt{s}=m_{a_1} - m_\pi$ and
this makes $\rho_V$ broadened.  When the system evolves toward chiral
symmetry restoration with temperature, it was shown that the chiral
mixing vanishes at a critical point $\Tc$ in the chiral
limit~\cite{Harada:2008hj}.  Therefore the ``maximal mixing''
scenario, $\epsilon=1/2$, does not happen but instead, the resonance
peak and the bump tend to be degenerate forming a single peaked
structure which obviously indicating the degenerate $m_\rho$ and
$m_{a_1}$.  It is also interesting that even with explicit symmetry
breaking the $\rho$ and $a_1$ masses are predicted to be well
degenerate around a pseudo-critical temperature.

At finite chemical potential chiral mixing is generated at tree level
since charge conjugation invariance is lost, which allows the
following term that mixes vector and axial-vector
fields~\cite{Harada:2009cn},
\begin{equation}
 \mathcal{L}_{\rm mix} = 2C\epsilon^{0\nu\lambda\sigma}
  \text{tr}\bigl[ \partial_\nu V_\lambda \cdot A_\sigma
   + \partial_\nu A_\lambda \cdot V_\sigma \bigr]
\label{eq:dmix}
\end{equation}
with the coupling $C$ and the total anti-symmetric tensor
$\epsilon^{0123}=1$.  One can deduce the above term from the
$\omega\,\rho\,a_1$ part in the Wess-Zumino-Witten term with replacing
$\omega_0$ with its expectation value,
\begin{equation}
 \langle \omega_0 \rangle = g_{\omega NN}
  \frac{\rho_{\rm B}}{m_\omega^2}\;.
\end{equation}
Therefore, the coupling $C$ is
\begin{equation}
 C = g_{\omega\rho a_1}\langle\omega_0\rangle\;,
\end{equation}
and at normal nuclear density $\rho_0$ one finds $C \sim 0.1\GeV$.
The mixing term is also derived from the reduction of five-dimensional
Chern-Simons action to four dimensions in a holographic QCD
model~\cite{Domokos:2007kt} where $C \sim 1\GeV$ at
$\rho_{\rm B}=\rho_0$ was found.  Such a strong mixing is, however,
unrealistic since otherwise it yields vector meson condensation
slightly above $\rho_0$.  Thus, a large $C$ is supposed to be an
artifact in large $\Nc$ and/or of the specific model construction
applied to a dense system.

The mixing term \eqref{eq:dmix} modifies the dispersion relation
resulting in
\begin{equation}
 p_0^2 - \vec{p}^{\;2} = \frac{1}{2} \biggl[ m_V^2 + m_A^2 \pm
  \sqrt{(m_A^2 - m_V^2)^2 + 16C^2\vec{p}^{\;2}} \biggr]\;,
\label{eq:disp}
\end{equation}
which describes the propagation of a mixture of the transverse $\rho$
and $a_1$ mesons.  The longitudinal polarizations follow the standard
dispersion relation.  Consequently, the spin-averaged current
correlator is superposition of the longitudinal mode peaked at $m_V$
and the transverse modes with two bumps peaked at shifted $m_V$ and
$m_A$ obeying Eq.~\eqref{eq:disp}.  The presence of mixing makes the
entire spectral function broadened and this influences over dilepton
production rates.  Its impact on the rates crucially depends on the
size of $C$.  Let us suppose that the lowest $\omega$ dominance giving
rise to $C \sim 0.1\GeV$ at $\rho_{\rm B}=\rho_0$ is favored.  Mixing
does not have significance at this density.  With increasing density,
i.e.\ increasing $C$, the structure of the spectral function starts to
show a change from the vacuum distribution, and such a modification
becomes distinct when density reaches
$\sim 3\rho_0$~\cite{Harada:2009cn}.

The mixing~\eqref{eq:dmix} is chirally symmetric and thus does not
vanish at chiral restoration in contrast to the vanishing chiral
mixing at finite temperature without baryon density.  In the presence
of both mixing effects, a tendency of chiral symmetry restoration
signaled by degenerating $\rho$ and $a_1$ mesons would become more
obscure.  Therefore, it seems a hard issue to observe a clear signal
of chiral symmetry restoration in dilepton measurements.  At least
from the theoretical side, the in-medium vector spectral functions
known to agree with data can be compared with the axial-vector spectra
calculated in a given model with those mixing.  Unless the description
for the hadrons involved in terms of local fields breaks down totally,
such a comparison could offer us an indirect evidence for the symmetry
restoration, with attention to the notion of in-medium mass.

\subsection{Astrophysical Implication}

Historically speaking, the research on dense nuclear matter and the
microscopic calculation of the EoS were motivated in the context of
the structures of compact stellar objects (see
e.g.\ Ref.~\cite{shapiro2008black} for a textbook).

There was an astonishing news in the year of 2010;  it was reported in
Ref.~\cite{Demorest:2010bx} that the Shapiro delay measurement for the
binary pulsar PSR~J1614-2230 results in the pulsar mass of
$(1.97\pm0.04) M_\odot$.  This discovery has a tremendous impact on
nuclear physics:  Indeed, the abstract of Ref.~\cite{Demorest:2010bx}
reads, ``\textit{effectively rules out the presence of hyperons,
  bosons, or free quarks at densities comparable to the nuclear
  saturation density}.''  Once the EoS is known, one
can plug $p(\rho)$, the pressure as a function of density, into the
Tolman-Oppenheimer-Volkoff (TOV) equation to solve the relation
between the mass and the radius of compact stars.  The mass cannot be
arbitrarily large but bounded by an upper limit corresponding to
$p(\rho)$ that should sustain the star against the gravitational
collapse.  The EoS should be harder, i.e.\ $p$ should be larger for a
given $\rho$, to support a larger mass.

The problem is that the EoS tends to be softer with exotic
compositions, with which it may not be consistent with the existence
of the neutron star as heavy as almost twice of the solar mass, though
it can be; see Refs.~\cite{Alford:2004pf,Klahn:2006iw}.  There is not
really a final statement on this issue, and actually it is quite
difficult to establish any reliable derivation of the EoS of quark
matter (see e.g.\ Ref.~\cite{Kurkela:2009gj} for a recent attempt),
and thus it can become harder than believed (see
Ref.~\cite{Masuda:2012kf} for example).  Also, if the repulsive
vector-interaction as in Eq.~\eqref{eq:NJL4} is strong enough, it
would easily make the EoS sufficiently hard.  In other words, to
explain the neutron star with $\sim 2M_\odot$, a substantially large
vector interaction should be expected, and this implies in turn that
the QCD critical point is disfavored from the QCD phase
diagram~\cite{Bratovic:2012qs}.

Without assuming the state of matter, it is also a feasible program to
compile astrophysical observations to constrain the EoS of dense
matter directly.  For recent progresses in this direction, see
Refs.~\cite{Ozel:2010fw,Lattimer:2010zz,Sotani:2012qc,Steiner:2012xt}.

The cooling rate is also informative on the state of matter in the
neutron stars~\cite{Yakovlev:2004iq}, and it is an interesting
question how to constrain the possibility of the pion
condensation~\cite{Umeda:1994it}.  We must keep in mind, however, that
the environment of the neutron star is not necessarily the same as
that we consider for the QCD phase diagram or for the heavy-ion
collision.  In the neutron star $\beta$ equilibrium is achieved, while
it is usually symmetric nuclear and quark matter that is relevant to
the phase diagram research.  Besides, even if exotic contents such as
the pion condensation, the kaon condensation, and quark matter were
ruled out from the neutron-star observation (which is difficult from
the mass-radius relation alone~\cite{Alford:2004pf}), they might be
still possible in symmetric matter and/or at higher densities.

\section{Summary and Outlook}
\label{sec:outlook}
 
\begin{figure}
 \begin{center}
 \includegraphics[width=0.475\textwidth]{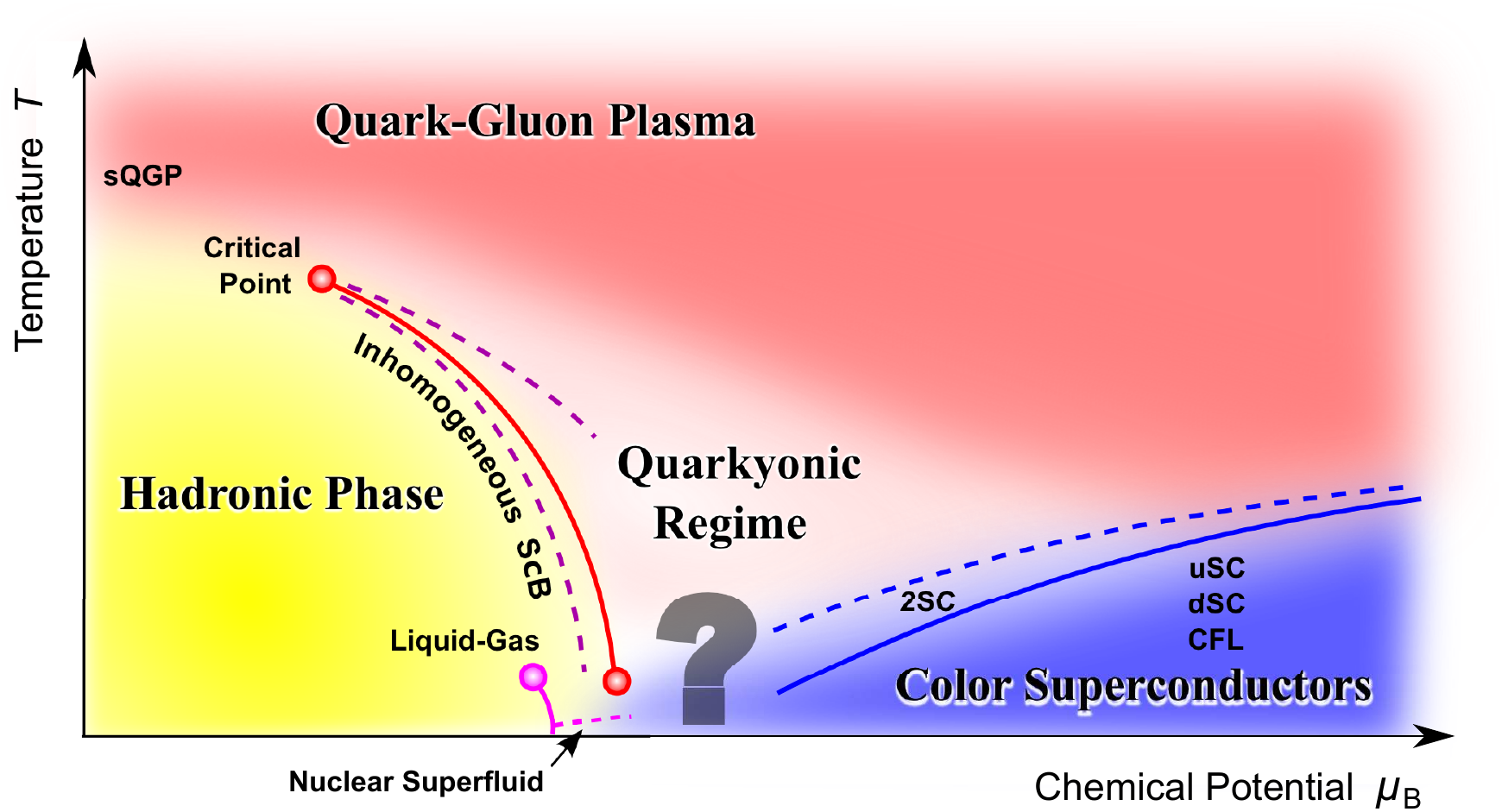} \hspace{1em}
 \includegraphics[width=0.475\textwidth]{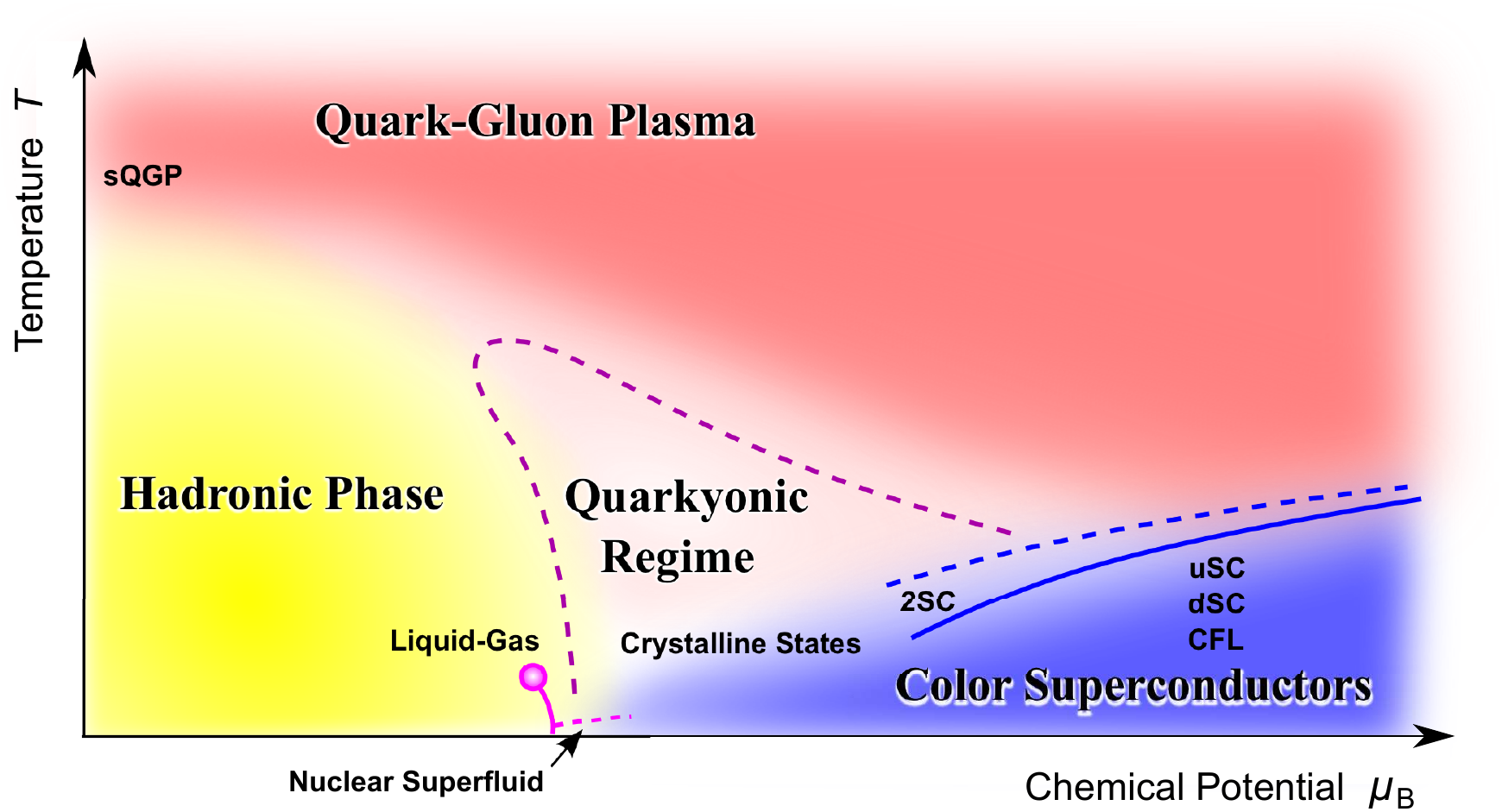}
 \end{center}
 \caption{Two representative scenarios for the QCD phase diagram; one
   with the QCD critical point(s) adapted from
   Ref.~\cite{Fukushima:2010bq} and the other without the first-order
   phase transition at all.  ``Quarkyonic Matter'' is intentionally
   replaced with ``Quarkyonic Regime'' (see an explanation in the
   text).}
 \label{fig:phase}
\end{figure}


In this review we have addressed physical properties of nuclear and
quark matter and given pedagogical descriptions on the tools used in
theoretical research.  On the phase diagram with two environmental
parameters, the temperature $T$ and the baryon chemical potential
$\muB$, only a small portion has been understood;  the QCD phase
transitions of deconfinement and chiral restoration at (nearly) zero
density, and the nuclear liquid-gas phase transition that is
inevitable from the saturation property of nuclear matter.  For the
zero-density crossover, a pile of experimental and lattice-QCD data
have been accumulated.  Also the hadron resonance gas model works
nicely, with which a physical picture has been established.  In the
nuclear physics territory the correct understanding is guided by
experimental data, but the direct application of QCD to nuclear matter
is still a big challenge.  In practice one cannot avoid modeling the
QCD dynamics in a form of the effective description.

In nuclear matter not far from the normal nuclear density chiral
symmetry cannot be totally restored, but partial restoration has been
confirmed both in theory and in experiment of the deeply bound pionic
states.  Mesons and baryons should be modified accordingly, and the
hidden local symmetry model provides us with the most consistent
description on these issues.  It is, however, still hard to make
conclusive statements on the changes of the hadron spectra and the
mixing pattern at finite density.  To grasp the qualitative character
of dense matter, it is useful to increase the number of colors to
infinity; $\Nc\to\infty$.  Then, the Skyrme model and also the most
promising holographic QCD model (Sakai-Sugimoto model) tell us that
the ground state of nuclear matter at $\Nc\to\infty$ takes a crystal
structure and spatial modulations should appear at high density, which
is reminiscent of the old idea of the $p$-wave pion condensation.

Interestingly enough, this sort of inhomogeneity is favored by a
robust mechanism in quark matter.  Using a generic quasi-quark
picture, we have demonstrated a simple calculation how the first-order
phase transition and the inhomogeneous states can emerge in such a way
induced by the density effects.  This simple argument is helpful in
understanding what part is robust and what part is not in theoretical
predictions.  We have demonstrated that the first-order phase
transition and thus the QCD critical point can disappear from the
phase diagram if the repulsive vector interaction is substantially
strong, which is also suggested from the existence of the
two-solar-mass neutron star.  With the same parameter set, on the
other hand, the inhomogeneous (chiral spiral) phase can persist, and
we can say that the appearance of inhomogeneity is quite robust unlike
the QCD critical point.

Then, it is conceivable that such inhomogeneous states of quark matter
may be connected to the $p$-wave pion condensation.  More generally,
it might be even possible that strongly interacting nuclear matter has
a dual description in terms of quark degrees of freedom.  In fact,
such ``duality'' is suggested in the large-$\Nc$ counting, which
defines a new regime on the phase diagram called quarkyonic matter.
It was once said that quarkyonic matter could be a state with
confinement but chiral symmetry restoration, but this characterization
was misleading.  The crucial aspects of quarkyonic matter are; (1)
strong interaction of baryons, (2) inhomogeneous chiral condensates,
and (3) quark degrees of freedom.  Model buidling with these
properties correctly equipped along the similar line to the hadron
resonance gas model is an urgent theory problem.\\


What should the QCD phase diagram look like?  In the market of the QCD
phase diagram research, there are many variants of the schematic
figures.  We pick two representative scenarios up here in
Fig.~\ref{fig:phase}.  The left figure (taken from
Ref.~\cite{Fukushima:2010bq}) is similar to the conventional type of
the QCD phase diagram.  It has a first-order boundary for the chiral
phase transition and accommodates the critical point(s).  The right
one is a rather unconventional but these two phase diagrams are both
viable enough.  The QCD phase transitions, namely, deconfinement
and chiral restoration, can be only crossovers and no critical point
(apart from the nuclear liquid-gas transition) exists on the whole QCD
phase diagram.  Even in this case the inhomogeneous states around the
regime of quarkyonic matter should be robust, as pictured with the
dashed curve, in a sense that no model has falsified yet.  In
particular, in this latter scenario, the quarkyonic chiral spiral, the
$p$-wave pion condensation in nuclear matter, and possibly the
crystalline color-superconductivity may well be linked without sharp
phase boundary, which is collectively referred to as
``crystalline state'' in the right panel of Fig.~\ref{fig:phase}.  In
these figures we intentionally use a term ``Quarkyonic Regime''
instead of quarkyonic matter because it is important to note that
quarkyonic matter is not a distinct state of matter.  Quarkyonic
matter should be understood as a transitional state between dense
nuclear matter and quark matter, having both aspects of them.

With available theoretical knowledge and experimental data, it is
still far from feasible to be able to select out the genuine one among
many proposed diagram.  We must solve QCD to establish a unique
picture of the QCD phase diagram, and at the same time, we need more
experimental data of the beam-energy scan from the relativistic
heavy-ion collision.  The sign problem inherent in the Monte-Carlo
simulation is extraordinarily difficult to handle in a satisfactory
manner, and the QCD phase diagram research awaits some other technical
breakthrough not relying on the importance sampling.

As an extrapolation from nuclear matter toward higher density, one
need to resort to effective approaches.  Applying effective chiral
Lagrangians in terms of a given set of mesons and baryons to dense
matter is based on the assumption that those hadrons keep their
particle identity and can still be expressed as local fields.  It is
not obvious whether this approximation is legitimate, but rather it
can be taken as a working hypothesis.  In fact, the spectroscopic
factor for various nuclei indicating a deviation from the fully
occupied mean-field orbits clearly shows that single-particle-ness of
excitations reaches $\sim70\%$~\cite{Dickhoff:2004xx}.  This would
encourage modeling dense matter in a quasi-particle picture.  In
particular, when partial restoration of chiral symmetry sets in,
\textit{light} degree(s) of freedom that will be degenerate with the
NG bosons must appear in a system.  Thus, if one chooses
``appropriate'' degrees of freedom, such effective Lagrangian
approaches with the local-field approximation would give a good
description.

This speculative consideration is also supported by our knowledge on
condensed matter physics where strongly correlated systems are well
described by effective field theories and the notion of
quasi-particles has been successful.  A considerable example is a
phase transition between the N\'{e}el (antiferromagnet) and valence
bond solid (VBS, paramagnet) phases.  Topology of those phases is well
captured by the non-linear sigma models and a skyrmion configuration
splitting into two half-skyrmions has been
known~\cite{Haldane:1988zz,Chakravarty:1989zz,Read:1989zz}. The
effective theory contains an emergent gauge field that mediates
interactions between effective degrees of freedom~\cite{Senthil}.
This is quite suggestive and a similarity would be expected in nuclear
many-body systems in high densities.  Indeed, the concept of induced
gauge fields, which arise when ``fast'' modes separated from ``soft''
modes are integrated out, have been recognized in various fields of
physics~\cite{Shapere:1989kp}.  In the context of QCD, non-Abelian
gauge potentials (Berry phases) appear naturally in topological chiral
bags under any adiabatic rotation~\cite{Lee:1993tg}.\\


On the QCD phase diagram we must widen our perspective outside of
nuclear matter toward higher temperature and/or density, and the
validity of effective approaches is questionable then.  To this end,
there is no bypass but solving QCD is the one-track way to reveal the
QCD phase diagram completely.  This used to be an impossible problem,
but nowadays, it is becoming tractable owing to the technical
developments in the functional methods.  The QCD application of the
Dyson-Schwinger equation has a long history and the mechanism of
color confinement has been understood based on the gluon and the ghost
propagators in the Landau gauge (see Ref.~\cite{Fischer:2008uz} for a
review).  Also, the Wetterich equation as we discussed in
Sec.~\ref{sec:quark-meson} is the exact formulation and has an
equivalent content as the Dyson-Schwinger
approach~\cite{Pawlowski:2003hq}.  The QCD-based understanding of
confinement is successfully transferred to the finite-$T$ studies on
the QCD phase
transitions~\cite{Braun:2007bx,Braun:2009gm,Fischer:2009wc,Fischer:2011mz}.
In these approaches toward QCD matter at high density, the most severe
obstacle lies in the description of baryons and quarks on the equal
footing.  There is no model known in the market that can cope with
baryons melting into quarks within a single framework.  In principle,
such a unified treatment of baryons and quarks may be formulated by
means of dynamical hadronization with quarks and diquarks.  Further
progresses in the Dyson-Schwinger and the functional RG methods with
inclusion of diquarks are indispensable for future studies.

The completion of the whole QCD phase diagram requires mutual
collaborations extending over nuclear physics, high-energy particle physics 
and super-string theories, assisted by  condensed-matter physics and
astrophysics.  This is such an interdisciplinary research area and new
ideas are always desired.

\section*{Acknowledgments}

The authors thank Bengt~Friman, Masayasu~Harada, Larry~McLerran,
Krzysztof~Redlich, Mannque~Rho, and Wolfram~Weise for fruitful
collaborations.
They also thank Marco~Panero, Giorgio~Torrieri, and Naoki~Yamamoto
for useful comments.
K.~F.\ is grateful to Jens~Braun,
Yoshimasa~Hidaka, Toru~Kojo, Pablo~Andres~Morales, Shin~Nakamura,
Jan~Pawlowski, Bernd-Jochen~Schaefer, and Vladimir~Skokov for useful
discussions.
C.~S.\ thanks Alexei~Larionov for useful discussions.
The work of C.~S.\ has been partially supported by the Hessian
LOEWE initiative through the Helmholtz International
Center for FAIR (HIC for FAIR).
K.~F.\ was supported by JSPS KAKENHI Grant Number 24740169.


\bibliographystyle{h-physrev4}
\bibliography{fuku,sasaki}

\end{document}